\documentclass[12pt,reprint,aip,jcp,floatfix]{revtex4-2}
\usepackage[english]{babel}
\usepackage[usenames,dvipsnames]{color}
\usepackage{graphics,graphicx}
\usepackage[symbol]{footmisc}
\usepackage{parskip}
\usepackage{amsmath}
\usepackage{amssymb}
\usepackage{physics}
\usepackage{float}
\usepackage{subcaption}
\usepackage[utf8]{inputenc}
\usepackage[title]{appendix}
\usepackage{hyperref}

\begin{document}

\title{A self-consistent field approach for the variational quantum eigensolver: orbital optimization goes adaptive}

\author{Aaron Fitzpatrick}
\affiliation{Algorithmiq Ltd, Kanavakatu 3C, FI-00160 Helsinki, Finland}
\author{Anton Nyk{\"a}nen}
\affiliation{Algorithmiq Ltd, Kanavakatu 3C, FI-00160 Helsinki, Finland}
\author{N. Walter Talarico}
\affiliation{Algorithmiq Ltd, Kanavakatu 3C, FI-00160 Helsinki, Finland}
\author{Alessandro Lunghi}
\affiliation{School of Physics, AMBER and CRANN Institute, Trinity College, Dublin 2, Ireland}
\author{Sabrina Maniscalco}
\affiliation{Algorithmiq Ltd, Kanavakatu 3C, FI-00160 Helsinki, Finland}
\author{Guillermo Garc{\'i}a-P{\'e}rez}
\affiliation{Algorithmiq Ltd, Kanavakatu 3C, FI-00160 Helsinki, Finland}
\author{Stefan Knecht}
\email{stefan@algorithmiq.fi}
\affiliation{Algorithmiq Ltd, Kanavakatu 3C, FI-00160 Helsinki, Finland}

\begin{abstract}
We present a self consistent field approach (SCF) within the Adaptive Derivative-Assembled Problem-Tailored Ansatz Variational Quantum Eigensolver (ADAPT-VQE) framework for efficient quantum simulations of chemical systems on near-term quantum computers. To this end, our ADAPT-VQE-SCF approach combines the idea of generating an ansatz with a small number of parameters, resulting in shallow-depth quantum circuits with a direct minimization of an energy expression which is correct to second order with respect to changes in the molecular orbital basis. 
Our numerical analysis, including calculations for the transition metal complex ferrocene (Fe$\rm (C_5H_5)_2$), indicates that convergence in the self-consistent orbital optimization loop can be reached without a considerable increase in the number of two-qubit gates in the quantum circuit by  comparison to a VQE optimization in the initial molecular orbital basis. Moreover, the orbital optimization can be carried out \textit{simultaneously} within each iteration of the ADAPT-VQE cycle.
ADAPT-VQE-SCF thus allows us to implement a routine analogous to CASSCF, a cornerstone of state-of-the-art computational chemistry, in a hardware-efficient manner on near-term quantum computers.
Hence, ADAPT-VQE-SCF paves the way towards a paradigm shift for quantitative quantum-chemistry simulations on quantum computers by requiring fewer qubits and opening up for the use of large and flexible atomic orbital basis sets in contrast to earlier methods that are predominantly based on the idea of full active spaces with minimal basis sets.
\end{abstract}

\date{\today}


\maketitle

\onecolumngrid

\section{Introduction}\label{sec:introduction}

Multiconfigurational self-consistent-field (MCSCF) theory \cite{szal12}
and, in particular its most popular subclass coined as complete active space SCF (CASSCF) \cite{,roos80a,olse11}, both represent cornerstones of
modern computational quantum chemistry \cite{roos16a} because they make it 
possible to efficiently account for the static and non-dynamical part
of the electron correlation energy in ground- and electronically excited 
states of molecular systems \cite{roos08a}\ formulated in either a 
state-specific or state-average formalism \cite{helm22a}.   

Conceptually, the central idea of the CASSCF \textit{ansatz} is as follows. At the outset, given a finite molecular orbital (MO) basis\ $\{\varphi_p\}$, expanded as linear combinations (with MO cofficients $C_{\mu p}$) of the atomic orbital basis
functions $\{\chi_\mu\}$\, 
\begin{equation}
    \varphi_p \equiv \left|p\right> = \sum_\mu C_{\mu p} \chi_\mu\ ,
    \label{eq:MObasis}
\end{equation}
one needs to identify a suitable active space (AS) consisting of $N$\ electrons in $L$ molecular orbitals, 
dubbed in shorthand notation as CAS($N,L$) within which we then seek to find an exact solution of the electronic Schr{\"o}dinger equation. To this end, one expands the CASSCF wave function $\left|0\right>$\ as a \textit{complete}, linear set of all possible (spin-adapted) Slater determinants $\{\phi_I\}$\ that can be constructed from a distribution of $N$\ electrons among those $L$ orbitals
\begin{equation}
    \left|0\right> = \sum_I c_I \left|\phi_I\right>\ ,
\label{eq:ci-ansatz}
\end{equation}
with $c_I$ being the so-called configuration interaction (CI) coefficients. We are denoting a single-electron wave function (that is the $p$-th MO) as $\left|p\right>$\ whereas we will refer to the $N$-electron wave function (consisting of\ $2L$\ spin-orbitals) by $\left|0\right>$. 
Then, based on the above full CI (FCI) wave function \textit{ansatz} within the CAS (Eq.~\eqref{eq:ci-ansatz}), solving the corresponding  Schr{\"o}dinger equation by means of a variational optimization of the total energy $E$\ yields simultaneously an optimal set of both MO and wave function coefficients tailored to the chosen CAS, 
\begin{equation}
    E = \min_{\boldsymbol{\kappa}, \boldsymbol{S}} \left<0\right|\hat{H}\left|0\right>\ ,
    \label{eq:cas-opt}
\end{equation}
where we describe variations in the MO and CI coefficients, respectively, by means of real, anti-symmetric orbital rotation and state rotation parameters $\boldsymbol{\kappa}$\ and 
$\boldsymbol{S}$, respectively. 
For example, by assuming that the initial set of MOs $\{\varphi_p\}$ is orthonormal, making use of a unitary transformation of the MOs   
\begin{equation}\label{eq:mo2mo_coeff}
\left|{q}^{}\right> = \sum_p \left|p\right> U_{pq}\ .
\end{equation}
where \textbf{U}=$\{U_{pq}\}$\ can be expressed in exponential form \cite{thou61,levy69a}
\begin{equation}\label{eR}
 {\rm \textbf{U}(\boldsymbol{\kappa})} = e^{\rm \boldsymbol{\kappa}} =1 +  {\rm \boldsymbol{\kappa}} + \frac{1}{2} {\rm \boldsymbol{\kappa}}{\rm \boldsymbol{\kappa}} + \cdots\ ,
\end{equation}
and $\boldsymbol{\kappa}$ is an antihermitean matrix (i.e. ${-\boldsymbol{\kappa}} = {\boldsymbol{\kappa}}^{\dagger}$)
comprising a set of independent orbital rotation parameters $\{\kappa_{pq}\}$ with $p > q$, will ensure that orthonormality of the (partially) optimized MOs is kept at every step throughout the optimization process. That is, we can express the orbital-optimized wave function $\left|\tilde{0}\right>$\ as,    
\begin{equation}
    \left|\tilde{0}\right> \equiv \boldsymbol{U}(\boldsymbol{{\kappa}})\left|{0}\right> =  \exp{\boldsymbol{{\kappa}}} \left|{0}\right>\ ,
    \label{eq:exp-MO}
\end{equation}
with elements $\kappa_{pq}$\ of the anti-hermitian matrix given by 
\begin{equation}
    \boldsymbol{{\kappa}} = \sum_{p>q} \kappa_{pq}\left(\hat{E}_{pq} - \hat{E}_{qp}\right)\ ,
    \label{eq:exp-MO-kappa}
\end{equation}
and the single-excitation operators $\hat{E}_{pq}$\ being defined as \cite{helg00},
\begin{equation}
    \hat{E}_{pq} = \hat{a}_{p\alpha}^{\dagger} \hat{a}_{q\alpha} + \hat{a}_{p\beta}^{\dagger} \hat{a}_{q\beta}\ .
\end{equation}
In passing, we note that similar considerations hold with respect to the state rotation parameters $\boldsymbol{S}$ that enter the optimization expression in Eq.~\eqref{eq:cas-opt}. In this context, besides an exponential parameterization \cite{yeag79,dalg79,olse85,helg86a,helg00} analogous to the one in Eq.~\eqref{eq:exp-MO}, the most common parameterization for the CI rotation employed in a multi-configurational wave function optimization algorithm is of a linear form \cite{helg00}.

The inherent exponential growth of the CI expansion in Eq.~\eqref{eq:ci-ansatz}\ with respect to increasing values of $N$\ and $L$\ in a traditional CAS($N$,$L$) framework has sparked in the past two decades the development of various ``classical" quantum-chemical approaches such as the density matrix renormalization group \cite{wout14,knec16a,chan16a,baia20a}, stochastic FCI Quantum Monte Carlo (FCIQMC) \cite{boot09,over14,lima16c,guth20a}, a deterministic FCIQMC alternative \cite{tubm16a}, (stochastic) selective CI \cite{liu2014_sds,liuw16a,scem18a,garn18a}, in particular Heat-Bath CI \cite{holm16a,smit17a,liju18a} to name just a few key players. Harnessing the combined full potential of some of the latter approaches, Umrigar, Chan\ and co-workers eventually succeeded in solving \emph{quantitatively} a long-standing challenge in quantum chemistry, namely the computation of the complex electronic structure and unusual potential energy curve of the chromium dimer with spectroscopic accuracy \cite{lars22a}. Despite such remarkable progress, reliable quantum-chemical simulations of, for example, (polynuclear) transition metal complexes, that are naturally occurring as co-factors in eucaryotic enzymes still pose a profound challenge for ``classical" multiconfigurational quantum chemistry \cite{goin22a,lees22a} while holding great promise for a successful application of quantum computing for quantum chemistry \cite{caoy19a,clau22a}.

Albeit still being in the era of near-term quantum computing devices equipped with at most a few hundred qubits that do not yet allow to implement robust error correction schemes \cite{dalt22a}, 
many promising quantum algorithms for reliably finding a ground or excited state energy of a given quantum (chemical) system have been proposed in recent years \cite{caoy19a}\ for quantum chemistry experiments on quantum computers. 
In order to circumvent prohibitive resource requirements associated with the most prominent quantum algorithm, namely the quantum phase estimation algorithm (QPE) \cite{kita95a,aspu05a}, 
the development of hybrid quantum-classical algorithms, notably of the Variational Quantum Eigensolver (VQE) \cite{peru14a,till22a,fedo22a}\ has received considerable attention both from theory and experimental communities. 
At the core, given a (quantum-chemical) Hamiltonian $\hat{H}$\ the aim of the variational VQE algorithm is to provide a(n approximate) solution to an eigenvalue problem by combining classical optimization techniques with a parametrized wave function approach, the latter being realized as a (shallow) quantum circuit on a quantum processor, 
commonly referred to as \textit{ansatz}. 
For example, by expanding our CAS wave function $\left|0\right>$\ in Eq.~\eqref{eq:ci-ansatz}\ on a quantum computer by means of a set of parameterised unitaries $\{\hat{U}_i(\theta_i)\}$,
\begin{equation}
    \ket{0}  = \prod_{i} \hat{U_{i}}(\theta_{i})[\bigotimes_{0}^{2L-1} \ket{\mathbf{0}}]\ ,
    \label{eq:qc-wave-ansatz}
\end{equation}
where $\ket{\mathbf{0}}$\ denotes the initial state of the $2*L$\ qubits and with $\{\theta_i\}$\ referring to a set of parameters for the unitary rotations taking values in $\left(-\pi,\pi\right]$, enables us to  
formulate the optimization problem in Eq.~\eqref{eq:cas-opt} as a variational optimization of the total energy $E$\ wrt\ the set of orbital rotation parameters $\boldsymbol{\kappa}$\ and the unitary qubit rotation angles $\boldsymbol{\theta}$, 
\begin{equation}\label{eq:vqe_cas_opt}
     E = \min_{\boldsymbol{\kappa}, \boldsymbol{\theta}} \left<0\right|U^\dagger(\boldsymbol{\theta})\hat{H}U(\boldsymbol{\theta})\left|0\right>\ .
\end{equation}
The VQE algorithm therefore relies on a suitable selection of a quantum circuit built from the parameterised quantum gates $\{\hat{U}_i(\theta_i)\}$\ where the rotation angles $\{\theta_i\}$\ are stepwise optimized on a classical computer 
while the evaluation of the energy $E$ and/or its derivative wrt wave function parameters needed for the classical optimizer 
is being realized through measurements on the quantum processor. 
In brief, the essential steps in a VQE approach for a quantum-chemical optimization problem are \cite{till22a,fedo22a}: 
(i) translate the Hamiltonian $\hat{H}$\ from a fermion representation to a qubit one, 
(ii) choose a suitable \textit{ansatz}, that is set up an (initial) quantum circuit,
(iii) perform the necessary measurements on the quantum computer to evaluate the expectation value in Eq.~\eqref{eq:vqe_cas_opt}, 
(iv) classically vary the parameters that define the action of the \textit{ansatz} on the qubit register to find the optimal parameters that minimize the expectation value in Eq.~\eqref{eq:vqe_cas_opt},  
where the variational principle then ensures that the energy in the optimized state $\ket{0}$\ is an upper bound to the ground state energy. 

Much work has been done in recent years on establishing a practical VQE framework focusing on finding \textit{adaptive}\ algorithms that entail minimal entangling gate requirements 
and lower circuit depths \cite{till22a,fedo22a}, such as, for example, the so-called ADAPT-VQE \cite{grim19a}, qubit-ADAPT-VQE \cite{tang21a} and TETRIS-ADAPT-VQE \cite{anas22a}.
In this work, we will add a pivotal milestone to the aforementioned \textit{adaptive}\ VQE algorithms\ by combining the adaptive expansion of the quantum circuit and ensuing optimization of its accompanying $\{\theta_i\}$\ parameter set 
with a simultaneous optimization of the orbital rotation parameters $\boldsymbol{\kappa}$\ in the variational energy minimization as outlined in Eq.~\eqref{eq:vqe_cas_opt}. 
In contrast to related recent works within the framework of near-term quantum-computing driven MCSCF-type approaches \cite{take20a,mizu20a,soko20a,yalo21a,PhysRevResearch.3.033230,goch22a,bierman2022improving,omiy22a}, our adaptive, self-consistent-field VQE algorithm, 
denoted in the following as ADAPT-VQE-SCF, greatly benefits from a considerable reduction in the number of two-qubit gates in the final ansatz that are required to reach convergence wrt chemical precision. 
Moreover, we not only provide a \textit{state-specific} energy minimization (Eq.~\eqref{eq:vqe_cas_opt}) ADAPT-VQE-SCF algorithm but also a state-average one that opens up for an optimization of a common set of MOs for a multitude of states $i$, that is,
\begin{equation}
E = \min_{\boldsymbol{\kappa}, \boldsymbol{\theta}} \sum_{i} \lambda_i \left<0_i\right|U^\dagger(\boldsymbol{\theta})\hat{H}U(\boldsymbol{\theta})\left|0_i\right>\ ,
    \label{eq:eq:vqe_cas_opt_sa}
\end{equation}
where $\lambda_i$\ is a pre-selected, fixed weighting factor for the $i$-th state that satisfy the constraint $\sum\limits_i \lambda_i = 1\ \forall \lambda_i \in [0,1]$\ but are arbitrary otherwise. Note that by making suitable choises for the set of states $i$, one could, for example, target a simultaneous optimization of 
the ground state as well as a number of excited states of the same spin/spatial symmetry or aim for an optimization of electronic states spanning over different spin/spatial symmetries.

This paper is organized as follows: On the basis of the work of Sun, Yang and Chan \cite{sunq17a}, the central
element of our algorithm is the coupled optimization of the (gradually growing) adaptive quantum circuit 
and the MOs to obtain a quadratically convergent ADAPT-VQE-SCF 
approach. 
In Section \ref{sec:theory:mcscf}, we briefly outline our orbital optimization
approach for a VQE-optimized \textit{ansatz}\ and discuss potential choices for an adaptive \textit{ansatz}\ in Section \ref{sec:theory:vqe}\ 
before providing in Section \ref{sec:theory:vqe-scf}\ a short summary of our proposed ADAPT-VQE-SCF approach. 
Numerical examples highlighting the potential of state-specific as well as state-averaged ADAPT-VQE-SCF \textit{ans\"atze}\ are presented in Section \ref{sec:results}.

\section{Theory}\label{sec:theory}

\subsection{Orbital Optimization}\label{sec:theory:mcscf}

\begin{figure}
    \centering
    \includegraphics[width = 1.0\textwidth]{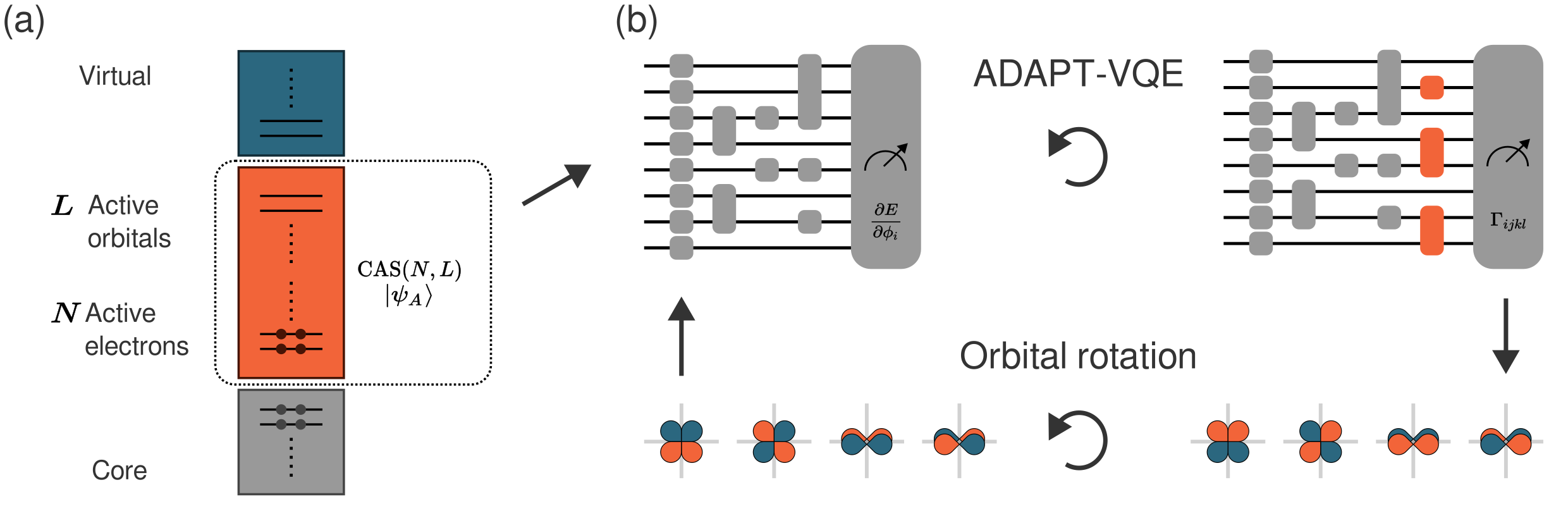}
    \caption{(a) MO energy level diagram for a generic CAS($N,L$) MO partitioning with index labels for the resulting inactive (core), active, and virtual (secondary) MO subspaces. (b) Graphical representation of the VQE-SCF algorithm.}
    \label{fig:vqe-scf}
\end{figure}

In this section, we briefly summarize the orbital optimzation algorithm for a general multiconfigurational CAS wave function \textit{ansatz}\ as proposed by Chan and co-workers \cite{sunq17a}\ and implemented in \textsc{PySCF} \cite{sunq17b,sunq20a}. Hence, our discussion in this work remains restricted to the essential steps necessary to illustrate the key features of our optimization scheme within a framework of adaptive VQE \textit{ans\"atze}. 
As depicted in Fig.~\ref{fig:vqe-scf}(a), we consider a partitioning of the MO space in ``inactive" or ``core"\ orbitals which remain doubly-occupied in any \textit{ansatz}\ considered whereas (partially) occupied MOs are denoted as ``active" MOs. The ``secondary" or ``virtual" orbitals are considered to be empty. Moreover, we assume that our quantum circuit maps a real wave function which enables us to restrict the orbitals to be real. In addition, in the ``classical" orbital optimizer we adopt a spin-restricted formalism such that each spatial orbital can be occupied by up to two electrons with opposite spin, that is $\tau = \{\alpha,\beta\}$.


Our starting point for the derivation of an orbital-optimization approach for an adapative VQE-SCF \textit{ansatz} is the full (non-relativistic) electronic Hamiltonian $\hat{H}$, 
\begin{align}\label{eq:Hami_op}
\hat{H} = & \sum_{p,q,\tau}^{} (p |h| q)\ \hat{a}_{p\tau}^{\dagger} \hat{a}_{q\tau}+ \frac{1}{2} \sum\limits_{\substack{p,q,r,s\\ \tau,\tau^{\prime}}}^{} (pq|rs)\ \hat{a}_{p\tau}^{\dagger} \hat{a}_{r\tau^{\prime}}^{\dagger} \hat{a}_{s\tau^{\prime}} \hat{a}_{q\tau}\ \nonumber \\
 = & \sum_{p,q}^{} h_{pq}\ \hat{E}_{pq} + \frac{1}{2} \sum\limits_{{p,q,r,s}}^{} (pq|rs)\ \left(\hat{E}_{pq}\hat{E}_{rs} - \delta_{qr}\hat{E}_{ps}\right)\,
\end{align}
where the parameters
\begin{equation}\label{eq:1eINT}
(p |h| q) = \int \varphi_p^*({\rm \textbf{r}}) \hat{h} \varphi_q({\rm \textbf{r}})d{\rm \textbf{r}}
\end{equation}
and 
\begin{equation}\label{eq:2eINT}
(pq|rs) = \int \varphi_p^*({\rm \textbf{r}}_1)\varphi_r^*({\rm \textbf{r}}_2) r_{12}^{-1} \varphi_q({\rm \textbf{r}}_1)\varphi_s({\rm \textbf{r}}_2)d{\rm \textbf{r}}_1 d{\rm \textbf{r}}_2
\end{equation}
are the usual one- and two-electron integrals expressed in orthonormal MO basis $\{\varphi_p\}$,  i.e., $\left<p\left.\right|q\right> = \delta_{pq}\ \forall\ p,q$, and the definition of the MOs $\{\varphi_p\}$ in terms of the underlying atomic orbital basis function $\{\chi_\mu\}$\ is given in Eq.~\eqref{eq:MObasis}.

For convenience, we further rewrite the Hamiltonian in Eq.~\eqref{eq:Hami_op} as 
\begin{equation}\label{eq:Hami_op_rw}
\hat{H} 
 =  \frac{1}{2} \sum\limits_{{p,q,r,s}}^{} {}^{2} K_{qs}^{pr}  \left(\hat{E}_{pq}\hat{E}_{rs} - \delta_{qr}\hat{E}_{ps}\right)\ ,
\end{equation}
where we introduced the two-particle reduced Hamiltonian matrix $^2\boldsymbol{K}$\ whose elements are given by, 
\begin{equation}\label{eq:K_red}
    {}^{2} K_{qs}^{pr} = (pq|rs) + \frac{1}{N-1} \left(
 h_{pq} \delta_{rs} + h_{rs}  \delta_{pq}\right)\ .
\end{equation}
Here, $N$\ specifies the number of active electrons considered in the CAS($N,L$)\ MO partitioning scheme (see also Fig.~\ref{fig:vqe-scf}(a)).
Since the resulting $N$-particle Hamiltonian $\hat{H}$\ comprises 
only up to two-electron interactions, the optimization problem stated in Eq.~\eqref{eq:vqe_cas_opt}\ can be reformulated such that it involves only the two-electron reduced density matrix $\boldsymbol{\Gamma}$\ (2-RDM)\ as well as the unitary orbital rotation matrices $\{U_{pi}\}$, 
\begin{eqnarray}
E &=& \min_{\boldsymbol{\kappa}, \boldsymbol{\theta}} \left<0\right|U^\dagger(\boldsymbol{\theta})\hat{H}U(\boldsymbol{\theta})\left|0\right>\ \nonumber \\
    & =& \min_{\boldsymbol{\kappa}, \boldsymbol{\theta}} \left(\sum_{i,j,k,l} {}^2K_{jl}^{ik}\ \Gamma_{ijkl} \right) \nonumber \\
    & =& \min_{\boldsymbol{\kappa}, \boldsymbol{\theta}} \left(\text{Tr}\left({}^2\boldsymbol{K}\ \boldsymbol{\Gamma} \right) \right)
\end{eqnarray}\label{eq:vqe_cas_opt_uu}
with the elements $\Gamma_{ijkl}$\ of the 2-RDM given by
\begin{equation}\label{eq:vqe_cas_2rdm}
     \Gamma_{ijkl} =  \frac{1}{2!}\left<0\right|U^\dagger(\boldsymbol{\theta})\left(\hat{E}_{ij}\hat{E}_{kl} - \delta_{jk}\hat{E}_{il}\right)U(\boldsymbol{\theta})\left|0\right>\ ,
\end{equation}
and elements of the (orbital-optimized) two-particle reduced Hamiltonian matrix expressed as 
\begin{equation}\label{eq:K_red_opt}
    {}^{2} K_{jl}^{ik} = \sum_{p,q,r,s} {}^{2} K_{qs}^{pr}  U_{pi}U_{qj}U_{rk}U_{sl}\ .
\end{equation}

Following Sun \textit{et al.}\ \cite{sunq17a}, we next define a Lagrangian $\mathcal{L}$\ with constraints on $\boldsymbol{\theta}$ and $\mathbf{U}$
\begin{equation}
    \begin{aligned}
\mathcal{L}(\boldsymbol{\kappa}, \boldsymbol{\theta}) &=E(\boldsymbol{\kappa}, \boldsymbol{\theta})-\mathcal{E}\left(\boldsymbol{\theta}^{\dagger} \boldsymbol{\theta}-1\right)\ ,
\end{aligned}
\end{equation}
such that the minimizing the energy $E$\ turns into a non-linear optimization problem for the orbital rotation parameters $\boldsymbol{\kappa}$\ and the unitary qubit rotation angles $\boldsymbol{\theta}$, respectively. The stationary conditions are then obtained when the gradient of the Lagrangian $\mathcal{L}$\ vanishes, that is 
\begin{eqnarray}
    \frac{\partial \mathcal{L}}{\partial \theta_i} &=& 0\ \\
    \frac{\partial \mathcal{L}}{\partial \kappa_{pq}} &=& 0\ \ .
\end{eqnarray}
In order to ensure a fast and optimum convergence, we strive for a second-order Newton-Raphson procedure to find the stationary points of $\mathcal{L}(\boldsymbol{\kappa}, \boldsymbol{\theta})$\ which takes into account a coupled optimization of the wave function and orbital rotation parameters. To this end, the resulting Newton-Raphson equations assume the form \cite{helg00,sunq17a},  
\begin{equation}
    \left(\begin{array}{ll}
\mathcal{H}^{c c} & \mathcal{H}^{c o} \\
\mathcal{H}^{o c} & \mathcal{H}^{o o}
\end{array}\right)\left(\begin{array}{l}
\boldsymbol{\theta}^1 \\
\boldsymbol{\kappa}^1
\end{array}\right)+\left(\begin{array}{l}
\mathcal{G}^c \\
\mathcal{G}^o
\end{array}\right)=0\ ,
\label{eq:NR-opt}
\end{equation}
where $\mathcal{G}^c \equiv \frac{\partial \mathcal{L}}{\partial \theta_i}$\ and $\mathcal{G}^o \equiv \frac{\partial \mathcal{L}}{\partial \kappa_{pq}}$\ denote the configurational ($c$)\ and orbital ($o$)\ gradient terms, respectively. Note, that the matrix of second derivatives (Hessian) $\mathcal{H}$\ in Eq.~\eqref{eq:NR-opt}\ comprises, besides the diagonal orbital-orbital ($oo$)\ and configurational-configurational $(cc)$\ 
parts, in addition off-diagonal $oc$ ($co$) terms consisting of mixed derivatives which read as 
\begin{equation}\label{eq:hessian-mixed}
    \mathcal{H}^{o c}_{I,pq} = \mathcal{H}^{c o}_{pq,I} =\frac{\partial^2\ \mathcal{L}}{\partial \kappa_{p q}\ \partial \theta_{I}} = \frac{\partial\ {}^2K_{j l}^{ik}}{\partial \kappa_{p q}} \frac{\partial\ \Gamma_{i j k l}}{\partial \theta_I}\ .
\end{equation} 
Proceeding with further steps as detailed in Ref.~\citenum{sunq17a} and in agreement with the related second-order orbital optimization approaches based on a CI \cite{know85a}\ as well as DMRG wave function expansion \cite{mayi16c}, the configurational correction vector $\boldsymbol{\theta^{1}}$\ of a Newton step in Eq.~\eqref{eq:NR-opt}\ can be approximated in an iterative, Davidson-type approach, that is, each configurational micro iteration reads as 
\begin{equation}\label{eq:theta-approx}
    \boldsymbol{\theta}^1 \approx - [\text{diag}(\boldsymbol{{}^2K} - E_0)]^{-1}\  \boldsymbol{{}^2K}^{\kappa} \boldsymbol{\theta}^0\ . 
\end{equation}
As indicated in Eq.~\eqref{eq:theta-approx}, the latter merely requires the application of one-index transformed elements of the two-particle reduced Hamiltonian ${}^2K_{jl}^{ik, \kappa}$, 
\begin{equation}\label{eq:Htilde-approx}
{}^2K_{jl, \kappa}^{ik} = 
{}^2K_{jl}^{pk} \kappa_{pi}^{1} + 
{}^2K_{pl}^{ik} \kappa_{pj}^{1} + 
{}^2K_{jl}^{ip} \kappa_{pk}^{1} + 
{}^2K_{jp}^{ik} \kappa_{pl}^{1}\ ,
\end{equation}
where the one-index transformation is based on the current approximation to the orbital rotation correction vector $\boldsymbol{\kappa}^1$. Since the coupling block $\mathcal{H}^{o c}$\ introduces a dependency for $\boldsymbol{\kappa}^1$\ on the first-order correction to the two-particle density matrix ${}^1\Gamma_{ijkl}$\ \cite{sunq17a},
\begin{equation}\label{eq:hoc-coupling}
    \mathcal{H}^{oc} \boldsymbol{\kappa}^1 = - \tilde{\mathcal{G}}^{o}\ ,
\end{equation}
with elements of $\tilde{\mathcal{G}}^{o}$\ given by 
\begin{equation}\label{eq:tilde-gopq}
\tilde{\mathcal{G}}^{o}_{pq} = {\mathcal{G}}^{o}_{pq} + \mathcal{H}^{oc} \boldsymbol{\theta}^1 = {\mathcal{G}}^{o}_{pq} + \frac{\partial\ {}^2K_{jl}^{ik}}{\partial \kappa_{pq}} \Gamma_{ijkl}^1\ ,
\end{equation}
$\Gamma_{ijkl}^1$\ can conveniently be approximated based on a finite difference evaluation as 
\begin{equation}\label{eq:gamma1}
    \Gamma_{ijkl}^1 \approx\ {}\Gamma_{ijkl}[\boldsymbol{\theta}^0 + \boldsymbol{\theta}^1] - [\boldsymbol{\theta}^0]\ ,
\end{equation}
following the iterative update of $\boldsymbol{\theta}^1$. 
Although the outlined procedure paves the way for a coupled configurational-orbital optimization 
which does not require to genuinely solve a CI-like response equation, the iterative solution of the coupled equations necessitate repeated calculations of (two-particle) RDMs as indicated in Eq.~\eqref{eq:gamma1}. Hence, in order to minimize a potential measurement overhead when sampling the RDMs based on the prepared circuit on the quantum computer, we take advantage of our adaptive informationally complete generalized measurement scheme described in full detail in Refs.~\citenum{garcia2021learning}\ and \citenum{glos22a}, respectively. 
In contrast to the above discussed coupled \textit{one-step} algorithm, neglecting the off-diagonal Hessian blocks leads to a so-called \textit{two-step} algorithm in which Newton steps for the orbital rotations and CI problem are carried out in an alternate fashion. 
Though the 2-step algorithm is computationally cheaper, it commonly suffers from slower convergence, and we did not consider it further in the molecular applications discussed in Section \ref{sec:results}. 

Given a (fermionic) electronic Hamiltonian $\hat{H}$\ (see Eq.~\eqref{eq:Hami_op}), the preparation of a suitable (CAS) wave function $\left|0\right>$\ on a quantum computer as shown in Eq.~\eqref{eq:qc-wave-ansatz}\ by means of a VQE\ algorithm (see Section \ref{sec:theory:vqe})\ as well as the corresponding calculation of the 2-RDM elements $\Gamma_{ijkl}$\ in Eq.~\eqref{eq:vqe_cas_2rdm}\ requires to map the fermionic Fock space operators 
onto the Hilbert space of a register of qubits. In practice, a number of such \textit{fermion-to-qubit}\ mappings exist for qubit representations of many-body problems, see for example Ref.~\citenum{baue20a}\ and references therein. 
In a forthcoming work, some of us present a formalism to design almost arbitrary, yet hardware-tailored valid mappings taking advantage of a ternary tree based algorithm \cite{mill22a}. 
In what follows, we employ two of the most common mappings, namely the Jordan-Wigner \cite{jordan1928pauli}\ as well as the Bravyi-Kitaev transformation \cite{bravyi2002fermionic}, respectively, 
both of which are implemented in the Qiskit open-source library for quantum computing\ \cite{qisk21a}. 
For a detailed discussion of the essential concepts that form the basis of the latter two popular mappings as well as their accompanying benefits/drawbacks\ in connection with the resulting qubit representation of the many-body quantum-chemical  Hamiltonian $\hat{H}$, we refer the interested reader to the work of Love and co-workers \cite{seel12a}. 

\subsection{VQE}\label{sec:theory:vqe}

In order to minimize the CAS energy expectation value in Eq.~\eqref{eq:vqe_cas_opt_uu}\ on a quantum computer, we represent our CAS wave function $\left|0\right>$\ by means of a set of parameterised unitaries ${\hat{U}_i(\theta_i)}$\ as outlined in Eq.~\eqref{eq:qc-wave-ansatz}\ where the latter are subsequently optimized in a VQE framework. As pointed out in the Introduction, the performance and rapid convergence of the VQE algorithm relies on being able to find a suitable selection of parameterised quantum gates ${\hat{U}_i(\theta_i)}$\ that form as a whole the quantum circuit or short \textit{ansatz}. In this regard, considerable efforts have been made in recent years on improving VQE in an attempt to find suitable ansätze with minimal entangling gate requirements and lower circuit depths \cite{till22a}. Out of the variety of resulting methods, the so-called ADAPT-VQE \cite{grim19a} and qubit-ADAPT-VQE \cite{tang21a} algorithms have shown particular promise and will be employed in the following for the construction of our quantum circuits. In passing we note, that some of us have combined very recently the ADAPT-VQE algorithm with informationally complete POVM measurements \cite{nyka22a} in our software platform \textsc{Aurora} \cite{aurora}. 
This allows us to efficiently mitigate the measurement overhead associated with traditional ADAPT-VQE algorithms \cite{gont22a}. 

The basic idea of the ADAPT algorithm is to build the ansatz iteratively by making use of a predefined pool of fermionic excitation operators $\{\hat{\tau}_i\}$, 
\begin{equation}\label{eq:pool}
    \{ \hat{\tau}_{i} \} = \{\left(a_{q}^{\dagger} a_{p}-a_{p}^{\dagger} a_{q}\right), \left(a_{p}^{\dagger} a_{q}^{\dagger} a_{r} a_{s}-a_{s}^{\dagger} a_{r}^{\dagger} a_{q} a_{p}\right)\}\ ,
\end{equation}
where the operator pool typically consists of all possible one- and two-electron terms defined by the system under consideration, which is mapped to qubit space with the same fermion-to-qubit mapping used for the Hamiltonian, yielding the qubit-space pool $\{ \hat{A}_i\}$. The determination of the next operator to be added to the ansatz necessitates to calculate the gradient with respect to the energy, i.e., measuring its effect on lowering the current approximation to the ground state energy, 
\begin{equation}\label{eq:energy_gradient}
   \frac{\partial {E}}{\partial \theta_{i}} = \bra{0} \comm{\hat{H}}{\hat{A}_{i}} \ket{0}\ .
\end{equation}
Here, $\theta_{i}$ are the variational parameters associated with the unitary operator $\hat{U}_i (\theta_i) = \exp(\theta_i \hat{A}_i)$ to be added to the quantum circuit at its current expansion point $\ket{0}$.
Consequently, one proceeds at this step with the operator $\hat{A}_i$\ that yields the largest gradient in Eq.~\eqref{eq:energy_gradient}. 
After the extension of the quantum circuit with the new operator $\hat{U}_i (\theta_i) = \exp(\theta_i \hat{A}_i)$, the parameters $\{\theta_i\}$\ of all operators in $\ket{0}$\ are re-optimized \cite{till22a}\ using a classical optimizer such that a new quantum circuit $\ket{0}$\ can be prepared in an ensuing step on the quantum computer. Closing the loop, the iterative procedure is continued until an exit criterion has been fulfilled. The latter could be met within preset numerical limits, for example, by a vanishing change in energy $\Delta E$\ between two consecutive parameter optimization steps or by vanishing total gradient norm of operators $\hat{A}_i$ in the pool (see Eq.~\eqref{eq:energy_gradient}).  
The hardware efficient version of ADAPT-VQE as proposed by Tang \textit{et al} \cite{tang21a}, proceeds along similar lines as the standard ADAPT-VQE outlined above. By contrast to the latter, the operator pool $\{\hat{Q}_{i}\}$\ in Qubit-ADAPT-VQE is defined in qubit rather than in fermionic space and consists of simple length $n \le 2L$\ Pauli strings $P_i$, 
\begin{equation}\label{eq:qadapt_pool}
    \hat{Q}_{i} \equiv iP_i = i \bigotimes_{j}^{n}\sigma^{p}_{j}, p \in \{x, y, z, 0\}\ ,
\end{equation}
where we have written the $i$-th Pauli string as a product of single-qubit Pauli matrices $\sigma$. 
On the one hand, qubit-ADAPT-VQE has been shown to commonly result in circuits that require fewer entangling two-qubit gates (CNOTs) to reach the ground state. On the other hand, the required  operator pool and hence the number of variational parameters is typically much larger compared to a fermionic pool in ADAPT-VQE, thus moving the computational cost away from the quantum computer towards the classical optimizer \cite{tang21a}. Further, the operators in the qubit pool have reduced physical interpretation and as such qubit-ADAPT loses some of the physical motivation inherent to the fermionic ADAPT version which may even lead to converging towards symmetry-broken solutions. 

\subsection{The ADAPT-VQE-SCF algorithm in a nutshell}\label{sec:theory:vqe-scf} 

In principle, one could use ADAPT-VQE to perform a CASSCF calculation by running the latter until convergence and then applying a single orbital optimization step. However, this would lead to very long execution times, since the costly ADAPT-VQE calculation would need to be implemented repeatedly.
Instead, our proposed ADAPT-VQE-SCF alternative circumvents this potential roadblock entirely by optimising the orbitals throughout the ansatz construction procedure.
The essential steps that define our ADAPT-VQE-SCF approach are as follows.

\begin{enumerate}
    \item Starting from a set of reference MOs, choose an active orbital space of $N$\ electrons in $L$\ orbitals, suitable to describe for the problem at hand, \textit{viz.} the electronic structure, as indicated in Fig.~\ref{fig:vqe-scf}(a). 
    \item Map the corresponding active-space Hamiltonian from fermionic to qubit space representation. 
    \item Perform an iteration of the ADAPT-VQE algorithm as described in Section \ref{sec:theory:vqe}\ and indicated on the top part of Fig.~\ref{fig:vqe-scf}(b) (that is, append $n$\ gate(s), denoted as $n$G-ADAPT\ in the following, and optimize the parameters in the ansatz). 
    \item Given the resulting quantum circuit $\ket{0}$, measure the 2-RDM elements $\Gamma_{ijkl} = \bra{0}\hat{a}_i^\dagger\hat{a}_k^\dagger\hat{a}_l\hat{a}_j\ket{0}$\ which are required as input to the orbital optimizer on the classical computer as discussed in Section \ref{sec:theory:mcscf}.
    \item After rotation of the molecular orbital basis, transform the active space Hamiltonian to this new MO basis and continue with step 2 until convergence is reached.  
\end{enumerate}

Interestingly, as we show below in Section \ref{sec:results}, even a minimal 1G-ADAPT-VQE scheme could lead in some cases not only to faster convergence but also to lower two-qubit (CNOTs) gate counts.  

\section{Results}\label{sec:results}

To validate the performance and the efficiency of our proposed ADAPT-VQE-SCF algorithm, we considered orbital optimization of CAS-type wave functions for a variety of molecular compounds. 
We assess the reliability and feasibility of our ADAPT-VQE-SCF approach for near-term quantum computers by taking into account both the resulting circuit complexity as well as accuracy by comparing for the latter to corresponding ``classical" CASSCF calculations.
Moreover, to ease the numerical analysis, we consider exclusively noiseless simulations where we assume an exact unitary matrix representation of the circuit and its application to the initial state vector of the qubit register without including any sampling or hardware noise. 

In Section \ref{sec:results:h4}, we first discuss state-specific optimizations for spin $S$\ singlet ($S=0$) as well as non-singlet ($S>0$) spin states, respectively, before turning to an analysis of state-average ADAPT-VQE-SCF data in Section \ref{sec:results:sa}. 
All calculations have been carried out within our \textsc{Aurora}\ software framework \cite{aurora}\ 
which steers the external \textsc{PySCF}~program~\cite{sunq17b,sunq20a} for the classical quantum-chemistry tasks while providing an interface to Qiskit \cite{qisk21a} for the quantum-computing tasks. If not stated otherwise, all calculations make use of a common computational setup: (i) Dunning's correlation-consistent atom-centered Gaussian-type basis sets of triple-$\zeta$ quality (cc-pVTZ, dubbed \texttt{tz}) for each unique atom type ~\cite{dunn89}, (ii) a case-by-case selected active orbital space CAS($N$,$L$), with $N$\ the number of active electrons and $L$\ the number of active spatial orbitals (see Fig.~\ref{fig:vqe-scf}(a)), and (iii) the Jordan-Wigner or Bravi-Kitaev mapping for the transformation of the fermionic active space Hamiltonian into a qubit-based representation.  

\subsection{State-specific ADAPT-VQE-SCF optimization}\label{sec:results:h4}

We begin our analysis by considering state-specific ADAPT-VQE-SCF optimizations for three representative molecular systems: 
a symmetrically stretched H$_4$\ chain 
with a common internuclear distance of $r_{\rm H-H} =1.5$ \AA\ for each connected pair of H atoms and a CAS(4,4)\ active orbital space for the multiconfigurational orbital optimization, 
LiH ($r_{\rm Li-H} = 1.6$\ \AA) with a CAS(4,4)  
as well as NO ($r_{\rm N-O} = 1.6$ \AA)\ with a CAS(4,5). 

\begin{figure}
    \begin{center}
        \includegraphics[width = 1.0\textwidth]{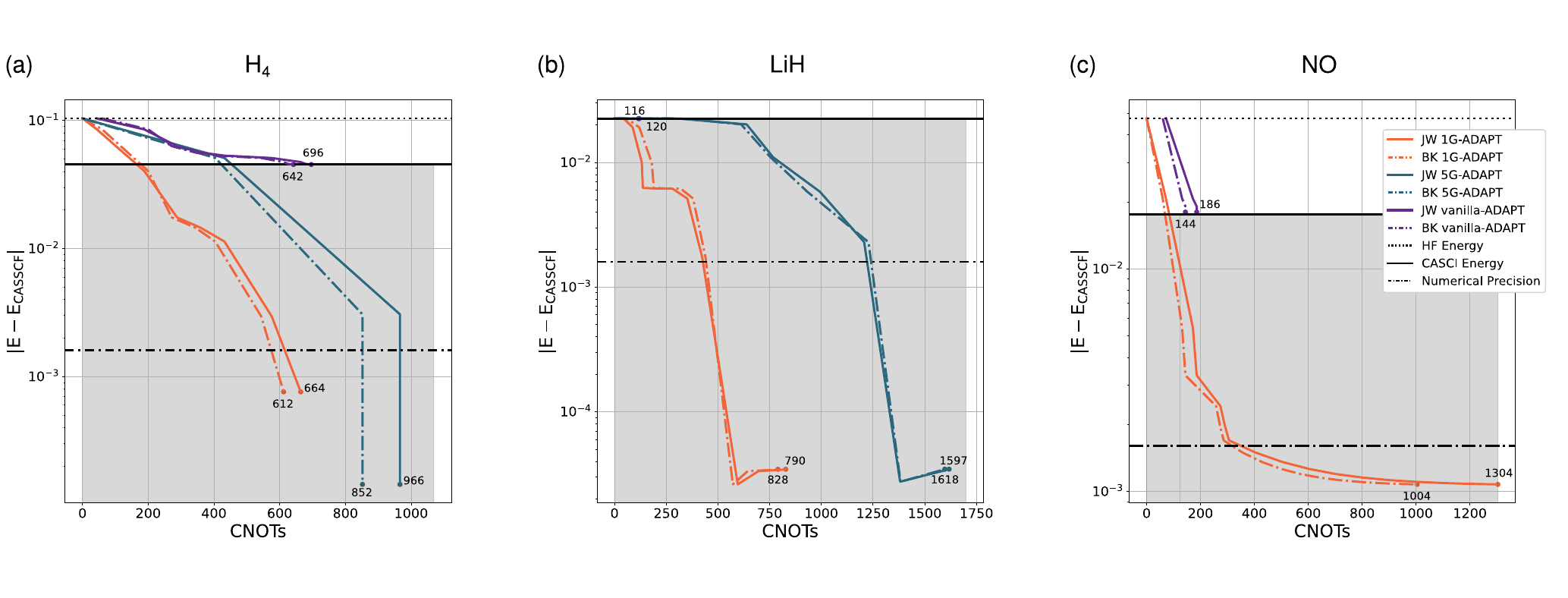}
    \end{center}
    \caption{Performance of the ADAPT-VQE-SCF algorithm for H$_4$ (panel (a)), LiH (panel (b)) and NO (panel (c)): 
    Absolute energy errors [Hartree] between the converged ADAPT-VQE-SCF energy and the reference CASSCF energy as a function of the number of CNOT gates in the converged ansatz. 
    Jordan-Wigner and Bravyi-Kitaev fermion-to-quibit mappings are denoted by solid and dashed lines, respectively. The color code indicates the method used for the VQE simulations: vanilla-ADAPT (purple), 1G-ADAPT: one gate added before carrying out a subsequent orbital rotation step (orange), 5G-ADAPT: as in 1G-ADAPT but with the addition of five gates (blue). Reference energies are indicated by a vertical dotted black (SCF energy), solid black (CASCI energy), and dashed-dotted black line (chemical precision wrt CASSCF), respectively.}
    \label{fig:fermionic_adapt}
\end{figure}

Fig.~\ref{fig:fermionic_adapt} highlights for all three molecular systems (panel (a): H$_4$, panel (b): LiH and panel (c): NO) 
the error in absolute energy (given in Hartree) between the final ADAPT-VQE-SCF energy and the corresponding reference CASSCF energy
as a function of the number of CNOT gates in the current expansion point of the ansatz. 
Starting either from a Jordan-Wigner (JW) or Bravyi-Kitaev (BK) mapping, we let the ADAPT algorithms run for a maximum of adding one (1G-ADAPT) or five gates (5G-ADAPT) before applying a unitary rotation of the MOs. Moreover, in the iteratively construction of the ansatz, 
we consider for LiH a spin-penalty scheme that 
allows us to disregard gates that do not preserve spin symmetry. 

As can be seen from Fig.~\ref{fig:fermionic_adapt}, we find that the ADAPT-VQE-SCF scheme converges within numerical precision to the classical CASSCF ground state energy for all the molecules, regardless of the underlying fermion-to-qubit mapping as well as $n$G-ADAPT scheme.
Interestingly, by allowing the ADAPT algorithm to append merely a single gate (1G-ADAPT) before performing an orbital rotation (orange lines in Fig.~\ref{fig:fermionic_adapt}) results not only in a shallower circuit but also leads to a faster convergence in contrast to appending multiple gates in the 5G-ADAPT optimization scheme (blue lines in Fig.~\ref{fig:fermionic_adapt}). 
This can be seen as a clear signature for the importance of relaxing the initial MO basis in a multiconfigurational wave function approach in a quantum computer experiment, which, although this is commonly known for ``classical" CAS-based methods, counterexamples have been discussed \cite{levi21a}\ for the particular case of CASCI. To underline our findings for ADAPT-VQE, we additionally performed vanilla-ADAPT-VQE optimizations (no MO optimisation, purple lines in Fig.~\ref{fig:fermionic_adapt}). Albeit converging correctly to the corresponding standard CASCI energies, indicated by black solid lines in the latter figure, significantly more CNOTs are required to achieve a similar accuracy as compared to the ADAPT-VQE-SCF cases to approach their CASSCF limit, in particular for the stretched H$_4$\ system. 
Note that for LiH, the initial SCF energy and corresponding CASCI energy for the considered CAS(4,4) are so close such that the convergence of the vanilla-ADAPT-VQE is barely visible on the logarithmic scale in Fig.~\ref{fig:fermionic_adapt}(b). 
Finally, we turn to a comparison of the choice of fermion-to-qubit mapping as the qubit-mode basis for the \textit{fermionic} ADAPT-VQE-SCF optimization scheme. Interestingly, for all three molecular systems under consideration, both mappings lead to a similar final CNOT count to achieve numerical convergence wrt the target CASSCF energy limit, despite their significantly different scaling in the Pauli weight of the qubit representation of the CAS Hamiltonian \cite{seel12a}.    

\begin{figure}
    \centering
    \includegraphics[width = 1.0\textwidth]{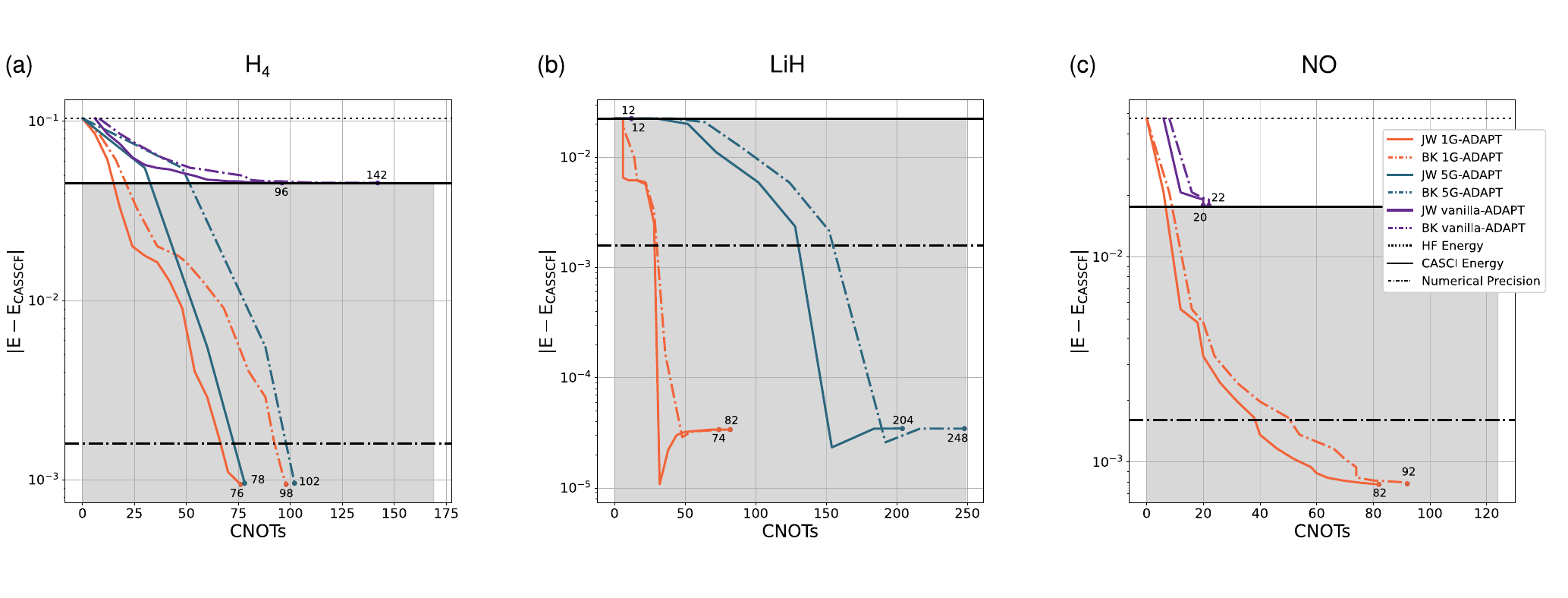}
    \caption{Performance of the Qubit-ADAPT-VQE-SCF algorithm for H$_4$ (panel (a)), LiH (panel (b)) and NO (panel (c)): 
    Absolute energy errors [Hartree] between the converged ADAPT-VQE-SCF energy and the reference CASSCF energy as a function of the number of CNOT gates in the converged ansatz. 
    Jordan-Wigner and Bravyi-Kitaev fermion-to-quibit mappings are denoted by solid and dashed lines, respectively. The color code indicates the method used for the VQE simulations: vanilla ADAPT (purple), 1G-ADAPT: one gate added before carrying out a subsequent orbital rotation step (orange), 5G-ADAPT: as in 1G-ADAPT but with the addition of five gates (blue). Reference energies are indicated by a vertical dotted black (SCF energy), solid black (CASCI energy), and dashed-dotted black line (chemical precision wrt CASSCF), respectively.}
    \label{fig:qubit_adapt}
\end{figure}

In analogy to Fig.~\ref{fig:fermionic_adapt}\ for the \textit{fermionic} ADAPT-VQE approach, we summarize in Fig.~\ref{fig:qubit_adapt} the absolute energy errors [given in Hartree] between the converged {Qubit}-ADAPT-VQE-SCF energies and their corresponding reference CASSCF energies as a function of the number of CNOT gates in the current expansion point of the ansatz. To ensure a fair comparison between both ADAPT schemes we considered for the analysis of the \textit{qubit} scheme the same set of molecules as in the previous discussion of the \textit{fermionic} optimization scheme. 
Considering first the case of a symmetrically stretched H$_4$\ molecule in panel (a) of Fig.~\ref{fig:qubit_adapt}, we note a faster convergence of the Qubit-ADAPT-VQE-SCF algorithm on the basis of a JW mapping in contrast to a BK mapping. Furthermore, similar to the fermionic ADAPT case, 
initially allowing to add only a single gate in the VQE step (1G-ADAPT)\ leads to an overall faster convergence of the orbital optimization algorithm wrt approaching
the CASSCF energy limit. Upon reaching a plateau with 1G-ADAPT, which is even more pronounced in the case of LiH (panel (b) in Fig.~\ref{fig:qubit_adapt}), 
a tighter convergence requires to switch to a 5G-ADAPT scheme. Hence, these findings suggest that for designing an optimal orbital optimization scheme within ADAPT-VQE-SCF 
it may be beneficial to first perform a few macro iterations in the SCF loop adopting a 1G-ADAPT strategy before turning to a 5G-ADAPT framework for the VQE part 
to ensure a smooth and numerically stable convergence to the (unknown) CASSCF limit. In passing we note that targeting the doubly-degenerate ${}^2\Pi$\ electronic ground state ($S=1/2$)\ within the Qubit-ADAPT-VQE-SCF scheme is not only feasible with a similar convergence pattern as in the fermionic case above but requires more than 10 times less CNOTs (entangling two-qubit gates) than in the latter. Moreover, such a quite striking reduction in the CNOTs count within Qubit-ADAPT-VQE-SCF in comparison to its fermionic counterpart can also be found for the other two molecular cases, namely H$_4$\ and LiH, respectively, both of which exhibit a singlet ($S=0$)\ ground state. As a result, Qubit-ADAPT-VQE seems to be superior to ADAPT-VQE based on a fermionic operator pool and, perhaps most importantly, 
a viable option to construct efficient ans\"atze with low CNOT counts within an orbital optimization framework 
that enables a state-specific optimization of (molecular) systems with (non-)degenerate target electronic states.

\subsection{State-average ADAPT-VQE-SCF optimization}\label{sec:results:sa}
In the previous section, we discussed the performance our ADAPT-VQE-SCF scheme to  optimize a single, selected electronic state of interest. 
In the following, we consider the simultaneous optimization of two (or in principle an arbitrary large number of) electronic states of different (or the same) spin symmetry, which not only is a frequently encountered problem in computational chemistry such as in the photochemistry polyenes \cite{curu17a} but also demonstrates the ability for ADAPT-VQE-SCF to treat degenerate (or quasi-degenerate) states on
the same footing\ similar to the algorithm proposed by Yalouz \textit{et al.}\ \cite{yalo21a}.  

\subsubsection{Oxygen dimer}

As a first example, we carried out 
state-average (SA) orbital optimizations for the $X {}^3\Sigma_g ^-$\ ($S=1$)\ ground and $a {}^1\Delta_g$\ ($S=0$) first excited state of the oxygen dimer O$_2$\ based on a CAS(6,4). In these SA-VQE-SCF/\texttt{TZ}\ calculations the energies of both spin states were obtained by running ADAPT-VQE\ twice, starting from two different reference states (with spin $S=0$ and $S=1$), and then using a spin-preserving pool. Note that, for each optimized quantum circuit, the 2-RDMs $\boldsymbol{\Gamma}_{\lambda_1}$\ and $\boldsymbol{\Gamma}_{\lambda_2}$, respectively, are measured \textit{state-specifically} in order to calculate an averaged first-order correction to the 2-RDM $\bar{\boldsymbol{\Gamma}}^{1}$(c.f.~Eq.~\eqref{eq:gamma1}\ as well as the corresponding total energy from an averaging of the state-specific energies (see also Eq.~\eqref{eq:eq:vqe_cas_opt_sa}). If not stated otherwise, we consider a democratic averaging scheme  with equal weights, that is, in a two-state ensemble we choose $\lambda_1 = \lambda_2 = 0.5$.

\begin{figure}
    \centering
    \includegraphics[width = 1.0\textwidth]{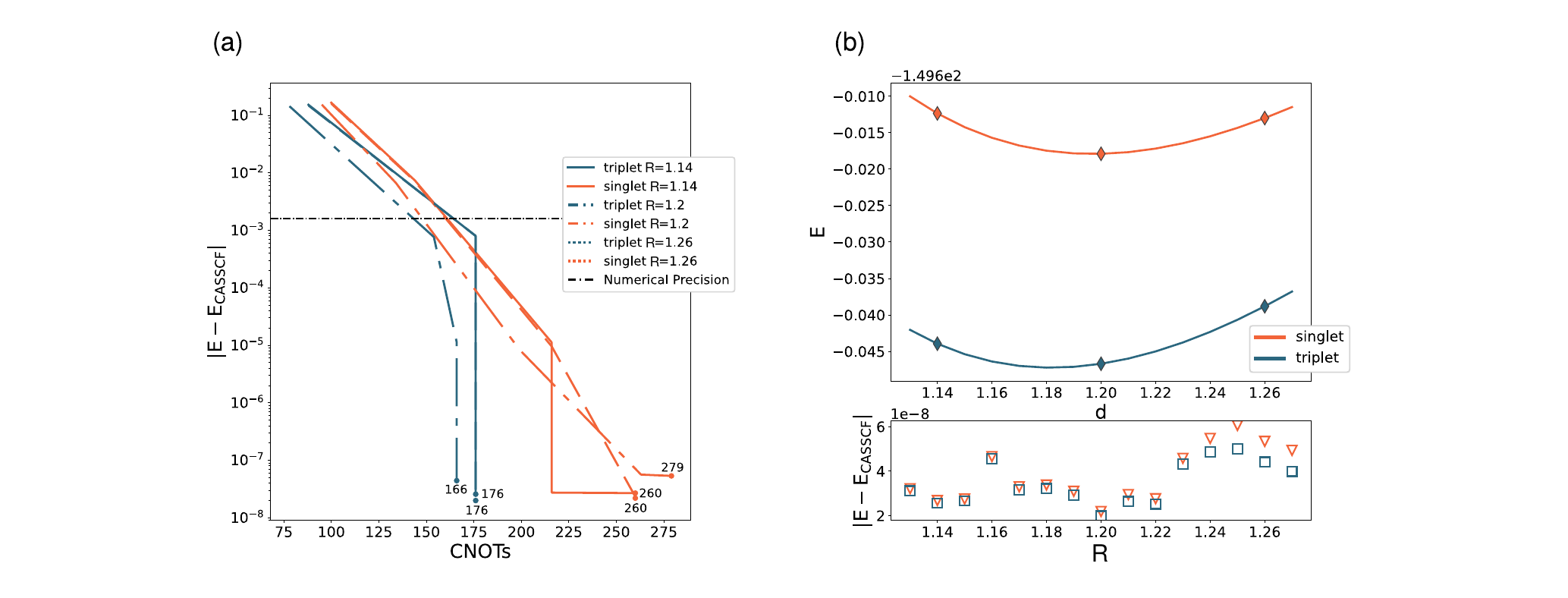}
    \caption{(a) Performance of the ADAPT-VQE-SCF for the oxygen dimer O$_2$: absolute energy errors between the converged ADAPT-VQE-SCF energy and the reference CASSCF energy as a function of the number of CNOT gates in the converged ansatz. Triplet oxygen ground state $X^3\Sigma_g^-$ (blue) and singlet first excited state $a^1\Delta_g$ (orange) at different inter-nuclear distances $R=1.14, 1.2, 1.26$ [\r{A}] (solid, dash, dash dotted). The black dash-dotted line represent chemical precision. (b) Upper panel: potential energy curve of the triplet oxygen (blue) and singlet oxygen (orange) for the converged run. Diamond symbols denote the distances considered in panel (a). Lower panel: absolute energy errors between the converged ADAPT-VQE-SCF energy and the reference CASSCF energy. All energies are given in Hartree.}
    \label{fig:fermionic_adapt_O2}
\end{figure}

Fig.~\ref{fig:fermionic_adapt_O2}(a) illustrates for a few selected O-O internuclear distances the convergence in absolute energy errors given in Hartree between our fermionic ADAPT-VQE-SCF based on a JW fermion-to-qubit mapping and a reference CASSCF approach as a function of the number of CNOT gates in the final constructed ansatz. 
To verify that our ADAPT-VQE-SCF approach allows us to obtain smooth potential energy curves (PECs) for the ground- as well as the excited state, 
we performed state-average orbital optimizations at various internuclear distances \textbf{R}, summarized in the upper panel of Fig.~\ref{fig:fermionic_adapt_O2}(b). The lower panel of Fig.~\ref{fig:fermionic_adapt_O2}(b)\ displays  the residual error of the individual ADAPT-VQE-SCF energies (in units of $10^-8$\ Hartree) for the triplet and singlet state, respectively, with respect to their CASSCF reference values as a function of the O-O internuclear distance. All calculations made use of the 1G-ADAPT scheme within the VQE step of the orbital optimization algorithm (\textit{vide supra}). 

Notably, as shown in Fig.~\ref{fig:fermionic_adapt_O2}(a), numerical convergence of the excited $a {}^1\Delta_g$\ singlet state requires more CNOTs 
as well as a deeper quantum circuit in contrast to the $X {}^3\Sigma_g^-$\ triplet ground state. This finding is hence in line with the common understanding that the singlet state requires in general a multiconfigurational description at any O-O internuclear distance 
whereas the triplet state is predominantly a single configurational problem close to the equilibrium structure of O$_2$. 
Setting out from the PEC data visualized in the upper panel of Fig.~\ref{fig:fermionic_adapt_O2}(b), 
we performed a polynomial fourth-order fit in order to extract spectroscopic properties such as the equilibrium internuclear distance $R_e$, the fundamental vibrational frequency  $\omega_e$ as well as the adiabatic electronic excitation energy $T_e$. 
The results for the latter are compiled in Table \ref{tab:o2-spec}. 
As can be seen from a comparison with the experimental data \cite{nistO2} (listed in parenthesis), our calculated equilibrium internuclear distances derived from our SA-VQE-SCF data are consistently too short by about 0.02-0.03 \AA. Similarly, we obtain an overestimation of the $\omega_e$\ with fundamental frequencies exceeding their experimental references by $\approx +12\%$. Finally, the adiabatic excitation energy T$_e$\ from the triplet ground to the first excited singlet state is underestimated in our CAS(6,4)-based SA-VQE-SCF model by $\approx$\ 1500 cm$^{-1}$, that is the energy gap between the minima of the two PECs are too close. These overall seemingly large discrepancies for the computed molecular properties are probably best explained by the lack of dynamical electron correlation in our VQE-SCF model 
which is, by construction, designed to only account for static electron correlation. 
To pave the way towards an efficient account of both static and dynamical correlation, we have very recently devised a multi-reference perturbation ansatz based on a VQE-type reference state which will be discussed elsewhere \cite{vqe_pt2}. 

\begin{table}
    \centering
    \begin{tabular}{l|c|c|c}\hline \hline
        state &  R$_e$ [\AA] & $\omega_e$ [cm$^{-1}$] & T$_e$ [cm$^{-1}$]\\ \hline
        $X {}^3\Sigma_g^{-}$ & 1.18 (1.208) & 1766 (1580) & 0\\
        $a {}^1\Delta_g^{}$ & 1.20 (1.216) & 1645 (1484) & 6435 (7918)\\\hline
    \end{tabular}
    \caption{Spectroscopic constants for the $X {}^3\Sigma_g^{-}$\ and $a {}^1\Delta_g^{}$\ electronic ground\ and first excited state of O$_2$, respectively, as obtained from state-average VQE-SCF calculations with a CAS(4,4) active orbital space. Experimental data taken from Ref.~\citenum{nistO2}\ is reported in parenthesis. For further computational details, see the main text.}
    \label{tab:o2-spec}
\end{table}
    
\subsubsection{Ferrocene}

\begin{figure}
\centering
         \includegraphics[width=0.75\textwidth]{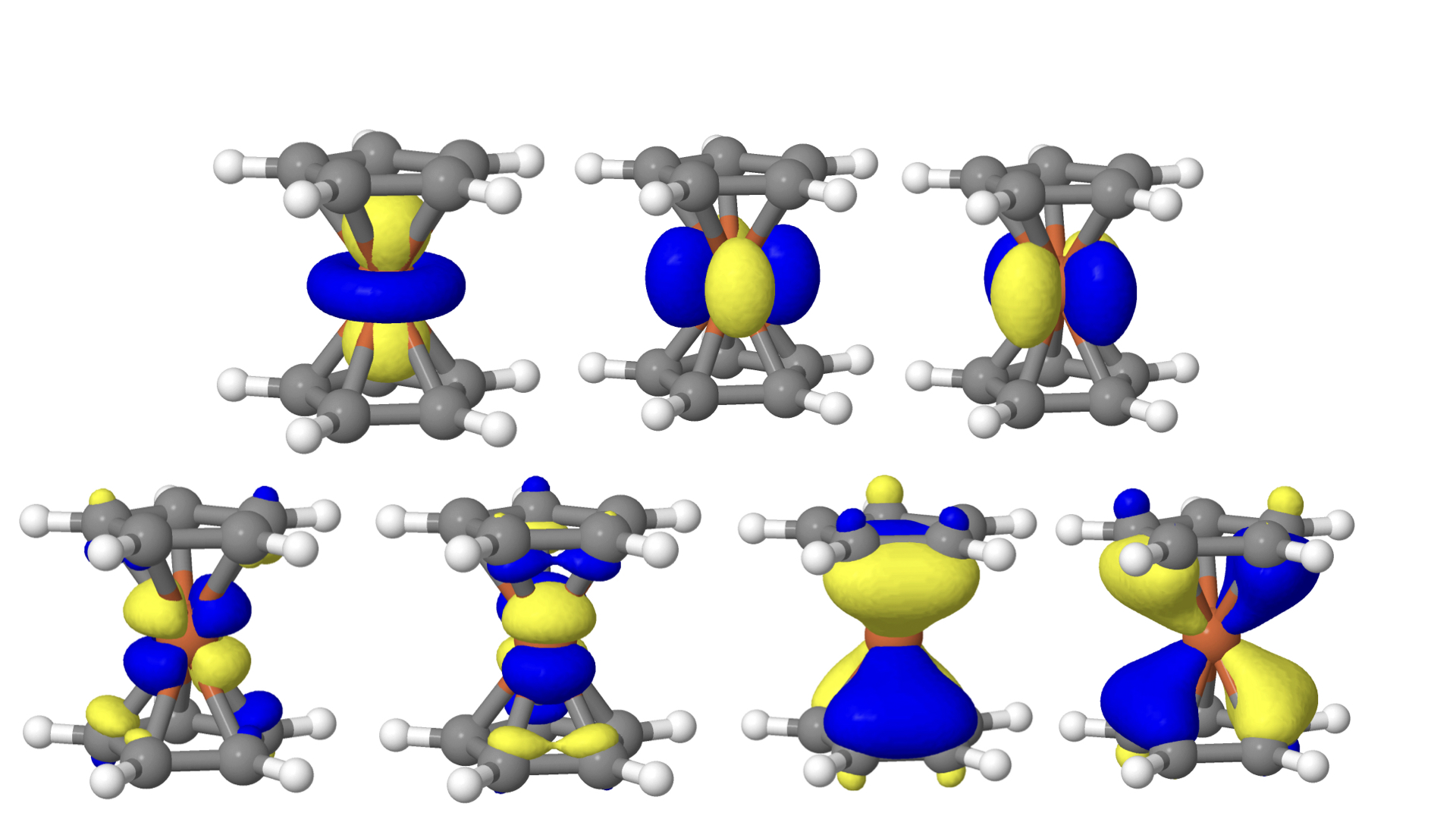}
         \caption{Active orbitals of Fe(Cp)$_2$\ within a CAS(10,7). For further details, see text.}
       \label{fig:CAS-fecp2}
\end{figure}

As a final example for our state-average ADAPT-VQE-SCF approach, we consider in the following the singlet ground and first excited triplet states of ferrocene (Fe(Cp)$_2$) in a staggered D$_{5\rm h}$\ geometry (see Fig.~\ref{fig:CAS-fecp2}) with the two planar cyclopentadienyl (Cp) rings and the $z$-axis in alignment with the Cp-Fe-Cp axis. The xyz coordinates for Fe(Cp)$_2$\ were taken from Ref.~\citenum{hard07a}. To pick a suitable active orbital space for a multiconfigurational description of the low-lying electronic state manifold, we followed the AVAS procedure outlined in Ref.~\citenum{sayf17a}\ and implemented in \textsc{PySCF}. The resulting CAS(10,7), considered in the following, was obtained by targeting a set of molecular orbitals with predominant Fe $3d$\ character in the AVAS procedure requesting a threshold of 0.1 for the overlap eigenvalue with a minimal Fe atomic orbital basis. As illustrated in Fig.~\ref{fig:CAS-fecp2}, besides the expected set of five Fe $3d$\ orbitals, the CAS comprises in addition two MOs that display a strong bonding character between the Cp rings and the Fe 3$d_{xz}$ 3$d_{yz}$\ orbitals such that the final active orbital space comprises 5+2=7 MOs, five of which are doubly occupied in the single-determinant SCF solution. Taking into account scalar-relativistic calculations on the basis of the spinfree exact-two-component Hamiltonian \cite{liu2006,ilia07}, all calculations made use of the all-electron cc-pVTZ-DK basis sets \cite{bala05a}\ (dubbed as \texttt{TZ}) for Fe, C and H. 
As discussed in detail by Sayfutyarova \textit{et al.}\ \cite{sayf17a}, although the ${}^1A_{1}^\prime$\ ground state of Fe(Cp)$_2$\ is predominantly of single-configurational character, the low-lying excited-state manifold exhibits a number of $d\rightarrow d$ singlet as well as triplet transitions. 
They are best characterized as excitations occurring from the set of non-bonding Fe $d$\ orbitals (upper row of Fig.~\ref{fig:CAS-fecp2}, Fe 3$d_{x2-y2}, 3d_{xy}$ and $3d_{z^2}$\ MOs) to the antibonding orbitals exhibiting mainly Fe $3d_{xz}$\ and $3d_{yz}$\ character (c.f.~the two left-most MOs in the second row of Fig.~\ref{fig:CAS-fecp2}), respectively. 
Most importantly, albeit being of single-excitation nature, 
even a \textit{qualitatively} appropriate description of these excited states requires a multi-configurational treatment \cite{sayf17a}. 

Although we make for our present calculations use of a smaller set of target states (one singlet and the spatially degenerate pair of lowest-lying triplet states) than compared to the 13-state manifold (consisting of seven singlet and six triplet states)  in Ref.~\citenum{sayf17a}, 
our CASSCF/\texttt{TZ}\ reference data for the lowest-lying $1 {}^3E^{\prime\prime}$\ excitation energy in Fe(Cp)$_2$\ of 0.93 eV compares favorably with their corresponding data of 0.97 eV, respectively. For the SA-ADAPT-VQE optimization, we considered a smaller active space (CAS(6,5))\ which comprises the five MOs of the preceeding CASSCF step with occupation number in the natural orbital basis deviating by more than 0.03 from a canonical occupation of 2 or 0. The excitation energy for the ${}^1A_{1}^\prime\ \rightarrow\ 1 {}^3E^{\prime\prime}$\ transition reduces in this case to 0.70 eV which is a clear indication that seemingly subtle correlation contributions from the now additional two core, \textit{viz.} doubly occupied, MOs\ have a significant contribution to a balanced description of the singlet ground and triplet excited states. 
Moreover, in order to achieve a full numerical convergence of the ADAPT-VQE energies requires for the singlet ground rather than the triplet excited states a considerable larger number of CNOTs, that is 72 and 8, respectively. The latter finding indicates that the origin of the discrepancy in the excitation energy likely originates from an unbalanced description of the electronic structure of the 
singlet ground state. 

Finally, let us stress that orbital relaxation has a substantial effect in the present case, i.e. it reduces the excitation energy for the lowest-lying triplet $d\rightarrow\ d$\ transition by almost 50\% (sic!), that is from 1.84 eV to 0.93 eV. Notably, the latter reduction is of \textit{equal} importance as the inclusion of dynamical electron correlation which, as reported in Table 1 of Ref.~\citenum{sayf17a}, leads to an increase in the excitation energy to the $1 {}^3E^{\prime\prime}$\ state to 1.88 eV (column CASSCF+PT2 in Table 1 of Ref.~\citenum{sayf17a}). 



\section{Conclusions and Outlook}\label{sec:conclu}
In this article we have presented the framework and implementation for a self-consistent field complete-active space 
orbital-optimization algorithm that exploits adaptively optimized quantum circuits 
for the variational quantum eigensolver serving as active space solver in replacement of a ``classical" full CI solver. Although our orbital optimization algorithm for quantum-computing driven quantum chemistry is by no means the first of its kind, see for example Refs.~ \citenum{take20a,mizu20a,soko20a,yalo21a,PhysRevResearch.3.033230,goch22a,bierman2022improving,omiy22a}, 
it is, to the best of our knowledge, the first one that exploits within the VQE solver framework quantum circuits based on the so-called ADAPT family of \textit{ans\"atze}. 
Dubbed as ADAPT-VQE-SCF approach, we demonstrated its capabilities by considering  various molecular quantum-chemical applications ranging from state-specific electronic ground-state optimizations to multi-root state-average optimizations simultaneously targeting electronic states of differing spin multiplicity. 
By validating with exact data obtained from classical CASSCF calculations, we showed that convergence for ADAPT-VQE-SCF below chemical (or even spectroscopic) precision can be achieved for a variety of molecular systems exhibiting electronic ground- and excited states of different spin multiplicity on the basis of typically considerably less than 10$^3$ CNOTs. 

Moreover, by opening up for a simultaneous optimization of electronic states with different symmetries (spin in this work 
but spatial symmetries could be considered as 
well), we were able to democratically describe the ground- and low-lying excited state manifold of O$_2$\ and ferrocene. These calculations also revealed that dynamical electron correlation effects cannot be neglected if one seeks for \textit{quantitative}\ results \cite{schl22a}. Work along these  lines are currently in progress in our team based on our ADAPT-VQE-SCF framework and will be addressed in a forthcoming work \cite{vqe_pt2}. In addition, we will also explore CAS-based hybrid algorithms such as the ones proposed in Refs.~\citenum{fromager07,hede15b,ross21a}\ which combine density functional theory with active space-based wave function theories in a rigorous manner.   

Finally, in the present work we exclusively assumed a noiseless implementation of quantum circuits which is, given the present near-term character of quantum computers, is an approximation. In future work, we will consider ADAPT-VQE-SCF applications on actual quantum devices where we will combine our current ADAPT-VQE-SCF algorithm with our variational tensor-network driven noise-mitigation post-processing algorithm \cite{garc22b,vilma22a}.

\begin{acknowledgments}
The ADAPT-VQE-SCF approach presented in this work is  integrated in Aurora, a proprietary quantum chemistry platform developed by Algorithmiq Ltd.
\end{acknowledgments}

\section*{Data availability}\label{sec:SIData}
Data is available from the authors upon reasonable request. 

\bibliography{bibliography.bib}

\begin{thebibliography}{83}%
\makeatletter
\providecommand \@ifxundefined [1]{%
 \@ifx{#1\undefined}
}%
\providecommand \@ifnum [1]{%
 \ifnum #1\expandafter \@firstoftwo
 \else \expandafter \@secondoftwo
 \fi
}%
\providecommand \@ifx [1]{%
 \ifx #1\expandafter \@firstoftwo
 \else \expandafter \@secondoftwo
 \fi
}%
\providecommand \natexlab [1]{#1}%
\providecommand \enquote  [1]{``#1''}%
\providecommand \bibnamefont  [1]{#1}%
\providecommand \bibfnamefont [1]{#1}%
\providecommand \citenamefont [1]{#1}%
\providecommand \href@noop [0]{\@secondoftwo}%
\providecommand \href [0]{\begingroup \@sanitize@url \@href}%
\providecommand \@href[1]{\@@startlink{#1}\@@href}%
\providecommand \@@href[1]{\endgroup#1\@@endlink}%
\providecommand \@sanitize@url [0]{\catcode `\\12\catcode `\$12\catcode
  `\&12\catcode `\#12\catcode `\^12\catcode `\_12\catcode `\%12\relax}%
\providecommand \@@startlink[1]{}%
\providecommand \@@endlink[0]{}%
\providecommand \url  [0]{\begingroup\@sanitize@url \@url }%
\providecommand \@url [1]{\endgroup\@href {#1}{\urlprefix }}%
\providecommand \urlprefix  [0]{URL }%
\providecommand \Eprint [0]{\href }%
\providecommand \doibase [0]{https://doi.org/}%
\providecommand \selectlanguage [0]{\@gobble}%
\providecommand \bibinfo  [0]{\@secondoftwo}%
\providecommand \bibfield  [0]{\@secondoftwo}%
\providecommand \translation [1]{[#1]}%
\providecommand \BibitemOpen [0]{}%
\providecommand \bibitemStop [0]{}%
\providecommand \bibitemNoStop [0]{.\EOS\space}%
\providecommand \EOS [0]{\spacefactor3000\relax}%
\providecommand \BibitemShut  [1]{\csname bibitem#1\endcsname}%
\let\auto@bib@innerbib\@empty
\bibitem [{\citenamefont {Szalay}\ \emph {et~al.}(2012)\citenamefont {Szalay},
  \citenamefont {M{\"u}ller}, \citenamefont {Gidofalvi}, \citenamefont
  {Lischka},\ and\ \citenamefont {Shepard}}]{szal12}%
  \BibitemOpen
  \bibfield  {author} {\bibinfo {author} {\bibfnamefont {P.~G.}\ \bibnamefont
  {Szalay}}, \bibinfo {author} {\bibfnamefont {T.}~\bibnamefont {M{\"u}ller}},
  \bibinfo {author} {\bibfnamefont {G.}~\bibnamefont {Gidofalvi}}, \bibinfo
  {author} {\bibfnamefont {H.}~\bibnamefont {Lischka}},\ and\ \bibinfo {author}
  {\bibfnamefont {R.}~\bibnamefont {Shepard}},\ }\bibfield  {title} {\enquote
  {\bibinfo {title} {Multiconfiguration self-consistent field and
  multireference configuration interaction methods and applications},}\ }\href
  {https://doi.org/10.1021/cr200137a} {\bibfield  {journal} {\bibinfo
  {journal} {Chem. Rev.}\ }\textbf {\bibinfo {volume} {112}},\ \bibinfo {pages}
  {108--181} (\bibinfo {year} {2012})}\BibitemShut {NoStop}%
\bibitem [{\citenamefont {Roos}, \citenamefont {Taylor},\ and\ \citenamefont
  {Siegbahn}(1980)}]{roos80a}%
  \BibitemOpen
  \bibfield  {author} {\bibinfo {author} {\bibfnamefont {B.}~\bibnamefont
  {Roos}}, \bibinfo {author} {\bibfnamefont {P.~R.}\ \bibnamefont {Taylor}},\
  and\ \bibinfo {author} {\bibfnamefont {P.~A.}\ \bibnamefont {Siegbahn}},\
  }\bibfield  {title} {\enquote {\bibinfo {title} {A complete active space
  {SCF} method-{(CASSCF)} using a density matrix formulated super-{CI}
  approach},}\ }\href {https://doi.org/10.1016/0301-0104(80)80045-0} {\bibfield
   {journal} {\bibinfo  {journal} {Chem. Phys.}\ }\textbf {\bibinfo {volume}
  {48}},\ \bibinfo {pages} {157--173} (\bibinfo {year} {1980})}\BibitemShut
  {NoStop}%
\bibitem [{\citenamefont {Olsen}(2011)}]{olse11}%
  \BibitemOpen
  \bibfield  {author} {\bibinfo {author} {\bibfnamefont {J.}~\bibnamefont
  {Olsen}},\ }\bibfield  {title} {\enquote {\bibinfo {title} {The {CASSCF}
  method: A perspective and commentary},}\ }\href
  {https://doi.org/10.1002/qua.23107} {\bibfield  {journal} {\bibinfo
  {journal} {Int. J. Quantum Chem.}\ }\textbf {\bibinfo {volume} {111}},\
  \bibinfo {pages} {3267--3272} (\bibinfo {year} {2011})}\BibitemShut {NoStop}%
\bibitem [{\citenamefont {Roos}\ \emph {et~al.}(2016)\citenamefont {Roos},
  \citenamefont {Lindh}, \citenamefont {Malmqvist}, \citenamefont {Veryazov},\
  and\ \citenamefont {Widmark}}]{roos16a}%
  \BibitemOpen
  \bibfield  {author} {\bibinfo {author} {\bibfnamefont {B.~O.}\ \bibnamefont
  {Roos}}, \bibinfo {author} {\bibfnamefont {R.}~\bibnamefont {Lindh}},
  \bibinfo {author} {\bibfnamefont {P.-{\AA}.}\ \bibnamefont {Malmqvist}},
  \bibinfo {author} {\bibfnamefont {V.}~\bibnamefont {Veryazov}},\ and\
  \bibinfo {author} {\bibfnamefont {P.-O.}\ \bibnamefont {Widmark}},\ }\enquote
  {\bibinfo {title} {Multiconfigurational quantum chemistry},}\ in\ \href@noop
  {} {\emph {\bibinfo {booktitle} {Multiconfigurational Quantum Chemistry}}}\
  (\bibinfo  {publisher} {John Wiley \& Sons, Inc.},\ \bibinfo {year} {2016})\
  Chap.\ \bibinfo {chapter} {{CASPT2/CASSCF Applications}}, pp.\ \bibinfo
  {pages} {157--219}\BibitemShut {NoStop}%
\bibitem [{\citenamefont {Roos}(2008)}]{roos08a}%
  \BibitemOpen
  \bibfield  {author} {\bibinfo {author} {\bibfnamefont {B.~O.}\ \bibnamefont
  {Roos}},\ }\bibfield  {title} {\enquote {\bibinfo {title}
  {Multiconfigurational quantum chemistry for ground and excited states},}\
  }in\ \href@noop {} {\emph {\bibinfo {booktitle} {Radiation Induced Molecular
  Phenomena in Nucleic Acids}}},\ \bibinfo {editor} {edited by\ \bibinfo
  {editor} {\bibfnamefont {M.~K.}\ \bibnamefont {Shukla}}\ and\ \bibinfo
  {editor} {\bibfnamefont {J.}~\bibnamefont {Leszczynski}}}\ (\bibinfo
  {publisher} {Springer},\ \bibinfo {year} {2008})\ pp.\ \bibinfo {pages}
  {125--156}\BibitemShut {NoStop}%
\bibitem [{\citenamefont {Helmich-Paris}(2022)}]{helm22a}%
  \BibitemOpen
  \bibfield  {author} {\bibinfo {author} {\bibfnamefont {B.}~\bibnamefont
  {Helmich-Paris}},\ }\bibfield  {title} {\enquote {\bibinfo {title} {A
  trust-region augmented hessian implementation for state-specific and
  state-averaged {{CASSCF}} wave functions},}\ }\href
  {https://doi.org/10.1063/5.0090447} {\bibfield  {journal} {\bibinfo
  {journal} {J.~Chem.~Phys.}\ }\textbf {\bibinfo {volume} {156}},\ \bibinfo
  {pages} {204104} (\bibinfo {year} {2022})}\BibitemShut {NoStop}%
\bibitem [{\citenamefont {Thouless}(1961)}]{thou61}%
  \BibitemOpen
  \bibfield  {author} {\bibinfo {author} {\bibfnamefont {D.~J.}\ \bibnamefont
  {Thouless}},\ }\href@noop {} {\emph {\bibinfo {title} {The Quantum Mechanics
  of Many Body Systems}}}\ (\bibinfo  {publisher} {Academic Press, New York},\
  \bibinfo {year} {1961})\BibitemShut {NoStop}%
\bibitem [{\citenamefont {Levy}(1969)}]{levy69a}%
  \BibitemOpen
  \bibfield  {author} {\bibinfo {author} {\bibfnamefont {B.}~\bibnamefont
  {Levy}},\ }\bibfield  {title} {\enquote {\bibinfo {title}
  {{Multi-configuration self-consistent wavefunctions of formaldehyde}},}\
  }\href {https://doi.org/10.1016/0009-2614(69)85022-0} {\bibfield  {journal}
  {\bibinfo  {journal} {Chem. Phys. Lett.}\ }\textbf {\bibinfo {volume} {4}},\
  \bibinfo {pages} {17--19} (\bibinfo {year} {1969})}\BibitemShut {NoStop}%
\bibitem [{\citenamefont {Helgaker}, \citenamefont {J{\o}rgensen},\ and\
  \citenamefont {Olsen}(2000)}]{helg00}%
  \BibitemOpen
  \bibinfo {editor} {\bibfnamefont {T.}~\bibnamefont {Helgaker}}, \bibinfo
  {editor} {\bibfnamefont {P.}~\bibnamefont {J{\o}rgensen}},\ and\ \bibinfo
  {editor} {\bibfnamefont {J.}~\bibnamefont {Olsen}},\ eds.,\ \href@noop {}
  {\emph {\bibinfo {title} {{M}olecular {E}lectronic-{S}tructure {T}heory}}},\
  \bibinfo {edition} {1st}\ ed.\ (\bibinfo  {publisher} {Wiley {\&} Sons,
  Chichester (England)},\ \bibinfo {year} {2000})\BibitemShut {NoStop}%
\bibitem [{\citenamefont {Yeager}\ and\ \citenamefont
  {J{\o}rgensen}(1979)}]{yeag79}%
  \BibitemOpen
  \bibfield  {author} {\bibinfo {author} {\bibfnamefont {D.~L.}\ \bibnamefont
  {Yeager}}\ and\ \bibinfo {author} {\bibfnamefont {P.}~\bibnamefont
  {J{\o}rgensen}},\ }\bibfield  {title} {\enquote {\bibinfo {title}
  {Convergency studies of second and approximate second order
  multiconfigurational hartree-fock procedures},}\ }\href
  {https://doi.org/10.1063/1.438363} {\bibfield  {journal} {\bibinfo  {journal}
  {J.~Chem.~Phys.}\ }\textbf {\bibinfo {volume} {71}},\ \bibinfo {pages}
  {755--760} (\bibinfo {year} {1979})}\BibitemShut {NoStop}%
\bibitem [{\citenamefont {Dalgaard}(1979)}]{dalg79}%
  \BibitemOpen
  \bibfield  {author} {\bibinfo {author} {\bibfnamefont {E.}~\bibnamefont
  {Dalgaard}},\ }\bibfield  {title} {\enquote {\bibinfo {title} {A
  quadratically convergent reference state optimization procedure},}\ }\href
  {https://doi.org/10.1016/0009-2614(79)80291-2} {\bibfield  {journal}
  {\bibinfo  {journal} {Chem.~Phys.~Lett.}\ }\textbf {\bibinfo {volume} {65}},\
  \bibinfo {pages} {559--563} (\bibinfo {year} {1979})}\BibitemShut {NoStop}%
\bibitem [{\citenamefont {Olsen}\ and\ \citenamefont
  {J{\o}rgensen}(1985)}]{olse85}%
  \BibitemOpen
  \bibfield  {author} {\bibinfo {author} {\bibfnamefont {J.}~\bibnamefont
  {Olsen}}\ and\ \bibinfo {author} {\bibfnamefont {P.}~\bibnamefont
  {J{\o}rgensen}},\ }\bibfield  {title} {\enquote {\bibinfo {title} {{Linear
  and nonlinear response functions for an exact state and for an MCSCF
  state}},}\ }\href {https://doi.org/10.1063/1.448223} {\bibfield  {journal}
  {\bibinfo  {journal} {J. Chem. Phys.}\ }\textbf {\bibinfo {volume} {82}},\
  \bibinfo {pages} {3235--3264} (\bibinfo {year} {1985})}\BibitemShut {NoStop}%
\bibitem [{\citenamefont {Helgaker}\ \emph {et~al.}(1986)\citenamefont
  {Helgaker}, \citenamefont {Alml\"{o}f}, \citenamefont {Jensen},\ and\
  \citenamefont {J{\o}rgensen}}]{helg86a}%
  \BibitemOpen
  \bibfield  {author} {\bibinfo {author} {\bibfnamefont {T.~U.}\ \bibnamefont
  {Helgaker}}, \bibinfo {author} {\bibfnamefont {J.}~\bibnamefont
  {Alml\"{o}f}}, \bibinfo {author} {\bibfnamefont {H.~J.~A.}\ \bibnamefont
  {Jensen}},\ and\ \bibinfo {author} {\bibfnamefont {P.}~\bibnamefont
  {J{\o}rgensen}},\ }\bibfield  {title} {\enquote {\bibinfo {title} {Molecular
  hessians for large-scale {MCSCF} wave functions},}\ }\href
  {https://doi.org/10.1063/1.450771} {\bibfield  {journal} {\bibinfo  {journal}
  {J.~Chem.~Phys.}\ }\textbf {\bibinfo {volume} {84}},\ \bibinfo {pages}
  {6266--6279} (\bibinfo {year} {1986})}\BibitemShut {NoStop}%
\bibitem [{\citenamefont {Wouters}\ and\ \citenamefont {van
  Neck}(2014)}]{wout14}%
  \BibitemOpen
  \bibfield  {author} {\bibinfo {author} {\bibfnamefont {S.}~\bibnamefont
  {Wouters}}\ and\ \bibinfo {author} {\bibfnamefont {D.}~\bibnamefont {van
  Neck}},\ }\bibfield  {title} {\enquote {\bibinfo {title} {The density matrix
  renormalization group for ab initio quantum chemistry},}\ }\href
  {https://doi.org/10.1140/epjd/e2014-50500-1} {\bibfield  {journal} {\bibinfo
  {journal} {Eur. Phys. J. D}\ }\textbf {\bibinfo {volume} {68}},\ \bibinfo
  {pages} {272} (\bibinfo {year} {2014})}\BibitemShut {NoStop}%
\bibitem [{\citenamefont {Knecht}\ \emph {et~al.}(2016)\citenamefont {Knecht},
  \citenamefont {Hedegaard}, \citenamefont {Keller}, \citenamefont {Kovyrshin},
  \citenamefont {Ma}, \citenamefont {Muolo}, \citenamefont {Stein},\ and\
  \citenamefont {Reiher}}]{knec16a}%
  \BibitemOpen
  \bibfield  {author} {\bibinfo {author} {\bibfnamefont {S.}~\bibnamefont
  {Knecht}}, \bibinfo {author} {\bibfnamefont {E.~D.}\ \bibnamefont
  {Hedegaard}}, \bibinfo {author} {\bibfnamefont {S.}~\bibnamefont {Keller}},
  \bibinfo {author} {\bibfnamefont {A.}~\bibnamefont {Kovyrshin}}, \bibinfo
  {author} {\bibfnamefont {Y.}~\bibnamefont {Ma}}, \bibinfo {author}
  {\bibfnamefont {A.}~\bibnamefont {Muolo}}, \bibinfo {author} {\bibfnamefont
  {C.~J.}\ \bibnamefont {Stein}},\ and\ \bibinfo {author} {\bibfnamefont
  {M.}~\bibnamefont {Reiher}},\ }\bibfield  {title} {\enquote {\bibinfo {title}
  {New approaches for ab initio calculations of molecules with strong electron
  correlation},}\ }\href {https://doi.org/10.2533/chimia.2016.244} {\bibfield
  {journal} {\bibinfo  {journal} {Chimia}\ }\textbf {\bibinfo {volume} {70}},\
  \bibinfo {pages} {244--251} (\bibinfo {year} {2016})}\BibitemShut {NoStop}%
\bibitem [{\citenamefont {Chan}\ \emph {et~al.}(2016)\citenamefont {Chan},
  \citenamefont {Keselman}, \citenamefont {Nakatani}, \citenamefont {Li},\ and\
  \citenamefont {White}}]{chan16a}%
  \BibitemOpen
  \bibfield  {author} {\bibinfo {author} {\bibfnamefont {G.~K.-L.}\
  \bibnamefont {Chan}}, \bibinfo {author} {\bibfnamefont {A.}~\bibnamefont
  {Keselman}}, \bibinfo {author} {\bibfnamefont {N.}~\bibnamefont {Nakatani}},
  \bibinfo {author} {\bibfnamefont {Z.}~\bibnamefont {Li}},\ and\ \bibinfo
  {author} {\bibfnamefont {S.~R.}\ \bibnamefont {White}},\ }\bibfield  {title}
  {\enquote {\bibinfo {title} {{Matrix product operators, matrix product
  states, and ab initio density matrix renormalization group algorithms}},}\
  }\href {https://doi.org/10.1063/1.4955108} {\bibfield  {journal} {\bibinfo
  {journal} {J. Chem. Phys.}\ }\textbf {\bibinfo {volume} {145}},\ \bibinfo
  {pages} {014102} (\bibinfo {year} {2016})}\BibitemShut {NoStop}%
\bibitem [{\citenamefont {Baiardi}\ and\ \citenamefont
  {Reiher}(2020)}]{baia20a}%
  \BibitemOpen
  \bibfield  {author} {\bibinfo {author} {\bibfnamefont {A.}~\bibnamefont
  {Baiardi}}\ and\ \bibinfo {author} {\bibfnamefont {M.}~\bibnamefont
  {Reiher}},\ }\bibfield  {title} {\enquote {\bibinfo {title} {The density
  matrix renormalization group in chemistry and molecular physics: Recent
  developments and new challenges},}\ }\href
  {https://doi.org/10.1063/1.5129672} {\bibfield  {journal} {\bibinfo
  {journal} {J. Chem. Phys.}\ }\textbf {\bibinfo {volume} {152}},\ \bibinfo
  {pages} {040903} (\bibinfo {year} {2020})}\BibitemShut {NoStop}%
\bibitem [{\citenamefont {Booth}, \citenamefont {Thom},\ and\ \citenamefont
  {Alavi}(2009)}]{boot09}%
  \BibitemOpen
  \bibfield  {author} {\bibinfo {author} {\bibfnamefont {G.~H.}\ \bibnamefont
  {Booth}}, \bibinfo {author} {\bibfnamefont {A.~J.~W.}\ \bibnamefont {Thom}},\
  and\ \bibinfo {author} {\bibfnamefont {A.}~\bibnamefont {Alavi}},\ }\bibfield
   {title} {\enquote {\bibinfo {title} {{Fermion Monte Carlo without fixed
  nodes: A game of life, death, and annihilation in Slater determinant
  space}},}\ }\href {https://doi.org/10.1063/1.3193710} {\bibfield  {journal}
  {\bibinfo  {journal} {J. Chem. Phys.}\ }\textbf {\bibinfo {volume} {131}},\
  \bibinfo {pages} {054106--11} (\bibinfo {year} {2009})}\BibitemShut {NoStop}%
\bibitem [{\citenamefont {Overy}\ \emph {et~al.}(2014)\citenamefont {Overy},
  \citenamefont {Booth}, \citenamefont {Blunt}, \citenamefont {Shepherd},
  \citenamefont {Cleland},\ and\ \citenamefont {Alavi}}]{over14}%
  \BibitemOpen
  \bibfield  {author} {\bibinfo {author} {\bibfnamefont {C.}~\bibnamefont
  {Overy}}, \bibinfo {author} {\bibfnamefont {G.~H.}\ \bibnamefont {Booth}},
  \bibinfo {author} {\bibfnamefont {N.~S.}\ \bibnamefont {Blunt}}, \bibinfo
  {author} {\bibfnamefont {J.~J.}\ \bibnamefont {Shepherd}}, \bibinfo {author}
  {\bibfnamefont {D.}~\bibnamefont {Cleland}},\ and\ \bibinfo {author}
  {\bibfnamefont {A.}~\bibnamefont {Alavi}},\ }\bibfield  {title} {\enquote
  {\bibinfo {title} {{Unbiased reduced density matrices and electronic
  properties from full configuration interaction quantum Monte Carlo}},}\
  }\href {https://doi.org/10.1063/1.4904313} {\bibfield  {journal} {\bibinfo
  {journal} {J. Chem. Phys.}\ }\textbf {\bibinfo {volume} {141}},\ \bibinfo
  {pages} {244117--12} (\bibinfo {year} {2014})}\BibitemShut {NoStop}%
\bibitem [{\citenamefont {Li~Manni}, \citenamefont {Smart},\ and\ \citenamefont
  {Alavi}(2016)}]{lima16c}%
  \BibitemOpen
  \bibfield  {author} {\bibinfo {author} {\bibfnamefont {G.}~\bibnamefont
  {Li~Manni}}, \bibinfo {author} {\bibfnamefont {S.~D.}\ \bibnamefont
  {Smart}},\ and\ \bibinfo {author} {\bibfnamefont {A.}~\bibnamefont {Alavi}},\
  }\bibfield  {title} {\enquote {\bibinfo {title} {{Combining the Complete
  Active Space Self-Consistent Field Method and the Full Configuration
  Interaction Quantum Monte Carlo within a Super-CI Framework, with Application
  to Challenging Metal-Porphyrins}},}\ }\href
  {https://doi.org/10.1021/acs.jctc.5b01190} {\bibfield  {journal} {\bibinfo
  {journal} {J. Chem. Theory Comput.}\ }\textbf {\bibinfo {volume} {12}},\
  \bibinfo {pages} {1245--1258} (\bibinfo {year} {2016})}\BibitemShut {NoStop}%
\bibitem [{\citenamefont {Guther}\ \emph {et~al.}(2020)\citenamefont {Guther},
  \citenamefont {Anderson}, \citenamefont {Blunt}, \citenamefont {Bogdanov},
  \citenamefont {Cleland}, \citenamefont {Dattani}, \citenamefont {Dobrautz},
  \citenamefont {Ghanem}, \citenamefont {Jeszenszki}, \citenamefont
  {Liebermann}, \citenamefont {Manni}, \citenamefont {Lozovoi}, \citenamefont
  {Luo}, \citenamefont {Ma}, \citenamefont {Merz}, \citenamefont {Overy},
  \citenamefont {Rampp}, \citenamefont {Samanta}, \citenamefont {Schwarz},
  \citenamefont {Shepherd}, \citenamefont {Smart}, \citenamefont {Vitale},
  \citenamefont {Weser}, \citenamefont {Booth},\ and\ \citenamefont
  {Alavi}}]{guth20a}%
  \BibitemOpen
  \bibfield  {author} {\bibinfo {author} {\bibfnamefont {K.}~\bibnamefont
  {Guther}}, \bibinfo {author} {\bibfnamefont {R.~J.}\ \bibnamefont
  {Anderson}}, \bibinfo {author} {\bibfnamefont {N.~S.}\ \bibnamefont {Blunt}},
  \bibinfo {author} {\bibfnamefont {N.~A.}\ \bibnamefont {Bogdanov}}, \bibinfo
  {author} {\bibfnamefont {D.}~\bibnamefont {Cleland}}, \bibinfo {author}
  {\bibfnamefont {N.}~\bibnamefont {Dattani}}, \bibinfo {author} {\bibfnamefont
  {W.}~\bibnamefont {Dobrautz}}, \bibinfo {author} {\bibfnamefont
  {K.}~\bibnamefont {Ghanem}}, \bibinfo {author} {\bibfnamefont
  {P.}~\bibnamefont {Jeszenszki}}, \bibinfo {author} {\bibfnamefont
  {N.}~\bibnamefont {Liebermann}}, \bibinfo {author} {\bibfnamefont {G.~L.}\
  \bibnamefont {Manni}}, \bibinfo {author} {\bibfnamefont {A.~Y.}\ \bibnamefont
  {Lozovoi}}, \bibinfo {author} {\bibfnamefont {H.}~\bibnamefont {Luo}},
  \bibinfo {author} {\bibfnamefont {D.}~\bibnamefont {Ma}}, \bibinfo {author}
  {\bibfnamefont {F.}~\bibnamefont {Merz}}, \bibinfo {author} {\bibfnamefont
  {C.}~\bibnamefont {Overy}}, \bibinfo {author} {\bibfnamefont
  {M.}~\bibnamefont {Rampp}}, \bibinfo {author} {\bibfnamefont {P.~K.}\
  \bibnamefont {Samanta}}, \bibinfo {author} {\bibfnamefont {L.~R.}\
  \bibnamefont {Schwarz}}, \bibinfo {author} {\bibfnamefont {J.~J.}\
  \bibnamefont {Shepherd}}, \bibinfo {author} {\bibfnamefont {S.~D.}\
  \bibnamefont {Smart}}, \bibinfo {author} {\bibfnamefont {E.}~\bibnamefont
  {Vitale}}, \bibinfo {author} {\bibfnamefont {O.}~\bibnamefont {Weser}},
  \bibinfo {author} {\bibfnamefont {G.~H.}\ \bibnamefont {Booth}},\ and\
  \bibinfo {author} {\bibfnamefont {A.}~\bibnamefont {Alavi}},\ }\bibfield
  {title} {\enquote {\bibinfo {title} {{NECI}: $n$-electron configuration
  interaction with an emphasis on state-of-the-art stochastic methods},}\
  }\href {https://doi.org/10.1063/5.0005754} {\bibfield  {journal} {\bibinfo
  {journal} {J.~Chem.~Phys.}\ }\textbf {\bibinfo {volume} {153}},\ \bibinfo
  {pages} {034107} (\bibinfo {year} {2020})}\BibitemShut {NoStop}%
\bibitem [{\citenamefont {Tubman}\ \emph {et~al.}(2016)\citenamefont {Tubman},
  \citenamefont {Lee}, \citenamefont {Takeshita}, \citenamefont {Head-Gordon},\
  and\ \citenamefont {Whaley}}]{tubm16a}%
  \BibitemOpen
  \bibfield  {author} {\bibinfo {author} {\bibfnamefont {N.~M.}\ \bibnamefont
  {Tubman}}, \bibinfo {author} {\bibfnamefont {J.}~\bibnamefont {Lee}},
  \bibinfo {author} {\bibfnamefont {T.~Y.}\ \bibnamefont {Takeshita}}, \bibinfo
  {author} {\bibfnamefont {M.}~\bibnamefont {Head-Gordon}},\ and\ \bibinfo
  {author} {\bibfnamefont {K.~B.}\ \bibnamefont {Whaley}},\ }\bibfield  {title}
  {\enquote {\bibinfo {title} {A deterministic alternative to the full
  configuration interaction quantum monte carlo method},}\ }\href
  {https://doi.org/10.1063/1.4955109} {\bibfield  {journal} {\bibinfo
  {journal} {J.~Chem.~Phys.}\ }\textbf {\bibinfo {volume} {145}},\ \bibinfo
  {pages} {044112} (\bibinfo {year} {2016})}\BibitemShut {NoStop}%
\bibitem [{\citenamefont {Liu}\ and\ \citenamefont
  {Hoffmann}(2014)}]{liu2014_sds}%
  \BibitemOpen
  \bibfield  {author} {\bibinfo {author} {\bibfnamefont {W.}~\bibnamefont
  {Liu}}\ and\ \bibinfo {author} {\bibfnamefont {M.~R.}\ \bibnamefont
  {Hoffmann}},\ }\bibfield  {title} {\enquote {\bibinfo {title} {Sds: the
  ``static-dynamic-static" framework for strongly correlated electrons},}\
  }\href@noop {} {\bibfield  {journal} {\bibinfo  {journal} {Theor. Chem.
  Acc.}\ }\textbf {\bibinfo {volume} {133}},\ \bibinfo {pages} {1481--xx}
  (\bibinfo {year} {2014})}\BibitemShut {NoStop}%
\bibitem [{\citenamefont {Liu}\ and\ \citenamefont {Hoffmann}(2016)}]{liuw16a}%
  \BibitemOpen
  \bibfield  {author} {\bibinfo {author} {\bibfnamefont {W.}~\bibnamefont
  {Liu}}\ and\ \bibinfo {author} {\bibfnamefont {M.~R.}\ \bibnamefont
  {Hoffmann}},\ }\bibfield  {title} {\enquote {\bibinfo {title} {{iCI}:
  Iterative {CI} toward full {CI}},}\ }\href
  {https://doi.org/10.1021/acs.jctc.5b01099} {\bibfield  {journal} {\bibinfo
  {journal} {J. Chem. Theory Comput.}\ }\textbf {\bibinfo {volume} {12}},\
  \bibinfo {pages} {1169–1178} (\bibinfo {year} {2016})}\BibitemShut
  {NoStop}%
\bibitem [{\citenamefont {Scemama}\ \emph {et~al.}(2018)\citenamefont
  {Scemama}, \citenamefont {Benali}, \citenamefont {Jacquemin}, \citenamefont
  {Caffarel},\ and\ \citenamefont {Loos}}]{scem18a}%
  \BibitemOpen
  \bibfield  {author} {\bibinfo {author} {\bibfnamefont {A.}~\bibnamefont
  {Scemama}}, \bibinfo {author} {\bibfnamefont {A.}~\bibnamefont {Benali}},
  \bibinfo {author} {\bibfnamefont {D.}~\bibnamefont {Jacquemin}}, \bibinfo
  {author} {\bibfnamefont {M.}~\bibnamefont {Caffarel}},\ and\ \bibinfo
  {author} {\bibfnamefont {P.-F.}\ \bibnamefont {Loos}},\ }\bibfield  {title}
  {\enquote {\bibinfo {title} {Excitation energies from diffusion monte carlo
  using selected configuration interaction nodes},}\ }\href
  {https://doi.org/10.1063/1.5041327} {\bibfield  {journal} {\bibinfo
  {journal} {J.~Chem.~Phys.}\ }\textbf {\bibinfo {volume} {149}},\ \bibinfo
  {pages} {034108} (\bibinfo {year} {2018})}\BibitemShut {NoStop}%
\bibitem [{\citenamefont {Garniron}\ \emph {et~al.}(2018)\citenamefont
  {Garniron}, \citenamefont {Scemama}, \citenamefont {Giner}, \citenamefont
  {Caffarel},\ and\ \citenamefont {Loos}}]{garn18a}%
  \BibitemOpen
  \bibfield  {author} {\bibinfo {author} {\bibfnamefont {Y.}~\bibnamefont
  {Garniron}}, \bibinfo {author} {\bibfnamefont {A.}~\bibnamefont {Scemama}},
  \bibinfo {author} {\bibfnamefont {E.}~\bibnamefont {Giner}}, \bibinfo
  {author} {\bibfnamefont {M.}~\bibnamefont {Caffarel}},\ and\ \bibinfo
  {author} {\bibfnamefont {P.-F.}\ \bibnamefont {Loos}},\ }\bibfield  {title}
  {\enquote {\bibinfo {title} {Selected configuration interaction dressed by
  perturbation},}\ }\href {https://doi.org/10.1063/1.5044503} {\bibfield
  {journal} {\bibinfo  {journal} {J.~Chem.~Phys.}\ }\textbf {\bibinfo {volume}
  {149}},\ \bibinfo {pages} {064103} (\bibinfo {year} {2018})}\BibitemShut
  {NoStop}%
\bibitem [{\citenamefont {Holmes}, \citenamefont {Tubman},\ and\ \citenamefont
  {Umrigar}(2016)}]{holm16a}%
  \BibitemOpen
  \bibfield  {author} {\bibinfo {author} {\bibfnamefont {A.~A.}\ \bibnamefont
  {Holmes}}, \bibinfo {author} {\bibfnamefont {N.~M.}\ \bibnamefont {Tubman}},\
  and\ \bibinfo {author} {\bibfnamefont {C.~J.}\ \bibnamefont {Umrigar}},\
  }\bibfield  {title} {\enquote {\bibinfo {title} {Heat-bath configuration
  interaction: An efficient selected configuration interaction algorithm
  inspired by heat-bath sampling},}\ }\href
  {https://doi.org/10.1021/acs.jctc.6b00407} {\bibfield  {journal} {\bibinfo
  {journal} {J.~Chem.~Theory~Comput.}\ }\textbf {\bibinfo {volume} {12}},\
  \bibinfo {pages} {3674--3680} (\bibinfo {year} {2016})}\BibitemShut {NoStop}%
\bibitem [{\citenamefont {Smith}\ \emph {et~al.}(2017)\citenamefont {Smith},
  \citenamefont {Mussard}, \citenamefont {Holmes},\ and\ \citenamefont
  {Sharma}}]{smit17a}%
  \BibitemOpen
  \bibfield  {author} {\bibinfo {author} {\bibfnamefont {J.~E.~T.}\
  \bibnamefont {Smith}}, \bibinfo {author} {\bibfnamefont {B.}~\bibnamefont
  {Mussard}}, \bibinfo {author} {\bibfnamefont {A.~A.}\ \bibnamefont
  {Holmes}},\ and\ \bibinfo {author} {\bibfnamefont {S.}~\bibnamefont
  {Sharma}},\ }\bibfield  {title} {\enquote {\bibinfo {title} {Cheap and near
  exact casscf with large active spaces},}\ }\href
  {https://doi.org/10.1021/acs.jctc.7b00900} {\bibfield  {journal} {\bibinfo
  {journal} {J. Chem. Theory Comput.}\ }\textbf {\bibinfo {volume} {13}},\
  \bibinfo {pages} {5468–--5478} (\bibinfo {year} {2017})}\BibitemShut
  {NoStop}%
\bibitem [{\citenamefont {Li}\ \emph {et~al.}(2018)\citenamefont {Li},
  \citenamefont {Otten}, \citenamefont {Holmes}, \citenamefont {Sharma},\ and\
  \citenamefont {Umrigar}}]{liju18a}%
  \BibitemOpen
  \bibfield  {author} {\bibinfo {author} {\bibfnamefont {J.}~\bibnamefont
  {Li}}, \bibinfo {author} {\bibfnamefont {M.}~\bibnamefont {Otten}}, \bibinfo
  {author} {\bibfnamefont {A.~A.}\ \bibnamefont {Holmes}}, \bibinfo {author}
  {\bibfnamefont {S.}~\bibnamefont {Sharma}},\ and\ \bibinfo {author}
  {\bibfnamefont {C.~J.}\ \bibnamefont {Umrigar}},\ }\bibfield  {title}
  {\enquote {\bibinfo {title} {Fast semistochastic heat-bath configuration
  interaction},}\ }\href {https://doi.org/10.1063/1.5055390} {\bibfield
  {journal} {\bibinfo  {journal} {J.~Chem.~Phys.}\ }\textbf {\bibinfo {volume}
  {149}},\ \bibinfo {pages} {214110} (\bibinfo {year} {2018})}\BibitemShut
  {NoStop}%
\bibitem [{\citenamefont {Larsson}\ \emph {et~al.}(2022)\citenamefont
  {Larsson}, \citenamefont {Zhai}, \citenamefont {Umrigar},\ and\ \citenamefont
  {Chan}}]{lars22a}%
  \BibitemOpen
  \bibfield  {author} {\bibinfo {author} {\bibfnamefont {H.~R.}\ \bibnamefont
  {Larsson}}, \bibinfo {author} {\bibfnamefont {H.}~\bibnamefont {Zhai}},
  \bibinfo {author} {\bibfnamefont {C.~J.}\ \bibnamefont {Umrigar}},\ and\
  \bibinfo {author} {\bibfnamefont {G.~K.-L.}\ \bibnamefont {Chan}},\
  }\bibfield  {title} {\enquote {\bibinfo {title} {The chromium dimer: Closing
  a chapter of quantum chemistry},}\ }\href
  {https://doi.org/10.1021/jacs.2c06357} {\bibfield  {journal} {\bibinfo
  {journal} {J.~Am.~Chem.~Soc.}\ }\textbf {\bibinfo {volume} {144}},\ \bibinfo
  {pages} {15932--15937} (\bibinfo {year} {2022})}\BibitemShut {NoStop}%
\bibitem [{\citenamefont {Goings}\ \emph {et~al.}(2022)\citenamefont {Goings},
  \citenamefont {White}, \citenamefont {Lee}, \citenamefont {Tautermann},
  \citenamefont {Degroote}, \citenamefont {Gidney}, \citenamefont {Shiozaki},
  \citenamefont {Babbush},\ and\ \citenamefont {Rubin}}]{goin22a}%
  \BibitemOpen
  \bibfield  {author} {\bibinfo {author} {\bibfnamefont {J.~J.}\ \bibnamefont
  {Goings}}, \bibinfo {author} {\bibfnamefont {A.}~\bibnamefont {White}},
  \bibinfo {author} {\bibfnamefont {J.}~\bibnamefont {Lee}}, \bibinfo {author}
  {\bibfnamefont {C.~S.}\ \bibnamefont {Tautermann}}, \bibinfo {author}
  {\bibfnamefont {M.}~\bibnamefont {Degroote}}, \bibinfo {author}
  {\bibfnamefont {C.}~\bibnamefont {Gidney}}, \bibinfo {author} {\bibfnamefont
  {T.}~\bibnamefont {Shiozaki}}, \bibinfo {author} {\bibfnamefont
  {R.}~\bibnamefont {Babbush}},\ and\ \bibinfo {author} {\bibfnamefont {N.~C.}\
  \bibnamefont {Rubin}},\ }\bibfield  {title} {\enquote {\bibinfo {title}
  {Reliably assessing the electronic structure of cytochrome p450 on today's
  classical computers and tomorrow's quantum computers},}\ }\href
  {https://doi.org/10.1073/pnas.2203533119} {\bibfield  {journal} {\bibinfo
  {journal} {Proc.~Nat.~Acad.~Sci.}\ }\textbf {\bibinfo {volume} {119}}
  (\bibinfo {year} {2022}),\ 10.1073/pnas.2203533119}\BibitemShut {NoStop}%
\bibitem [{\citenamefont {Lee}\ \emph {et~al.}(2022)\citenamefont {Lee},
  \citenamefont {Lee}, \citenamefont {Zhai}, \citenamefont {Tong},
  \citenamefont {Dalzell}, \citenamefont {Kumar}, \citenamefont {Helms},
  \citenamefont {Gray}, \citenamefont {Cui}, \citenamefont {Liu}, \citenamefont
  {Kastoryano}, \citenamefont {Babbush}, \citenamefont {Preskill},
  \citenamefont {Reichman}, \citenamefont {Campbell}, \citenamefont {Valeev},
  \citenamefont {Lin},\ and\ \citenamefont {Chan}}]{lees22a}%
  \BibitemOpen
  \bibfield  {author} {\bibinfo {author} {\bibfnamefont {S.}~\bibnamefont
  {Lee}}, \bibinfo {author} {\bibfnamefont {J.}~\bibnamefont {Lee}}, \bibinfo
  {author} {\bibfnamefont {H.}~\bibnamefont {Zhai}}, \bibinfo {author}
  {\bibfnamefont {Y.}~\bibnamefont {Tong}}, \bibinfo {author} {\bibfnamefont
  {A.~M.}\ \bibnamefont {Dalzell}}, \bibinfo {author} {\bibfnamefont
  {A.}~\bibnamefont {Kumar}}, \bibinfo {author} {\bibfnamefont
  {P.}~\bibnamefont {Helms}}, \bibinfo {author} {\bibfnamefont
  {J.}~\bibnamefont {Gray}}, \bibinfo {author} {\bibfnamefont {Z.-H.}\
  \bibnamefont {Cui}}, \bibinfo {author} {\bibfnamefont {W.}~\bibnamefont
  {Liu}}, \bibinfo {author} {\bibfnamefont {M.}~\bibnamefont {Kastoryano}},
  \bibinfo {author} {\bibfnamefont {R.}~\bibnamefont {Babbush}}, \bibinfo
  {author} {\bibfnamefont {J.}~\bibnamefont {Preskill}}, \bibinfo {author}
  {\bibfnamefont {D.~R.}\ \bibnamefont {Reichman}}, \bibinfo {author}
  {\bibfnamefont {E.~T.}\ \bibnamefont {Campbell}}, \bibinfo {author}
  {\bibfnamefont {E.~F.}\ \bibnamefont {Valeev}}, \bibinfo {author}
  {\bibfnamefont {L.}~\bibnamefont {Lin}},\ and\ \bibinfo {author}
  {\bibfnamefont {G.~K.-L.}\ \bibnamefont {Chan}},\ }\href
  {https://doi.org/10.48550/arxiv.2208.02199} {\enquote {\bibinfo {title} {Is
  there evidence for exponential quantum advantage in quantum chemistry?}}\ }
  (\bibinfo {year} {2022})\BibitemShut {NoStop}%
\bibitem [{\citenamefont {Cao}\ \emph {et~al.}(2019)\citenamefont {Cao},
  \citenamefont {Romero}, \citenamefont {Olson}, \citenamefont {Degroote},
  \citenamefont {Johnson}, \citenamefont {Kieferov{\'{a}}}, \citenamefont
  {Kivlichan}, \citenamefont {Menke}, \citenamefont {Peropadre}, \citenamefont
  {Sawaya}, \citenamefont {Sim}, \citenamefont {Veis},\ and\ \citenamefont
  {Aspuru-Guzik}}]{caoy19a}%
  \BibitemOpen
  \bibfield  {author} {\bibinfo {author} {\bibfnamefont {Y.}~\bibnamefont
  {Cao}}, \bibinfo {author} {\bibfnamefont {J.}~\bibnamefont {Romero}},
  \bibinfo {author} {\bibfnamefont {J.~P.}\ \bibnamefont {Olson}}, \bibinfo
  {author} {\bibfnamefont {M.}~\bibnamefont {Degroote}}, \bibinfo {author}
  {\bibfnamefont {P.~D.}\ \bibnamefont {Johnson}}, \bibinfo {author}
  {\bibfnamefont {M.}~\bibnamefont {Kieferov{\'{a}}}}, \bibinfo {author}
  {\bibfnamefont {I.~D.}\ \bibnamefont {Kivlichan}}, \bibinfo {author}
  {\bibfnamefont {T.}~\bibnamefont {Menke}}, \bibinfo {author} {\bibfnamefont
  {B.}~\bibnamefont {Peropadre}}, \bibinfo {author} {\bibfnamefont {N.~P.~D.}\
  \bibnamefont {Sawaya}}, \bibinfo {author} {\bibfnamefont {S.}~\bibnamefont
  {Sim}}, \bibinfo {author} {\bibfnamefont {L.}~\bibnamefont {Veis}},\ and\
  \bibinfo {author} {\bibfnamefont {A.}~\bibnamefont {Aspuru-Guzik}},\
  }\bibfield  {title} {\enquote {\bibinfo {title} {Quantum chemistry in the age
  of quantum computing},}\ }\href {https://doi.org/10.1021/acs.chemrev.8b00803}
  {\bibfield  {journal} {\bibinfo  {journal} {Chem.~Rev.}\ }\textbf {\bibinfo
  {volume} {119}},\ \bibinfo {pages} {10856--10915} (\bibinfo {year}
  {2019})}\BibitemShut {NoStop}%
\bibitem [{\citenamefont {Claudino}(2022)}]{clau22a}%
  \BibitemOpen
  \bibfield  {author} {\bibinfo {author} {\bibfnamefont {D.}~\bibnamefont
  {Claudino}},\ }\bibfield  {title} {\enquote {\bibinfo {title} {The basics of
  quantum computing for chemists},}\ }\href {https://doi.org/10.1002/qua.26990}
  {\bibfield  {journal} {\bibinfo  {journal} {Int.~J.~Quant.~Chem.}\ }\textbf
  {\bibinfo {volume} {122}} (\bibinfo {year} {2022}),\
  10.1002/qua.26990}\BibitemShut {NoStop}%
\bibitem [{\citenamefont {Dalton}\ \emph {et~al.}(2022)\citenamefont {Dalton},
  \citenamefont {Long}, \citenamefont {Yordanov}, \citenamefont {Smith},
  \citenamefont {Barnes}, \citenamefont {Mertig},\ and\ \citenamefont
  {Arvidsson-Shukur}}]{dalt22a}%
  \BibitemOpen
  \bibfield  {author} {\bibinfo {author} {\bibfnamefont {K.}~\bibnamefont
  {Dalton}}, \bibinfo {author} {\bibfnamefont {C.~K.}\ \bibnamefont {Long}},
  \bibinfo {author} {\bibfnamefont {Y.~S.}\ \bibnamefont {Yordanov}}, \bibinfo
  {author} {\bibfnamefont {C.~G.}\ \bibnamefont {Smith}}, \bibinfo {author}
  {\bibfnamefont {C.~H.~W.}\ \bibnamefont {Barnes}}, \bibinfo {author}
  {\bibfnamefont {N.}~\bibnamefont {Mertig}},\ and\ \bibinfo {author}
  {\bibfnamefont {D.~R.~M.}\ \bibnamefont {Arvidsson-Shukur}},\ }\href
  {https://doi.org/10.48550/ARXIV.2211.04505} {\enquote {\bibinfo {title}
  {Variational quantum chemistry requires gate-error probabilities below the
  fault-tolerance threshold},}\ } (\bibinfo {year} {2022}),\ \bibinfo {note}
  {arXiv:2211.04505}\BibitemShut {NoStop}%
\bibitem [{\citenamefont {Kitaev}(1995)}]{kita95a}%
  \BibitemOpen
  \bibfield  {author} {\bibinfo {author} {\bibfnamefont {A.~Y.}\ \bibnamefont
  {Kitaev}},\ }\href {https://doi.org/10.48550/ARXIV.QUANT-PH/9511026}
  {\enquote {\bibinfo {title} {Quantum measurements and the abelian stabilizer
  problem},}\ } (\bibinfo {year} {1995})\BibitemShut {NoStop}%
\bibitem [{\citenamefont {Aspuru-Guzik}\ \emph {et~al.}(2005)\citenamefont
  {Aspuru-Guzik}, \citenamefont {Dutoi}, \citenamefont {Love},\ and\
  \citenamefont {Head-Gordon}}]{aspu05a}%
  \BibitemOpen
  \bibfield  {author} {\bibinfo {author} {\bibfnamefont {A.}~\bibnamefont
  {Aspuru-Guzik}}, \bibinfo {author} {\bibfnamefont {A.~D.}\ \bibnamefont
  {Dutoi}}, \bibinfo {author} {\bibfnamefont {P.~J.}\ \bibnamefont {Love}},\
  and\ \bibinfo {author} {\bibfnamefont {M.}~\bibnamefont {Head-Gordon}},\
  }\bibfield  {title} {\enquote {\bibinfo {title} {Simulated quantum
  computation of molecular energies},}\ }\href
  {https://doi.org/10.1126/science.1113479} {\bibfield  {journal} {\bibinfo
  {journal} {Science}\ }\textbf {\bibinfo {volume} {309}},\ \bibinfo {pages}
  {1704--1707} (\bibinfo {year} {2005})}\BibitemShut {NoStop}%
\bibitem [{\citenamefont {Peruzzo}\ \emph {et~al.}(2014)\citenamefont
  {Peruzzo}, \citenamefont {McClean}, \citenamefont {Shadbolt}, \citenamefont
  {Yung}, \citenamefont {Zhou}, \citenamefont {Love}, \citenamefont
  {Aspuru-Guzik},\ and\ \citenamefont {O’Brien}}]{peru14a}%
  \BibitemOpen
  \bibfield  {author} {\bibinfo {author} {\bibfnamefont {A.}~\bibnamefont
  {Peruzzo}}, \bibinfo {author} {\bibfnamefont {J.}~\bibnamefont {McClean}},
  \bibinfo {author} {\bibfnamefont {P.}~\bibnamefont {Shadbolt}}, \bibinfo
  {author} {\bibfnamefont {M.-H.}\ \bibnamefont {Yung}}, \bibinfo {author}
  {\bibfnamefont {X.-Q.}\ \bibnamefont {Zhou}}, \bibinfo {author}
  {\bibfnamefont {P.~J.}\ \bibnamefont {Love}}, \bibinfo {author}
  {\bibfnamefont {A.}~\bibnamefont {Aspuru-Guzik}},\ and\ \bibinfo {author}
  {\bibfnamefont {J.~L.}\ \bibnamefont {O’Brien}},\ }\bibfield  {title}
  {\enquote {\bibinfo {title} {A variational eigenvalue solver on a photonic
  quantum processor},}\ }\href {https://doi.org/10.1038/ncomms5213} {\bibfield
  {journal} {\bibinfo  {journal} {Nat.~Commun.}\ }\textbf {\bibinfo {volume}
  {5}},\ \bibinfo {pages} {1--7} (\bibinfo {year} {2014})}\BibitemShut
  {NoStop}%
\bibitem [{\citenamefont {Tilly}\ \emph {et~al.}(2022)\citenamefont {Tilly},
  \citenamefont {Chen}, \citenamefont {Cao}, \citenamefont {Picozzi},
  \citenamefont {Setia}, \citenamefont {Li}, \citenamefont {Grant},
  \citenamefont {Wossnig}, \citenamefont {Rungger}, \citenamefont {Booth},\
  and\ \citenamefont {Tennyson}}]{till22a}%
  \BibitemOpen
  \bibfield  {author} {\bibinfo {author} {\bibfnamefont {J.}~\bibnamefont
  {Tilly}}, \bibinfo {author} {\bibfnamefont {H.}~\bibnamefont {Chen}},
  \bibinfo {author} {\bibfnamefont {S.}~\bibnamefont {Cao}}, \bibinfo {author}
  {\bibfnamefont {D.}~\bibnamefont {Picozzi}}, \bibinfo {author} {\bibfnamefont
  {K.}~\bibnamefont {Setia}}, \bibinfo {author} {\bibfnamefont
  {Y.}~\bibnamefont {Li}}, \bibinfo {author} {\bibfnamefont {E.}~\bibnamefont
  {Grant}}, \bibinfo {author} {\bibfnamefont {L.}~\bibnamefont {Wossnig}},
  \bibinfo {author} {\bibfnamefont {I.}~\bibnamefont {Rungger}}, \bibinfo
  {author} {\bibfnamefont {G.~H.}\ \bibnamefont {Booth}},\ and\ \bibinfo
  {author} {\bibfnamefont {J.}~\bibnamefont {Tennyson}},\ }\bibfield  {title}
  {\enquote {\bibinfo {title} {{The Variational Quantum Eigensolver: A review
  of methods and best practices}},}\ }\href
  {https://doi.org/10.1016/j.physrep.2022.08.003} {\bibfield  {journal}
  {\bibinfo  {journal} {Phys.~Rep.}\ }\textbf {\bibinfo {volume} {986}},\
  \bibinfo {pages} {1--128} (\bibinfo {year} {2022})}\BibitemShut {NoStop}%
\bibitem [{\citenamefont {Fedorov}\ \emph {et~al.}(2022)\citenamefont
  {Fedorov}, \citenamefont {Peng}, \citenamefont {Govind},\ and\ \citenamefont
  {Alexeev}}]{fedo22a}%
  \BibitemOpen
  \bibfield  {author} {\bibinfo {author} {\bibfnamefont {D.~A.}\ \bibnamefont
  {Fedorov}}, \bibinfo {author} {\bibfnamefont {B.}~\bibnamefont {Peng}},
  \bibinfo {author} {\bibfnamefont {N.}~\bibnamefont {Govind}},\ and\ \bibinfo
  {author} {\bibfnamefont {Y.}~\bibnamefont {Alexeev}},\ }\bibfield  {title}
  {\enquote {\bibinfo {title} {{VQE} method: A short survey and recent
  developments},}\ }\href {https://doi.org/10.1186/s41313-021-00032-6}
  {\bibfield  {journal} {\bibinfo  {journal} {Mat.~Theory}\ }\textbf {\bibinfo
  {volume} {6}},\ \bibinfo {pages} {1--21} (\bibinfo {year}
  {2022})}\BibitemShut {NoStop}%
\bibitem [{\citenamefont {Grimsley}\ \emph {et~al.}(2019)\citenamefont
  {Grimsley}, \citenamefont {Economou}, \citenamefont {Barnes},\ and\
  \citenamefont {Mayhall}}]{grim19a}%
  \BibitemOpen
  \bibfield  {author} {\bibinfo {author} {\bibfnamefont {H.~R.}\ \bibnamefont
  {Grimsley}}, \bibinfo {author} {\bibfnamefont {S.~E.}\ \bibnamefont
  {Economou}}, \bibinfo {author} {\bibfnamefont {E.}~\bibnamefont {Barnes}},\
  and\ \bibinfo {author} {\bibfnamefont {N.~J.}\ \bibnamefont {Mayhall}},\
  }\bibfield  {title} {\enquote {\bibinfo {title} {An adaptive variational
  algorithm for exact molecular simulations on a quantum computer},}\ }\href
  {https://doi.org/10.1038/s41467-019-10988-2} {\bibfield  {journal} {\bibinfo
  {journal} {Nat.~Commun.}\ }\textbf {\bibinfo {volume} {10}} (\bibinfo {year}
  {2019}),\ 10.1038/s41467-019-10988-2}\BibitemShut {NoStop}%
\bibitem [{\citenamefont {Tang}\ \emph {et~al.}(2021)\citenamefont {Tang},
  \citenamefont {Shkolnikov}, \citenamefont {Barron}, \citenamefont {Grimsley},
  \citenamefont {Mayhall}, \citenamefont {Barnes},\ and\ \citenamefont
  {Economou}}]{tang21a}%
  \BibitemOpen
  \bibfield  {author} {\bibinfo {author} {\bibfnamefont {H.~L.}\ \bibnamefont
  {Tang}}, \bibinfo {author} {\bibfnamefont {V.}~\bibnamefont {Shkolnikov}},
  \bibinfo {author} {\bibfnamefont {G.~S.}\ \bibnamefont {Barron}}, \bibinfo
  {author} {\bibfnamefont {H.~R.}\ \bibnamefont {Grimsley}}, \bibinfo {author}
  {\bibfnamefont {N.~J.}\ \bibnamefont {Mayhall}}, \bibinfo {author}
  {\bibfnamefont {E.}~\bibnamefont {Barnes}},\ and\ \bibinfo {author}
  {\bibfnamefont {S.~E.}\ \bibnamefont {Economou}},\ }\bibfield  {title}
  {\enquote {\bibinfo {title} {Qubit-{ADAPT}-{VQE}: An adaptive algorithm for
  constructing hardware-efficient ansätze on a quantum processor},}\ }\href
  {https://doi.org/10.1103/prxquantum.2.020310} {\bibfield  {journal} {\bibinfo
   {journal} {{PRX} Quantum}\ }\textbf {\bibinfo {volume} {2}} (\bibinfo {year}
  {2021}),\ 10.1103/prxquantum.2.020310}\BibitemShut {NoStop}%
\bibitem [{\citenamefont {Anastasiou}\ \emph {et~al.}(2022)\citenamefont
  {Anastasiou}, \citenamefont {Chen}, \citenamefont {Mayhall}, \citenamefont
  {Barnes},\ and\ \citenamefont {Economou}}]{anas22a}%
  \BibitemOpen
  \bibfield  {author} {\bibinfo {author} {\bibfnamefont {P.~G.}\ \bibnamefont
  {Anastasiou}}, \bibinfo {author} {\bibfnamefont {Y.}~\bibnamefont {Chen}},
  \bibinfo {author} {\bibfnamefont {N.~J.}\ \bibnamefont {Mayhall}}, \bibinfo
  {author} {\bibfnamefont {E.}~\bibnamefont {Barnes}},\ and\ \bibinfo {author}
  {\bibfnamefont {S.~E.}\ \bibnamefont {Economou}},\ }\href
  {https://doi.org/10.48550/arxiv.2209.10562} {\enquote {\bibinfo {title}
  {{TETRIS-ADAPT-VQE}: An adaptive algorithm that yields shallower, denser
  circuit ans\"{a}tze},}\ } (\bibinfo {year} {2022}),\ \bibinfo {note}
  {arXiv:2209.10562}\BibitemShut {NoStop}%
\bibitem [{\citenamefont {Takeshita}\ \emph {et~al.}(2020)\citenamefont
  {Takeshita}, \citenamefont {Rubin}, \citenamefont {Jiang}, \citenamefont
  {Lee}, \citenamefont {Babbush},\ and\ \citenamefont {McClean}}]{take20a}%
  \BibitemOpen
  \bibfield  {author} {\bibinfo {author} {\bibfnamefont {T.}~\bibnamefont
  {Takeshita}}, \bibinfo {author} {\bibfnamefont {N.~C.}\ \bibnamefont
  {Rubin}}, \bibinfo {author} {\bibfnamefont {Z.}~\bibnamefont {Jiang}},
  \bibinfo {author} {\bibfnamefont {E.}~\bibnamefont {Lee}}, \bibinfo {author}
  {\bibfnamefont {R.}~\bibnamefont {Babbush}},\ and\ \bibinfo {author}
  {\bibfnamefont {J.~R.}\ \bibnamefont {McClean}},\ }\bibfield  {title}
  {\enquote {\bibinfo {title} {Increasing the representation accuracy of
  quantum simulations of chemistry without extra quantum resources},}\ }\href
  {https://doi.org/10.1103/physrevx.10.011004} {\bibfield  {journal} {\bibinfo
  {journal} {Phys.~Rev.~X}\ }\textbf {\bibinfo {volume} {10}} (\bibinfo {year}
  {2020}),\ 10.1103/physrevx.10.011004}\BibitemShut {NoStop}%
\bibitem [{\citenamefont {Mizukami}\ \emph {et~al.}(2020)\citenamefont
  {Mizukami}, \citenamefont {Mitarai}, \citenamefont {Nakagawa}, \citenamefont
  {Yamamoto}, \citenamefont {Yan},\ and\ \citenamefont {ya~Ohnishi}}]{mizu20a}%
  \BibitemOpen
  \bibfield  {author} {\bibinfo {author} {\bibfnamefont {W.}~\bibnamefont
  {Mizukami}}, \bibinfo {author} {\bibfnamefont {K.}~\bibnamefont {Mitarai}},
  \bibinfo {author} {\bibfnamefont {Y.~O.}\ \bibnamefont {Nakagawa}}, \bibinfo
  {author} {\bibfnamefont {T.}~\bibnamefont {Yamamoto}}, \bibinfo {author}
  {\bibfnamefont {T.}~\bibnamefont {Yan}},\ and\ \bibinfo {author}
  {\bibfnamefont {Y.}~\bibnamefont {ya~Ohnishi}},\ }\bibfield  {title}
  {\enquote {\bibinfo {title} {Orbital optimized unitary coupled cluster theory
  for quantum computer},}\ }\href
  {https://doi.org/10.1103/physrevresearch.2.033421} {\bibfield  {journal}
  {\bibinfo  {journal} {Phys.~Rev.~Res.}\ }\textbf {\bibinfo {volume} {2}}
  (\bibinfo {year} {2020}),\ 10.1103/physrevresearch.2.033421}\BibitemShut
  {NoStop}%
\bibitem [{\citenamefont {Sokolov}\ \emph {et~al.}(2020)\citenamefont
  {Sokolov}, \citenamefont {Barkoutsos}, \citenamefont {Ollitrault},
  \citenamefont {Greenberg}, \citenamefont {Rice}, \citenamefont {Pistoia},\
  and\ \citenamefont {Tavernelli}}]{soko20a}%
  \BibitemOpen
  \bibfield  {author} {\bibinfo {author} {\bibfnamefont {I.~O.}\ \bibnamefont
  {Sokolov}}, \bibinfo {author} {\bibfnamefont {P.~K.}\ \bibnamefont
  {Barkoutsos}}, \bibinfo {author} {\bibfnamefont {P.~J.}\ \bibnamefont
  {Ollitrault}}, \bibinfo {author} {\bibfnamefont {D.}~\bibnamefont
  {Greenberg}}, \bibinfo {author} {\bibfnamefont {J.}~\bibnamefont {Rice}},
  \bibinfo {author} {\bibfnamefont {M.}~\bibnamefont {Pistoia}},\ and\ \bibinfo
  {author} {\bibfnamefont {I.}~\bibnamefont {Tavernelli}},\ }\bibfield  {title}
  {\enquote {\bibinfo {title} {Quantum orbital-optimized unitary coupled
  cluster methods in the strongly correlated regime: {C}an quantum algorithms
  outperform their classical equivalents?}}\ }\href
  {https://doi.org/10.1063/1.5141835} {\bibfield  {journal} {\bibinfo
  {journal} {J.~Chem.~Phys.}\ }\textbf {\bibinfo {volume} {152}},\ \bibinfo
  {pages} {124107} (\bibinfo {year} {2020})}\BibitemShut {NoStop}%
\bibitem [{\citenamefont {Yalouz}\ \emph {et~al.}(2021)\citenamefont {Yalouz},
  \citenamefont {Senjean}, \citenamefont {G\"{u}nther}, \citenamefont {Buda},
  \citenamefont {O'Brien},\ and\ \citenamefont {Visscher}}]{yalo21a}%
  \BibitemOpen
  \bibfield  {author} {\bibinfo {author} {\bibfnamefont {S.}~\bibnamefont
  {Yalouz}}, \bibinfo {author} {\bibfnamefont {B.}~\bibnamefont {Senjean}},
  \bibinfo {author} {\bibfnamefont {J.}~\bibnamefont {G\"{u}nther}}, \bibinfo
  {author} {\bibfnamefont {F.}~\bibnamefont {Buda}}, \bibinfo {author}
  {\bibfnamefont {T.~E.}\ \bibnamefont {O'Brien}},\ and\ \bibinfo {author}
  {\bibfnamefont {L.}~\bibnamefont {Visscher}},\ }\bibfield  {title} {\enquote
  {\bibinfo {title} {A state-averaged orbital-optimized hybrid
  quantum-classical algorithm for a democratic description of ground and
  excited states},}\ }\href {https://doi.org/10.1088/2058-9565/abd334}
  {\bibfield  {journal} {\bibinfo  {journal} {Quantum~Sci.~Technol.}\ }\textbf
  {\bibinfo {volume} {6}},\ \bibinfo {pages} {024004} (\bibinfo {year}
  {2021})}\BibitemShut {NoStop}%
\bibitem [{\citenamefont {Tilly}\ \emph {et~al.}(2021)\citenamefont {Tilly},
  \citenamefont {Sriluckshmy}, \citenamefont {Patel}, \citenamefont {Fontana},
  \citenamefont {Rungger}, \citenamefont {Grant}, \citenamefont {Anderson},
  \citenamefont {Tennyson},\ and\ \citenamefont
  {Booth}}]{PhysRevResearch.3.033230}%
  \BibitemOpen
  \bibfield  {author} {\bibinfo {author} {\bibfnamefont {J.}~\bibnamefont
  {Tilly}}, \bibinfo {author} {\bibfnamefont {P.~V.}\ \bibnamefont
  {Sriluckshmy}}, \bibinfo {author} {\bibfnamefont {A.}~\bibnamefont {Patel}},
  \bibinfo {author} {\bibfnamefont {E.}~\bibnamefont {Fontana}}, \bibinfo
  {author} {\bibfnamefont {I.}~\bibnamefont {Rungger}}, \bibinfo {author}
  {\bibfnamefont {E.}~\bibnamefont {Grant}}, \bibinfo {author} {\bibfnamefont
  {R.}~\bibnamefont {Anderson}}, \bibinfo {author} {\bibfnamefont
  {J.}~\bibnamefont {Tennyson}},\ and\ \bibinfo {author} {\bibfnamefont
  {G.~H.}\ \bibnamefont {Booth}},\ }\bibfield  {title} {\enquote {\bibinfo
  {title} {Reduced density matrix sampling: Self-consistent embedding and
  multiscale electronic structure on current generation quantum computers},}\
  }\href {https://doi.org/10.1103/PhysRevResearch.3.033230} {\bibfield
  {journal} {\bibinfo  {journal} {Phys. Rev. Research}\ }\textbf {\bibinfo
  {volume} {3}},\ \bibinfo {pages} {033230} (\bibinfo {year}
  {2021})}\BibitemShut {NoStop}%
\bibitem [{\citenamefont {Gocho}\ \emph {et~al.}(2021)\citenamefont {Gocho},
  \citenamefont {Gao}, \citenamefont {Kobayashi}, \citenamefont {Inagaki},\
  and\ \citenamefont {Hatanaka}}]{goch22a}%
  \BibitemOpen
  \bibfield  {author} {\bibinfo {author} {\bibfnamefont {S.}~\bibnamefont
  {Gocho}}, \bibinfo {author} {\bibfnamefont {Q.}~\bibnamefont {Gao}}, \bibinfo
  {author} {\bibfnamefont {T.}~\bibnamefont {Kobayashi}}, \bibinfo {author}
  {\bibfnamefont {T.}~\bibnamefont {Inagaki}},\ and\ \bibinfo {author}
  {\bibfnamefont {M.}~\bibnamefont {Hatanaka}},\ }\href
  {https://doi.org/10.48550/ARXIV.2110.14448} {\enquote {\bibinfo {title}
  {Enhancement of complete active space self-consistent field ({CASSCF})
  calculation accuracy on noisy quantum device: Use of spin-restricted ansatz
  and constrained optimizer for photo-excitation calculations of ethylene and
  phenol blue dye},}\ } (\bibinfo {year} {2021}),\ \bibinfo {note} {arXiv:
  2110.14448}\BibitemShut {NoStop}%
\bibitem [{\citenamefont {Bierman}, \citenamefont {Li},\ and\ \citenamefont
  {Lu}(2022)}]{bierman2022improving}%
  \BibitemOpen
  \bibfield  {author} {\bibinfo {author} {\bibfnamefont {J.}~\bibnamefont
  {Bierman}}, \bibinfo {author} {\bibfnamefont {Y.}~\bibnamefont {Li}},\ and\
  \bibinfo {author} {\bibfnamefont {J.}~\bibnamefont {Lu}},\ }\href
  {https://doi.org/10.48550/arXiv.2208.14431} {\enquote {\bibinfo {title}
  {Improving the accuracy of variational quantum eigensolvers with fewer qubits
  using orbital optimization},}\ } (\bibinfo {year} {2022}),\ \bibinfo {note}
  {arXiv: 2208.14431}\BibitemShut {NoStop}%
\bibitem [{\citenamefont {Omiya}\ \emph {et~al.}(2022)\citenamefont {Omiya},
  \citenamefont {Nakagawa}, \citenamefont {Koh}, \citenamefont {Mizukami},
  \citenamefont {Gao},\ and\ \citenamefont {Kobayashi}}]{omiy22a}%
  \BibitemOpen
  \bibfield  {author} {\bibinfo {author} {\bibfnamefont {K.}~\bibnamefont
  {Omiya}}, \bibinfo {author} {\bibfnamefont {Y.~O.}\ \bibnamefont {Nakagawa}},
  \bibinfo {author} {\bibfnamefont {S.}~\bibnamefont {Koh}}, \bibinfo {author}
  {\bibfnamefont {W.}~\bibnamefont {Mizukami}}, \bibinfo {author}
  {\bibfnamefont {Q.}~\bibnamefont {Gao}},\ and\ \bibinfo {author}
  {\bibfnamefont {T.}~\bibnamefont {Kobayashi}},\ }\bibfield  {title} {\enquote
  {\bibinfo {title} {Analytical energy gradient for state-averaged
  orbital-optimized variational quantum eigensolvers and its application to a
  photochemical reaction},}\ }\href {https://doi.org/10.1021/acs.jctc.1c00877}
  {\bibfield  {journal} {\bibinfo  {journal} {J.~Chem.~Theory~Comput.}\
  }\textbf {\bibinfo {volume} {18}},\ \bibinfo {pages} {741--748} (\bibinfo
  {year} {2022})}\BibitemShut {NoStop}%
\bibitem [{\citenamefont {Sun}, \citenamefont {Yang},\ and\ \citenamefont
  {Chan}(2017)}]{sunq17a}%
  \BibitemOpen
  \bibfield  {author} {\bibinfo {author} {\bibfnamefont {Q.}~\bibnamefont
  {Sun}}, \bibinfo {author} {\bibfnamefont {J.}~\bibnamefont {Yang}},\ and\
  \bibinfo {author} {\bibfnamefont {G.~K.-L.}\ \bibnamefont {Chan}},\
  }\bibfield  {title} {\enquote {\bibinfo {title} {A general second order
  complete active space self-consistent-field solver for large-scale
  systems},}\ }\href {https://doi.org/10.1016/j.cplett.2017.03.004} {\bibfield
  {journal} {\bibinfo  {journal} {Chem. Phys. Lett.}\ }\textbf {\bibinfo
  {volume} {683}},\ \bibinfo {pages} {291--299} (\bibinfo {year}
  {2017})}\BibitemShut {NoStop}%
\bibitem [{\citenamefont {Sun}\ \emph {et~al.}(2017)\citenamefont {Sun},
  \citenamefont {Berkelbach}, \citenamefont {Blunt}, \citenamefont {Booth},
  \citenamefont {Guo}, \citenamefont {Li}, \citenamefont {Liu}, \citenamefont
  {McClain}, \citenamefont {Sayfutyarova}, \citenamefont {Sharma},
  \citenamefont {Wouters},\ and\ \citenamefont {Chan}}]{sunq17b}%
  \BibitemOpen
  \bibfield  {author} {\bibinfo {author} {\bibfnamefont {Q.}~\bibnamefont
  {Sun}}, \bibinfo {author} {\bibfnamefont {T.~C.}\ \bibnamefont {Berkelbach}},
  \bibinfo {author} {\bibfnamefont {N.~S.}\ \bibnamefont {Blunt}}, \bibinfo
  {author} {\bibfnamefont {G.~H.}\ \bibnamefont {Booth}}, \bibinfo {author}
  {\bibfnamefont {S.}~\bibnamefont {Guo}}, \bibinfo {author} {\bibfnamefont
  {Z.}~\bibnamefont {Li}}, \bibinfo {author} {\bibfnamefont {J.}~\bibnamefont
  {Liu}}, \bibinfo {author} {\bibfnamefont {J.~D.}\ \bibnamefont {McClain}},
  \bibinfo {author} {\bibfnamefont {E.~R.}\ \bibnamefont {Sayfutyarova}},
  \bibinfo {author} {\bibfnamefont {S.}~\bibnamefont {Sharma}}, \bibinfo
  {author} {\bibfnamefont {S.}~\bibnamefont {Wouters}},\ and\ \bibinfo {author}
  {\bibfnamefont {G.~K.-L.}\ \bibnamefont {Chan}},\ }\bibfield  {title}
  {\enquote {\bibinfo {title} {\textsc{PySCF}: the python-based simulations of
  chemistry framework},}\ }\href {https://doi.org/10.1002/wcms.1340} {\bibfield
   {journal} {\bibinfo  {journal} {{WIREs~Comput.~Mol.~Sci}}\ }\textbf
  {\bibinfo {volume} {8}} (\bibinfo {year} {2017}),\
  10.1002/wcms.1340}\BibitemShut {NoStop}%
\bibitem [{\citenamefont {Sun}\ \emph {et~al.}(2020)\citenamefont {Sun},
  \citenamefont {Zhang}, \citenamefont {Banerjee}, \citenamefont {Bao},
  \citenamefont {Barbry}, \citenamefont {Blunt}, \citenamefont {Bogdanov},
  \citenamefont {Booth}, \citenamefont {Chen}, \citenamefont {Cui},
  \citenamefont {Eriksen}, \citenamefont {Gao}, \citenamefont {Guo},
  \citenamefont {Hermann}, \citenamefont {Hermes}, \citenamefont {Koh},
  \citenamefont {Koval}, \citenamefont {Lehtola}, \citenamefont {Li},
  \citenamefont {Liu}, \citenamefont {Mardirossian}, \citenamefont {McClain},
  \citenamefont {Motta}, \citenamefont {Mussard}, \citenamefont {Pham},
  \citenamefont {Pulkin}, \citenamefont {Purwanto}, \citenamefont {Robinson},
  \citenamefont {Ronca}, \citenamefont {Sayfutyarova}, \citenamefont
  {Scheurer}, \citenamefont {Schurkus}, \citenamefont {Smith}, \citenamefont
  {Sun}, \citenamefont {Sun}, \citenamefont {Upadhyay}, \citenamefont {Wagner},
  \citenamefont {Wang}, \citenamefont {White}, \citenamefont {Whitfield},
  \citenamefont {Williamson}, \citenamefont {Wouters}, \citenamefont {Yang},
  \citenamefont {Yu}, \citenamefont {Zhu}, \citenamefont {Berkelbach},
  \citenamefont {Sharma}, \citenamefont {Sokolov},\ and\ \citenamefont
  {Chan}}]{sunq20a}%
  \BibitemOpen
  \bibfield  {author} {\bibinfo {author} {\bibfnamefont {Q.}~\bibnamefont
  {Sun}}, \bibinfo {author} {\bibfnamefont {X.}~\bibnamefont {Zhang}}, \bibinfo
  {author} {\bibfnamefont {S.}~\bibnamefont {Banerjee}}, \bibinfo {author}
  {\bibfnamefont {P.}~\bibnamefont {Bao}}, \bibinfo {author} {\bibfnamefont
  {M.}~\bibnamefont {Barbry}}, \bibinfo {author} {\bibfnamefont {N.~S.}\
  \bibnamefont {Blunt}}, \bibinfo {author} {\bibfnamefont {N.~A.}\ \bibnamefont
  {Bogdanov}}, \bibinfo {author} {\bibfnamefont {G.~H.}\ \bibnamefont {Booth}},
  \bibinfo {author} {\bibfnamefont {J.}~\bibnamefont {Chen}}, \bibinfo {author}
  {\bibfnamefont {Z.-H.}\ \bibnamefont {Cui}}, \bibinfo {author} {\bibfnamefont
  {J.~J.}\ \bibnamefont {Eriksen}}, \bibinfo {author} {\bibfnamefont
  {Y.}~\bibnamefont {Gao}}, \bibinfo {author} {\bibfnamefont {S.}~\bibnamefont
  {Guo}}, \bibinfo {author} {\bibfnamefont {J.}~\bibnamefont {Hermann}},
  \bibinfo {author} {\bibfnamefont {M.~R.}\ \bibnamefont {Hermes}}, \bibinfo
  {author} {\bibfnamefont {K.}~\bibnamefont {Koh}}, \bibinfo {author}
  {\bibfnamefont {P.}~\bibnamefont {Koval}}, \bibinfo {author} {\bibfnamefont
  {S.}~\bibnamefont {Lehtola}}, \bibinfo {author} {\bibfnamefont
  {Z.}~\bibnamefont {Li}}, \bibinfo {author} {\bibfnamefont {J.}~\bibnamefont
  {Liu}}, \bibinfo {author} {\bibfnamefont {N.}~\bibnamefont {Mardirossian}},
  \bibinfo {author} {\bibfnamefont {J.~D.}\ \bibnamefont {McClain}}, \bibinfo
  {author} {\bibfnamefont {M.}~\bibnamefont {Motta}}, \bibinfo {author}
  {\bibfnamefont {B.}~\bibnamefont {Mussard}}, \bibinfo {author} {\bibfnamefont
  {H.~Q.}\ \bibnamefont {Pham}}, \bibinfo {author} {\bibfnamefont
  {A.}~\bibnamefont {Pulkin}}, \bibinfo {author} {\bibfnamefont
  {W.}~\bibnamefont {Purwanto}}, \bibinfo {author} {\bibfnamefont {P.~J.}\
  \bibnamefont {Robinson}}, \bibinfo {author} {\bibfnamefont {E.}~\bibnamefont
  {Ronca}}, \bibinfo {author} {\bibfnamefont {E.~R.}\ \bibnamefont
  {Sayfutyarova}}, \bibinfo {author} {\bibfnamefont {M.}~\bibnamefont
  {Scheurer}}, \bibinfo {author} {\bibfnamefont {H.~F.}\ \bibnamefont
  {Schurkus}}, \bibinfo {author} {\bibfnamefont {J.~E.~T.}\ \bibnamefont
  {Smith}}, \bibinfo {author} {\bibfnamefont {C.}~\bibnamefont {Sun}}, \bibinfo
  {author} {\bibfnamefont {S.-N.}\ \bibnamefont {Sun}}, \bibinfo {author}
  {\bibfnamefont {S.}~\bibnamefont {Upadhyay}}, \bibinfo {author}
  {\bibfnamefont {L.~K.}\ \bibnamefont {Wagner}}, \bibinfo {author}
  {\bibfnamefont {X.}~\bibnamefont {Wang}}, \bibinfo {author} {\bibfnamefont
  {A.}~\bibnamefont {White}}, \bibinfo {author} {\bibfnamefont {J.~D.}\
  \bibnamefont {Whitfield}}, \bibinfo {author} {\bibfnamefont {M.~J.}\
  \bibnamefont {Williamson}}, \bibinfo {author} {\bibfnamefont
  {S.}~\bibnamefont {Wouters}}, \bibinfo {author} {\bibfnamefont
  {J.}~\bibnamefont {Yang}}, \bibinfo {author} {\bibfnamefont {J.~M.}\
  \bibnamefont {Yu}}, \bibinfo {author} {\bibfnamefont {T.}~\bibnamefont
  {Zhu}}, \bibinfo {author} {\bibfnamefont {T.~C.}\ \bibnamefont {Berkelbach}},
  \bibinfo {author} {\bibfnamefont {S.}~\bibnamefont {Sharma}}, \bibinfo
  {author} {\bibfnamefont {A.~Y.}\ \bibnamefont {Sokolov}},\ and\ \bibinfo
  {author} {\bibfnamefont {G.~K.-L.}\ \bibnamefont {Chan}},\ }\bibfield
  {title} {\enquote {\bibinfo {title} {Recent developments in the
  \textsc{PySCF} program package},}\ }\href {https://doi.org/10.1063/5.0006074}
  {\bibfield  {journal} {\bibinfo  {journal} {J.~Chem.~Phys.}\ }\textbf
  {\bibinfo {volume} {153}},\ \bibinfo {pages} {024109} (\bibinfo {year}
  {2020})}\BibitemShut {NoStop}%
\bibitem [{\citenamefont {Knowles}\ and\ \citenamefont
  {Werner}(1985)}]{know85a}%
  \BibitemOpen
  \bibfield  {author} {\bibinfo {author} {\bibfnamefont {P.~J.}\ \bibnamefont
  {Knowles}}\ and\ \bibinfo {author} {\bibfnamefont {H.-J.}\ \bibnamefont
  {Werner}},\ }\bibfield  {title} {\enquote {\bibinfo {title} {An efficient
  second order {MCSCF} method for long configuration expansions},}\ }\href
  {https://doi.org/10.1016/0009-2614(85)80025-7} {\bibfield  {journal}
  {\bibinfo  {journal} {Chem. Phys. Lett.}\ }\textbf {\bibinfo {volume}
  {115}},\ \bibinfo {pages} {259--267} (\bibinfo {year} {1985})}\BibitemShut
  {NoStop}%
\bibitem [{\citenamefont {Ma}\ \emph {et~al.}(2017)\citenamefont {Ma},
  \citenamefont {Knecht}, \citenamefont {Keller},\ and\ \citenamefont
  {Reiher}}]{mayi16c}%
  \BibitemOpen
  \bibfield  {author} {\bibinfo {author} {\bibfnamefont {Y.}~\bibnamefont
  {Ma}}, \bibinfo {author} {\bibfnamefont {S.}~\bibnamefont {Knecht}}, \bibinfo
  {author} {\bibfnamefont {S.}~\bibnamefont {Keller}},\ and\ \bibinfo {author}
  {\bibfnamefont {M.}~\bibnamefont {Reiher}},\ }\bibfield  {title} {\enquote
  {\bibinfo {title} {Second-order self-consistent-field density-matrix
  renormalization group},}\ }\href {https://doi.org/10.1021/acs.jctc.6b01118}
  {\bibfield  {journal} {\bibinfo  {journal} {J.~Chem. Theory~Comput.}\
  }\textbf {\bibinfo {volume} {13}},\ \bibinfo {pages} {2533--2549} (\bibinfo
  {year} {2017})}\BibitemShut {NoStop}%
\bibitem [{\citenamefont {Garc{\'\i}a-P{\'e}rez}\ \emph
  {et~al.}(2021)\citenamefont {Garc{\'\i}a-P{\'e}rez}, \citenamefont {Rossi},
  \citenamefont {Sokolov}, \citenamefont {Tacchino}, \citenamefont
  {Barkoutsos}, \citenamefont {Mazzola}, \citenamefont {Tavernelli},\ and\
  \citenamefont {Maniscalco}}]{garcia2021learning}%
  \BibitemOpen
  \bibfield  {author} {\bibinfo {author} {\bibfnamefont {G.}~\bibnamefont
  {Garc{\'\i}a-P{\'e}rez}}, \bibinfo {author} {\bibfnamefont {M.~A.}\
  \bibnamefont {Rossi}}, \bibinfo {author} {\bibfnamefont {B.}~\bibnamefont
  {Sokolov}}, \bibinfo {author} {\bibfnamefont {F.}~\bibnamefont {Tacchino}},
  \bibinfo {author} {\bibfnamefont {P.~K.}\ \bibnamefont {Barkoutsos}},
  \bibinfo {author} {\bibfnamefont {G.}~\bibnamefont {Mazzola}}, \bibinfo
  {author} {\bibfnamefont {I.}~\bibnamefont {Tavernelli}},\ and\ \bibinfo
  {author} {\bibfnamefont {S.}~\bibnamefont {Maniscalco}},\ }\bibfield  {title}
  {\enquote {\bibinfo {title} {Learning to measure: Adaptive informationally
  complete generalized measurements for quantum algorithms},}\ }\href
  {https://doi.org/10.1103/PRXQuantum.2.040342} {\bibfield  {journal} {\bibinfo
   {journal} {PRX Quantum}\ }\textbf {\bibinfo {volume} {2}},\ \bibinfo {pages}
  {040342} (\bibinfo {year} {2021})}\BibitemShut {NoStop}%
\bibitem [{\citenamefont {Glos}\ \emph {et~al.}(2022)\citenamefont {Glos},
  \citenamefont {Nyk{\"a}nen}, \citenamefont {Borrelli}, \citenamefont
  {Maniscalco}, \citenamefont {Rossi}, \citenamefont {Zimbor{\'a}s},\ and\
  \citenamefont {Garc{\'\i}a-P{\'e}rez}}]{glos22a}%
  \BibitemOpen
  \bibfield  {author} {\bibinfo {author} {\bibfnamefont {A.}~\bibnamefont
  {Glos}}, \bibinfo {author} {\bibfnamefont {A.}~\bibnamefont {Nyk{\"a}nen}},
  \bibinfo {author} {\bibfnamefont {E.-M.}\ \bibnamefont {Borrelli}}, \bibinfo
  {author} {\bibfnamefont {S.}~\bibnamefont {Maniscalco}}, \bibinfo {author}
  {\bibfnamefont {M.~A.}\ \bibnamefont {Rossi}}, \bibinfo {author}
  {\bibfnamefont {Z.}~\bibnamefont {Zimbor{\'a}s}},\ and\ \bibinfo {author}
  {\bibfnamefont {G.}~\bibnamefont {Garc{\'\i}a-P{\'e}rez}},\ }\href
  {https://doi.org/10.48550/ARXIV.2208.07817} {\enquote {\bibinfo {title}
  {Adaptive povm implementations and measurement error mitigation strategies
  for near-term quantum devices},}\ } (\bibinfo {year} {2022})\BibitemShut
  {NoStop}%
\bibitem [{\citenamefont {Bauer}\ \emph {et~al.}(2020)\citenamefont {Bauer},
  \citenamefont {Bravyi}, \citenamefont {Motta},\ and\ \citenamefont
  {Chan}}]{baue20a}%
  \BibitemOpen
  \bibfield  {author} {\bibinfo {author} {\bibfnamefont {B.}~\bibnamefont
  {Bauer}}, \bibinfo {author} {\bibfnamefont {S.}~\bibnamefont {Bravyi}},
  \bibinfo {author} {\bibfnamefont {M.}~\bibnamefont {Motta}},\ and\ \bibinfo
  {author} {\bibfnamefont {G.~K.-L.}\ \bibnamefont {Chan}},\ }\bibfield
  {title} {\enquote {\bibinfo {title} {Quantum algorithms for quantum chemistry
  and quantum materials science},}\ }\href
  {https://doi.org/10.1021/acs.chemrev.9b00829} {\bibfield  {journal} {\bibinfo
   {journal} {Chem.~Rev.}\ }\textbf {\bibinfo {volume} {120}},\ \bibinfo
  {pages} {12685--12717} (\bibinfo {year} {2020})}\BibitemShut {NoStop}%
\bibitem [{\citenamefont {Miller}\ \emph {et~al.}(2022)\citenamefont {Miller},
  \citenamefont {Zimbor{\'a}s}, \citenamefont {Knecht}, \citenamefont
  {Maniscalco},\ and\ \citenamefont {Garc{\'i}a-P{\'e}rez}}]{mill22a}%
  \BibitemOpen
  \bibfield  {author} {\bibinfo {author} {\bibfnamefont {A.}~\bibnamefont
  {Miller}}, \bibinfo {author} {\bibfnamefont {Z.}~\bibnamefont
  {Zimbor{\'a}s}}, \bibinfo {author} {\bibfnamefont {S.}~\bibnamefont
  {Knecht}}, \bibinfo {author} {\bibfnamefont {S.}~\bibnamefont {Maniscalco}},\
  and\ \bibinfo {author} {\bibfnamefont {G.}~\bibnamefont
  {Garc{\'i}a-P{\'e}rez}},\ }\href {https://doi.org/10.48550/ARXIV.2212.09731}
  {\enquote {\bibinfo {title} {The {Bonsai} algorithm: grow your own
  fermion-to-qubit mapping},}\ } (\bibinfo {year} {2022}),\ \bibinfo {note}
  {arXiv: 2212.09731}\BibitemShut {NoStop}%
\bibitem [{\citenamefont {Jordan}\ and\ \citenamefont
  {Wigner}(1928)}]{jordan1928pauli}%
  \BibitemOpen
  \bibfield  {author} {\bibinfo {author} {\bibfnamefont {P.}~\bibnamefont
  {Jordan}}\ and\ \bibinfo {author} {\bibfnamefont {E.~P.}\ \bibnamefont
  {Wigner}},\ }\bibfield  {title} {\enquote {\bibinfo {title} {About the pauli
  exclusion principle},}\ }\href@noop {} {\bibfield  {journal} {\bibinfo
  {journal} {Z.~Phys.}\ }\textbf {\bibinfo {volume} {47}},\ \bibinfo {pages}
  {14--75} (\bibinfo {year} {1928})}\BibitemShut {NoStop}%
\bibitem [{\citenamefont {Bravyi}\ and\ \citenamefont
  {Kitaev}(2002)}]{bravyi2002fermionic}%
  \BibitemOpen
  \bibfield  {author} {\bibinfo {author} {\bibfnamefont {S.~B.}\ \bibnamefont
  {Bravyi}}\ and\ \bibinfo {author} {\bibfnamefont {A.~Y.}\ \bibnamefont
  {Kitaev}},\ }\bibfield  {title} {\enquote {\bibinfo {title} {Fermionic
  quantum computation},}\ }\href@noop {} {\bibfield  {journal} {\bibinfo
  {journal} {Ann.~Phys.}\ }\textbf {\bibinfo {volume} {298}},\ \bibinfo {pages}
  {210--226} (\bibinfo {year} {2002})}\BibitemShut {NoStop}%
\bibitem [{\citenamefont {tA~v}\ \emph {et~al.}(2021)\citenamefont {tA~v},
  \citenamefont {Anis}, \citenamefont {Abby-Mitchell}, \citenamefont {Abraham},
  \citenamefont {AduOffei}, \citenamefont {Agarwal}, \citenamefont {Agliardi},
  \citenamefont {Aharoni}, \citenamefont {Ajith}, \citenamefont {Akhalwaya},
  \citenamefont {Aleksandrowicz}, \citenamefont {Alexander}, \citenamefont
  {Amy}, \citenamefont {Anagolum}, \citenamefont {Anthony-Gandon},
  \citenamefont {Araujo}, \citenamefont {Arbel}, \citenamefont {Asfaw},
  \citenamefont {Athalye}, \citenamefont {Avkhadiev}, \citenamefont {Azaustre},
  \citenamefont {BHOLE}, \citenamefont {Bajpe}, \citenamefont {Banerjee},
  \citenamefont {Banerjee}, \citenamefont {Bang}, \citenamefont {Bansal},
  \citenamefont {Barkoutsos}, \citenamefont {Barnawal}, \citenamefont {Barron},
  \citenamefont {Barron}, \citenamefont {Bello}, \citenamefont {Ben-Haim},
  \citenamefont {Bennett}, \citenamefont {Bevenius}, \citenamefont {Bhatnagar},
  \citenamefont {Bhatnagar}, \citenamefont {Bhobe}, \citenamefont {Bianchini},
  \citenamefont {Bishop}, \citenamefont {Blank}, \citenamefont {Bolos},
  \citenamefont {Bopardikar}, \citenamefont {Bosch}, \citenamefont
  {Brandhofer}, \citenamefont {Brandon}, \citenamefont {Bravyi}, \citenamefont
  {Bryce-Fuller}, \citenamefont {Bucher}, \citenamefont {Burov}, \citenamefont
  {Cabrera}, \citenamefont {Calpin}, \citenamefont {Capelluto}, \citenamefont
  {Carballo}, \citenamefont {Carrascal}, \citenamefont {Carriker},
  \citenamefont {Carvalho}, \citenamefont {Chakrabarti}, \citenamefont {Chen},
  \citenamefont {Chen}, \citenamefont {Chen}, \citenamefont {Chen},
  \citenamefont {Chen}, \citenamefont {Chevallier}, \citenamefont {Chinda},
  \citenamefont {Cholarajan}, \citenamefont {Chow}, \citenamefont {Churchill},
  \citenamefont {CisterMoke}, \citenamefont {Claus}, \citenamefont {Clauss},
  \citenamefont {Clothier}, \citenamefont {Cocking}, \citenamefont {Cocuzzo},
  \citenamefont {Connor}, \citenamefont {Correa}, \citenamefont {Crockett},
  \citenamefont {Cross}, \citenamefont {Cross}, \citenamefont {Cross},
  \citenamefont {Cruz-Benito}, \citenamefont {Culver}, \citenamefont
  {C{\'o}rcoles-Gonzales}, \citenamefont {D}, \citenamefont {Dague},
  \citenamefont {Dandachi}, \citenamefont {Dangwal}, \citenamefont {Daniel},
  \citenamefont {Daniels}, \citenamefont {Dartiailh}, \citenamefont {Davila},
  \citenamefont {Debouni}, \citenamefont {Dekusar}, \citenamefont {Deshmukh},
  \citenamefont {Deshpande}, \citenamefont {Ding}, \citenamefont {Doi},
  \citenamefont {Dow}, \citenamefont {Downing}, \citenamefont {Drechsler},
  \citenamefont {Drudis}, \citenamefont {Dumitrescu}, \citenamefont {Dumon},
  \citenamefont {Duran}, \citenamefont {EL-Safty}, \citenamefont {Eastman},
  \citenamefont {Eberle}, \citenamefont {Ebrahimi}, \citenamefont {Eendebak},
  \citenamefont {Egger}, \citenamefont {ElePT}, \citenamefont {Elsayed},
  \citenamefont {Emilio}, \citenamefont {Espiricueta}, \citenamefont {Everitt},
  \citenamefont {Facoetti}, \citenamefont {Farida}, \citenamefont
  {Fern{\'a}ndez}, \citenamefont {Ferracin}, \citenamefont {Ferrari},
  \citenamefont {Ferrera}, \citenamefont {Fouilland}, \citenamefont {Frisch},
  \citenamefont {Fuhrer}, \citenamefont {Fuller}, \citenamefont {GEORGE},
  \citenamefont {Gacon}, \citenamefont {Gago}, \citenamefont {Gambella},
  \citenamefont {Gambetta}, \citenamefont {Gammanpila}, \citenamefont {Garcia},
  \citenamefont {Garg}, \citenamefont {Garion}, \citenamefont {Garrison},
  \citenamefont {Garrison}, \citenamefont {Gates}, \citenamefont {Gentinetta},
  \citenamefont {Georgiev}, \citenamefont {Gil}, \citenamefont {Gilliam},
  \citenamefont {Giridharan}, \citenamefont {Glen}, \citenamefont
  {Gomez-Mosquera}, \citenamefont {Gonzalo}, \citenamefont {de~la
  Puente~Gonz{\'a}lez}, \citenamefont {Gorzinski}, \citenamefont {Gould},
  \citenamefont {Greenberg}, \citenamefont {Grinko}, \citenamefont {Guan},
  \citenamefont {Guijo}, \citenamefont {Guillermo-Mijares-Vilarino},
  \citenamefont {Gunnels}, \citenamefont {Gupta}, \citenamefont {Gupta},
  \citenamefont {G{\"u}nther}, \citenamefont {Haglund}, \citenamefont {Haide},
  \citenamefont {Hamamura}, \citenamefont {Hamido}, \citenamefont {Harkins},
  \citenamefont {Hartman}, \citenamefont {Hasan}, \citenamefont {Havlicek},
  \citenamefont {Hellmers}, \citenamefont {Herok}, \citenamefont {Hill},
  \citenamefont {Hillmich}, \citenamefont {Hong}, \citenamefont {Horii},
  \citenamefont {Howington}, \citenamefont {Hu}, \citenamefont {Hu},
  \citenamefont {Huang}, \citenamefont {Huang}, \citenamefont {Huisman},
  \citenamefont {Imai}, \citenamefont {Imamichi}, \citenamefont {Ishizaki},
  \citenamefont {Ishwor}, \citenamefont {Iten}, \citenamefont {Itoko},
  \citenamefont {Ivrii}, \citenamefont {Javadi}, \citenamefont {Javadi-Abhari},
  \citenamefont {Javed}, \citenamefont {Jianhua}, \citenamefont {Jivrajani},
  \citenamefont {Johns}, \citenamefont {Johnstun}, \citenamefont
  {Jonathan-Shoemaker}, \citenamefont {JosDenmark}, \citenamefont {JoshDumo},
  \citenamefont {Judge}, \citenamefont {Kachmann}, \citenamefont {Kale},
  \citenamefont {Kanazawa}, \citenamefont {Kane}, \citenamefont {Kang-Bae},
  \citenamefont {Kapila}, \citenamefont {Karazeev}, \citenamefont {Kassebaum},
  \citenamefont {Kehrer}, \citenamefont {Kelso}, \citenamefont {Kelso},
  \citenamefont {van Kemenade}, \citenamefont {Khanderao}, \citenamefont
  {King}, \citenamefont {Kobayashi}, \citenamefont {Kovi11Day}, \citenamefont
  {Kovyrshin}, \citenamefont {Krishnakumar}, \citenamefont {Krishnamurthy},
  \citenamefont {Krishnan}, \citenamefont {Krsulich}, \citenamefont {Kumkar},
  \citenamefont {Kus}, \citenamefont {LaRose}, \citenamefont {Lacal},
  \citenamefont {Lambert}, \citenamefont {Landa}, \citenamefont {Lapeyre},
  \citenamefont {Latone}, \citenamefont {Lawrence}, \citenamefont {Lee},
  \citenamefont {Li}, \citenamefont {Liang}, \citenamefont {Lishman},
  \citenamefont {Liu}, \citenamefont {Liu}, \citenamefont {Lolcroc},
  \citenamefont {M}, \citenamefont {Madden}, \citenamefont {Maeng},
  \citenamefont {Maheshkar}, \citenamefont {Majmudar}, \citenamefont
  {Malyshev}, \citenamefont {Mandouh}, \citenamefont {Manela}, \citenamefont
  {Manjula}, \citenamefont {Marecek}, \citenamefont {Marques}, \citenamefont
  {Marwaha}, \citenamefont {Maslov}, \citenamefont {Maszota}, \citenamefont
  {Mathews}, \citenamefont {Matsuo}, \citenamefont {Mazhandu}, \citenamefont
  {McClure}, \citenamefont {McElaney}, \citenamefont {McElroy}, \citenamefont
  {McGarry}, \citenamefont {McKay}, \citenamefont {McPherson}, \citenamefont
  {Meesala}, \citenamefont {Meirom}, \citenamefont {Mendell}, \citenamefont
  {Metcalfe}, \citenamefont {Mevissen}, \citenamefont {Meyer}, \citenamefont
  {Mezzacapo}, \citenamefont {Midha}, \citenamefont {Millar}, \citenamefont
  {Miller}, \citenamefont {Miller}, \citenamefont {Minev}, \citenamefont
  {Mitchell}, \citenamefont {Moll}, \citenamefont {Montanez}, \citenamefont
  {Monteiro}, \citenamefont {Mooring}, \citenamefont {Morales}, \citenamefont
  {Moran}, \citenamefont {Morcuende}, \citenamefont {Mostafa}, \citenamefont
  {Motta}, \citenamefont {Moyard}, \citenamefont {Murali}, \citenamefont
  {Murata}, \citenamefont {M{\"u}ggenburg}, \citenamefont {NEMOZ},
  \citenamefont {Nadlinger}, \citenamefont {Nakanishi}, \citenamefont
  {Nannicini}, \citenamefont {Nation}, \citenamefont {Navarro}, \citenamefont
  {Naveh}, \citenamefont {Neagle}, \citenamefont {Neuweiler}, \citenamefont
  {Ngoueya}, \citenamefont {Nguyen}, \citenamefont {Nicander}, \citenamefont
  {Nick-Singstock}, \citenamefont {Niroula}, \citenamefont {Norlen},
  \citenamefont {NuoWenLei}, \citenamefont {O'Riordan}, \citenamefont
  {Ogunbayo}, \citenamefont {Ollitrault}, \citenamefont {Onodera},
  \citenamefont {Otaolea}, \citenamefont {Oud}, \citenamefont {Padilha},
  \citenamefont {Paik}, \citenamefont {Pal}, \citenamefont {Pang},
  \citenamefont {Panigrahi}, \citenamefont {Pascuzzi}, \citenamefont
  {Perriello}, \citenamefont {Peterson}, \citenamefont {Phan}, \citenamefont
  {Pilch}, \citenamefont {Piro}, \citenamefont {Pistoia}, \citenamefont
  {Piveteau}, \citenamefont {Plewa}, \citenamefont {Pocreau}, \citenamefont
  {Possel}, \citenamefont {Pozas-Kerstjens}, \citenamefont {Pracht},
  \citenamefont {Prokop}, \citenamefont {Prutyanov}, \citenamefont {Puri},
  \citenamefont {Puzzuoli}, \citenamefont {Pythonix}, \citenamefont
  {P{\'e}rez}, \citenamefont {Quant02}, \citenamefont {Quintiii}, \citenamefont
  {Rahman}, \citenamefont {Raja}, \citenamefont {Rajeev}, \citenamefont
  {Rajput}, \citenamefont {Ramagiri}, \citenamefont {Rao}, \citenamefont
  {Raymond}, \citenamefont {Reardon-Smith}, \citenamefont {Redondo},
  \citenamefont {Reuter}, \citenamefont {Rice}, \citenamefont {Riedemann},
  \citenamefont {Rietesh}, \citenamefont {Risinger}, \citenamefont {Rivero},
  \citenamefont {Rocca}, \citenamefont {Rodr{\'\i}guez}, \citenamefont
  {RohithKarur}, \citenamefont {Rosand}, \citenamefont {Rossmannek},
  \citenamefont {Ryu}, \citenamefont {SAPV}, \citenamefont {Sa}, \citenamefont
  {Saha}, \citenamefont {Ash-Saki}, \citenamefont {Salman}, \citenamefont
  {Sanand}, \citenamefont {Sandberg}, \citenamefont {Sandesara}, \citenamefont
  {Sapra}, \citenamefont {Sargsyan}, \citenamefont {Sarkar}, \citenamefont
  {Sathaye}, \citenamefont {Savola}, \citenamefont {Schmitt}, \citenamefont
  {Schnabel}, \citenamefont {Schoenfeld}, \citenamefont {Scholten},
  \citenamefont {Schoute}, \citenamefont {Schulterbrandt}, \citenamefont
  {Schwarm}, \citenamefont {Schweigert}, \citenamefont {Seaward}, \citenamefont
  {Sergi}, \citenamefont {Sertage}, \citenamefont {Setia}, \citenamefont
  {Shah}, \citenamefont {Shammah}, \citenamefont {Shanks}, \citenamefont
  {Sharma}, \citenamefont {Shaw}, \citenamefont {Shi}, \citenamefont
  {Shoemaker}, \citenamefont {Silva}, \citenamefont {Simonetto}, \citenamefont
  {Singh}, \citenamefont {Singh}, \citenamefont {Singh}, \citenamefont
  {Singkanipa}, \citenamefont {Siraichi}, \citenamefont {Siri}, \citenamefont
  {Sistos}, \citenamefont {Sistos}, \citenamefont {Sitdikov}, \citenamefont
  {Sivarajah}, \citenamefont {Slavikmew}, \citenamefont {Sletfjerding},
  \citenamefont {Smolin}, \citenamefont {Soeken}, \citenamefont {Sokolov},
  \citenamefont {Sokolov}, \citenamefont {Soloviev}, \citenamefont
  {SooluThomas}, \citenamefont {Starfish}, \citenamefont {Steenken},
  \citenamefont {Stypulkoski}, \citenamefont {Suau}, \citenamefont {Sun},
  \citenamefont {Sung}, \citenamefont {Suwama}, \citenamefont {S{\l}owik},
  \citenamefont {Taeja}, \citenamefont {Takahashi}, \citenamefont {Takawale},
  \citenamefont {Tavernelli}, \citenamefont {Taylor}, \citenamefont {Taylour},
  \citenamefont {Thomas}, \citenamefont {Tian}, \citenamefont {Tillet},
  \citenamefont {Tod}, \citenamefont {Tomasik}, \citenamefont {Tornow},
  \citenamefont {de~la Torre}, \citenamefont {Toural}, \citenamefont {Trabing},
  \citenamefont {Treinish}, \citenamefont {Trenev}, \citenamefont {TrishaPe},
  \citenamefont {Truger}, \citenamefont {Tsilimigkounakis}, \citenamefont
  {Tulsi}, \citenamefont {Tuna}, \citenamefont {Turner}, \citenamefont
  {Vaknin}, \citenamefont {Valcarce}, \citenamefont {Varchon}, \citenamefont
  {Vartak}, \citenamefont {Vazquez}, \citenamefont {Vijaywargiya},
  \citenamefont {Villar}, \citenamefont {Vishnu}, \citenamefont {Vogt-Lee},
  \citenamefont {Vuillot}, \citenamefont {WQ}, \citenamefont {Weaver},
  \citenamefont {Weidenfeller}, \citenamefont {Wieczorek}, \citenamefont
  {Wildstrom}, \citenamefont {Wilson}, \citenamefont {Winston}, \citenamefont
  {WinterSoldier}, \citenamefont {Woehr}, \citenamefont {Woerner},
  \citenamefont {Woo}, \citenamefont {Wood}, \citenamefont {Wood},
  \citenamefont {Wood}, \citenamefont {Wootton}, \citenamefont {Wright},
  \citenamefont {Xing}, \citenamefont {YU}, \citenamefont {Yaiza},
  \citenamefont {Yang}, \citenamefont {Yang}, \citenamefont {Yao},
  \citenamefont {Yeralin}, \citenamefont {Yonekura}, \citenamefont
  {Yonge-Mallo}, \citenamefont {Yoshida}, \citenamefont {Young}, \citenamefont
  {Yu}, \citenamefont {Yu}, \citenamefont {Yuma-Nakamura}, \citenamefont
  {Zachow}, \citenamefont {Zdanski}, \citenamefont {Zhang}, \citenamefont
  {Zidaru}, \citenamefont {Zimmermann}, \citenamefont {Zoufal}, \citenamefont
  {aeddins ibm}, \citenamefont {alexzhang13}, \citenamefont {b63},
  \citenamefont {bartek bartlomiej}, \citenamefont {bcamorrison}, \citenamefont
  {brandhsn}, \citenamefont {nick bronn}, \citenamefont {chetmurthy},
  \citenamefont {choerst ibm}, \citenamefont {comet}, \citenamefont {dalin27},
  \citenamefont {deeplokhande}, \citenamefont {dekel.meirom}, \citenamefont
  {derwind}, \citenamefont {dime10}, \citenamefont {dlasecki}, \citenamefont
  {ehchen}, \citenamefont {ewinston}, \citenamefont {fanizzamarco},
  \citenamefont {fs1132429}, \citenamefont {gadial}, \citenamefont
  {galeinston}, \citenamefont {georgezhou20}, \citenamefont {georgios ts},
  \citenamefont {gruu}, \citenamefont {hhorii}, \citenamefont {hhyap},
  \citenamefont {hykavitha}, \citenamefont {itoko}, \citenamefont
  {jeppevinkel}, \citenamefont {jessica angel7}, \citenamefont {jezerjojo14},
  \citenamefont {jliu45}, \citenamefont {johannesgreiner}, \citenamefont
  {jscott2}, \citenamefont {kUmezawa}, \citenamefont {klinvill}, \citenamefont
  {krutik2966}, \citenamefont {ma5x}, \citenamefont {michelle4654},
  \citenamefont {msuwama}, \citenamefont {nico lgrs}, \citenamefont
  {nrhawkins}, \citenamefont {ntgiwsvp}, \citenamefont {ordmoj}, \citenamefont
  {sagar pahwa}, \citenamefont {pritamsinha2304}, \citenamefont {rithikaadiga},
  \citenamefont {ryancocuzzo}, \citenamefont {saktar unr}, \citenamefont
  {saswati qiskit}, \citenamefont {sebastian mair}, \citenamefont {septembrr},
  \citenamefont {sethmerkel}, \citenamefont {sg495}, \citenamefont {shaashwat},
  \citenamefont {smturro2}, \citenamefont {sternparky}, \citenamefont
  {strickroman}, \citenamefont {tigerjack}, \citenamefont {tsura crisaldo},
  \citenamefont {upsideon}, \citenamefont {vadebayo49}, \citenamefont {welien},
  \citenamefont {willhbang}, \citenamefont {wmurphy collabstar}, \citenamefont
  {yang.luh}, \citenamefont {yuri@FreeBSD},\ and\ \citenamefont
  {{\v{C}}epulkovskis}}]{qisk21a}%
  \BibitemOpen
  \bibfield  {author} {\bibinfo {author} {\bibfnamefont {A.}~\bibnamefont
  {tA~v}}, \bibinfo {author} {\bibfnamefont {M.~S.}\ \bibnamefont {Anis}},
  \bibinfo {author} {\bibnamefont {Abby-Mitchell}}, \bibinfo {author}
  {\bibfnamefont {H.}~\bibnamefont {Abraham}}, \bibinfo {author} {\bibnamefont
  {AduOffei}}, \bibinfo {author} {\bibfnamefont {R.}~\bibnamefont {Agarwal}},
  \bibinfo {author} {\bibfnamefont {G.}~\bibnamefont {Agliardi}}, \bibinfo
  {author} {\bibfnamefont {M.}~\bibnamefont {Aharoni}}, \bibinfo {author}
  {\bibfnamefont {V.}~\bibnamefont {Ajith}}, \bibinfo {author} {\bibfnamefont
  {I.~Y.}\ \bibnamefont {Akhalwaya}}, \bibinfo {author} {\bibfnamefont
  {G.}~\bibnamefont {Aleksandrowicz}}, \bibinfo {author} {\bibfnamefont
  {T.}~\bibnamefont {Alexander}}, \bibinfo {author} {\bibfnamefont
  {M.}~\bibnamefont {Amy}}, \bibinfo {author} {\bibfnamefont {S.}~\bibnamefont
  {Anagolum}}, \bibinfo {author} {\bibnamefont {Anthony-Gandon}}, \bibinfo
  {author} {\bibfnamefont {I.~F.}\ \bibnamefont {Araujo}}, \bibinfo {author}
  {\bibfnamefont {E.}~\bibnamefont {Arbel}}, \bibinfo {author} {\bibfnamefont
  {A.}~\bibnamefont {Asfaw}}, \bibinfo {author} {\bibfnamefont
  {A.}~\bibnamefont {Athalye}}, \bibinfo {author} {\bibfnamefont
  {A.}~\bibnamefont {Avkhadiev}}, \bibinfo {author} {\bibfnamefont
  {C.}~\bibnamefont {Azaustre}}, \bibinfo {author} {\bibfnamefont
  {P.}~\bibnamefont {BHOLE}}, \bibinfo {author} {\bibfnamefont
  {V.}~\bibnamefont {Bajpe}}, \bibinfo {author} {\bibfnamefont
  {A.}~\bibnamefont {Banerjee}}, \bibinfo {author} {\bibfnamefont
  {S.}~\bibnamefont {Banerjee}}, \bibinfo {author} {\bibfnamefont
  {W.}~\bibnamefont {Bang}}, \bibinfo {author} {\bibfnamefont {A.}~\bibnamefont
  {Bansal}}, \bibinfo {author} {\bibfnamefont {P.}~\bibnamefont {Barkoutsos}},
  \bibinfo {author} {\bibfnamefont {A.}~\bibnamefont {Barnawal}}, \bibinfo
  {author} {\bibfnamefont {G.}~\bibnamefont {Barron}}, \bibinfo {author}
  {\bibfnamefont {G.~S.}\ \bibnamefont {Barron}}, \bibinfo {author}
  {\bibfnamefont {L.}~\bibnamefont {Bello}}, \bibinfo {author} {\bibfnamefont
  {Y.}~\bibnamefont {Ben-Haim}}, \bibinfo {author} {\bibfnamefont {M.~C.}\
  \bibnamefont {Bennett}}, \bibinfo {author} {\bibfnamefont {D.}~\bibnamefont
  {Bevenius}}, \bibinfo {author} {\bibfnamefont {D.}~\bibnamefont {Bhatnagar}},
  \bibinfo {author} {\bibfnamefont {P.}~\bibnamefont {Bhatnagar}}, \bibinfo
  {author} {\bibfnamefont {A.}~\bibnamefont {Bhobe}}, \bibinfo {author}
  {\bibfnamefont {P.}~\bibnamefont {Bianchini}}, \bibinfo {author}
  {\bibfnamefont {L.~S.}\ \bibnamefont {Bishop}}, \bibinfo {author}
  {\bibfnamefont {C.}~\bibnamefont {Blank}}, \bibinfo {author} {\bibfnamefont
  {S.}~\bibnamefont {Bolos}}, \bibinfo {author} {\bibfnamefont
  {S.}~\bibnamefont {Bopardikar}}, \bibinfo {author} {\bibfnamefont
  {S.}~\bibnamefont {Bosch}}, \bibinfo {author} {\bibfnamefont
  {S.}~\bibnamefont {Brandhofer}}, \bibinfo {author} {\bibnamefont {Brandon}},
  \bibinfo {author} {\bibfnamefont {S.}~\bibnamefont {Bravyi}}, \bibinfo
  {author} {\bibnamefont {Bryce-Fuller}}, \bibinfo {author} {\bibfnamefont
  {D.}~\bibnamefont {Bucher}}, \bibinfo {author} {\bibfnamefont
  {A.}~\bibnamefont {Burov}}, \bibinfo {author} {\bibfnamefont
  {F.}~\bibnamefont {Cabrera}}, \bibinfo {author} {\bibfnamefont
  {P.}~\bibnamefont {Calpin}}, \bibinfo {author} {\bibfnamefont
  {L.}~\bibnamefont {Capelluto}}, \bibinfo {author} {\bibfnamefont
  {J.}~\bibnamefont {Carballo}}, \bibinfo {author} {\bibfnamefont
  {G.}~\bibnamefont {Carrascal}}, \bibinfo {author} {\bibfnamefont
  {A.}~\bibnamefont {Carriker}}, \bibinfo {author} {\bibfnamefont
  {I.}~\bibnamefont {Carvalho}}, \bibinfo {author} {\bibfnamefont
  {R.}~\bibnamefont {Chakrabarti}}, \bibinfo {author} {\bibfnamefont
  {A.}~\bibnamefont {Chen}}, \bibinfo {author} {\bibfnamefont {C.-F.}\
  \bibnamefont {Chen}}, \bibinfo {author} {\bibfnamefont {E.}~\bibnamefont
  {Chen}}, \bibinfo {author} {\bibfnamefont {J.~C.}\ \bibnamefont {Chen}},
  \bibinfo {author} {\bibfnamefont {R.}~\bibnamefont {Chen}}, \bibinfo {author}
  {\bibfnamefont {F.}~\bibnamefont {Chevallier}}, \bibinfo {author}
  {\bibfnamefont {K.}~\bibnamefont {Chinda}}, \bibinfo {author} {\bibfnamefont
  {R.}~\bibnamefont {Cholarajan}}, \bibinfo {author} {\bibfnamefont {J.~M.}\
  \bibnamefont {Chow}}, \bibinfo {author} {\bibfnamefont {S.}~\bibnamefont
  {Churchill}}, \bibinfo {author} {\bibnamefont {CisterMoke}}, \bibinfo
  {author} {\bibfnamefont {C.}~\bibnamefont {Claus}}, \bibinfo {author}
  {\bibfnamefont {C.}~\bibnamefont {Clauss}}, \bibinfo {author} {\bibfnamefont
  {C.}~\bibnamefont {Clothier}}, \bibinfo {author} {\bibfnamefont
  {R.}~\bibnamefont {Cocking}}, \bibinfo {author} {\bibfnamefont
  {R.}~\bibnamefont {Cocuzzo}}, \bibinfo {author} {\bibfnamefont
  {J.}~\bibnamefont {Connor}}, \bibinfo {author} {\bibfnamefont
  {F.}~\bibnamefont {Correa}}, \bibinfo {author} {\bibfnamefont
  {Z.}~\bibnamefont {Crockett}}, \bibinfo {author} {\bibfnamefont {A.~J.}\
  \bibnamefont {Cross}}, \bibinfo {author} {\bibfnamefont {A.~W.}\ \bibnamefont
  {Cross}}, \bibinfo {author} {\bibfnamefont {S.}~\bibnamefont {Cross}},
  \bibinfo {author} {\bibfnamefont {J.}~\bibnamefont {Cruz-Benito}}, \bibinfo
  {author} {\bibfnamefont {C.}~\bibnamefont {Culver}}, \bibinfo {author}
  {\bibfnamefont {A.~D.}\ \bibnamefont {C{\'o}rcoles-Gonzales}}, \bibinfo
  {author} {\bibfnamefont {N.}~\bibnamefont {D}}, \bibinfo {author}
  {\bibfnamefont {S.}~\bibnamefont {Dague}}, \bibinfo {author} {\bibfnamefont
  {T.~E.}\ \bibnamefont {Dandachi}}, \bibinfo {author} {\bibfnamefont {A.~N.}\
  \bibnamefont {Dangwal}}, \bibinfo {author} {\bibfnamefont {J.}~\bibnamefont
  {Daniel}}, \bibinfo {author} {\bibfnamefont {M.}~\bibnamefont {Daniels}},
  \bibinfo {author} {\bibfnamefont {M.}~\bibnamefont {Dartiailh}}, \bibinfo
  {author} {\bibfnamefont {A.~R.}\ \bibnamefont {Davila}}, \bibinfo {author}
  {\bibfnamefont {F.}~\bibnamefont {Debouni}}, \bibinfo {author} {\bibfnamefont
  {A.}~\bibnamefont {Dekusar}}, \bibinfo {author} {\bibfnamefont
  {A.}~\bibnamefont {Deshmukh}}, \bibinfo {author} {\bibfnamefont
  {M.}~\bibnamefont {Deshpande}}, \bibinfo {author} {\bibfnamefont
  {D.}~\bibnamefont {Ding}}, \bibinfo {author} {\bibfnamefont {J.}~\bibnamefont
  {Doi}}, \bibinfo {author} {\bibfnamefont {E.~M.}\ \bibnamefont {Dow}},
  \bibinfo {author} {\bibfnamefont {P.}~\bibnamefont {Downing}}, \bibinfo
  {author} {\bibfnamefont {E.}~\bibnamefont {Drechsler}}, \bibinfo {author}
  {\bibfnamefont {M.~S.}\ \bibnamefont {Drudis}}, \bibinfo {author}
  {\bibfnamefont {E.}~\bibnamefont {Dumitrescu}}, \bibinfo {author}
  {\bibfnamefont {K.}~\bibnamefont {Dumon}}, \bibinfo {author} {\bibfnamefont
  {I.}~\bibnamefont {Duran}}, \bibinfo {author} {\bibfnamefont
  {K.}~\bibnamefont {EL-Safty}}, \bibinfo {author} {\bibfnamefont
  {E.}~\bibnamefont {Eastman}}, \bibinfo {author} {\bibfnamefont
  {G.}~\bibnamefont {Eberle}}, \bibinfo {author} {\bibfnamefont
  {A.}~\bibnamefont {Ebrahimi}}, \bibinfo {author} {\bibfnamefont
  {P.}~\bibnamefont {Eendebak}}, \bibinfo {author} {\bibfnamefont
  {D.}~\bibnamefont {Egger}}, \bibinfo {author} {\bibnamefont {ElePT}},
  \bibinfo {author} {\bibfnamefont {I.}~\bibnamefont {Elsayed}}, \bibinfo
  {author} {\bibnamefont {Emilio}}, \bibinfo {author} {\bibfnamefont
  {A.}~\bibnamefont {Espiricueta}}, \bibinfo {author} {\bibfnamefont
  {M.}~\bibnamefont {Everitt}}, \bibinfo {author} {\bibfnamefont
  {D.}~\bibnamefont {Facoetti}}, \bibinfo {author} {\bibnamefont {Farida}},
  \bibinfo {author} {\bibfnamefont {P.~M.}\ \bibnamefont {Fern{\'a}ndez}},
  \bibinfo {author} {\bibfnamefont {S.}~\bibnamefont {Ferracin}}, \bibinfo
  {author} {\bibfnamefont {D.}~\bibnamefont {Ferrari}}, \bibinfo {author}
  {\bibfnamefont {A.~H.}\ \bibnamefont {Ferrera}}, \bibinfo {author}
  {\bibfnamefont {R.}~\bibnamefont {Fouilland}}, \bibinfo {author}
  {\bibfnamefont {A.}~\bibnamefont {Frisch}}, \bibinfo {author} {\bibfnamefont
  {A.}~\bibnamefont {Fuhrer}}, \bibinfo {author} {\bibfnamefont
  {B.}~\bibnamefont {Fuller}}, \bibinfo {author} {\bibfnamefont
  {M.}~\bibnamefont {GEORGE}}, \bibinfo {author} {\bibfnamefont
  {J.}~\bibnamefont {Gacon}}, \bibinfo {author} {\bibfnamefont {B.~G.}\
  \bibnamefont {Gago}}, \bibinfo {author} {\bibfnamefont {C.}~\bibnamefont
  {Gambella}}, \bibinfo {author} {\bibfnamefont {J.~M.}\ \bibnamefont
  {Gambetta}}, \bibinfo {author} {\bibfnamefont {A.}~\bibnamefont
  {Gammanpila}}, \bibinfo {author} {\bibfnamefont {L.}~\bibnamefont {Garcia}},
  \bibinfo {author} {\bibfnamefont {T.}~\bibnamefont {Garg}}, \bibinfo {author}
  {\bibfnamefont {S.}~\bibnamefont {Garion}}, \bibinfo {author} {\bibfnamefont
  {J.~R.}\ \bibnamefont {Garrison}}, \bibinfo {author} {\bibfnamefont
  {J.}~\bibnamefont {Garrison}}, \bibinfo {author} {\bibfnamefont
  {T.}~\bibnamefont {Gates}}, \bibinfo {author} {\bibfnamefont
  {G.}~\bibnamefont {Gentinetta}}, \bibinfo {author} {\bibfnamefont
  {H.}~\bibnamefont {Georgiev}}, \bibinfo {author} {\bibfnamefont
  {L.}~\bibnamefont {Gil}}, \bibinfo {author} {\bibfnamefont {A.}~\bibnamefont
  {Gilliam}}, \bibinfo {author} {\bibfnamefont {A.}~\bibnamefont {Giridharan}},
  \bibinfo {author} {\bibnamefont {Glen}}, \bibinfo {author} {\bibfnamefont
  {J.}~\bibnamefont {Gomez-Mosquera}}, \bibinfo {author} {\bibnamefont
  {Gonzalo}}, \bibinfo {author} {\bibfnamefont {S.}~\bibnamefont {de~la
  Puente~Gonz{\'a}lez}}, \bibinfo {author} {\bibfnamefont {J.}~\bibnamefont
  {Gorzinski}}, \bibinfo {author} {\bibfnamefont {I.}~\bibnamefont {Gould}},
  \bibinfo {author} {\bibfnamefont {D.}~\bibnamefont {Greenberg}}, \bibinfo
  {author} {\bibfnamefont {D.}~\bibnamefont {Grinko}}, \bibinfo {author}
  {\bibfnamefont {W.}~\bibnamefont {Guan}}, \bibinfo {author} {\bibfnamefont
  {D.}~\bibnamefont {Guijo}}, \bibinfo {author} {\bibnamefont
  {Guillermo-Mijares-Vilarino}}, \bibinfo {author} {\bibfnamefont {J.~A.}\
  \bibnamefont {Gunnels}}, \bibinfo {author} {\bibfnamefont {H.}~\bibnamefont
  {Gupta}}, \bibinfo {author} {\bibfnamefont {N.}~\bibnamefont {Gupta}},
  \bibinfo {author} {\bibfnamefont {J.~M.}\ \bibnamefont {G{\"u}nther}},
  \bibinfo {author} {\bibfnamefont {M.}~\bibnamefont {Haglund}}, \bibinfo
  {author} {\bibfnamefont {I.}~\bibnamefont {Haide}}, \bibinfo {author}
  {\bibfnamefont {I.}~\bibnamefont {Hamamura}}, \bibinfo {author}
  {\bibfnamefont {O.~C.}\ \bibnamefont {Hamido}}, \bibinfo {author}
  {\bibfnamefont {F.}~\bibnamefont {Harkins}}, \bibinfo {author} {\bibfnamefont
  {K.}~\bibnamefont {Hartman}}, \bibinfo {author} {\bibfnamefont
  {A.}~\bibnamefont {Hasan}}, \bibinfo {author} {\bibfnamefont
  {V.}~\bibnamefont {Havlicek}}, \bibinfo {author} {\bibfnamefont
  {J.}~\bibnamefont {Hellmers}}, \bibinfo {author} {\bibfnamefont
  {{\L}.}~\bibnamefont {Herok}}, \bibinfo {author} {\bibfnamefont
  {R.}~\bibnamefont {Hill}}, \bibinfo {author} {\bibfnamefont {S.}~\bibnamefont
  {Hillmich}}, \bibinfo {author} {\bibfnamefont {C.}~\bibnamefont {Hong}},
  \bibinfo {author} {\bibfnamefont {H.}~\bibnamefont {Horii}}, \bibinfo
  {author} {\bibfnamefont {C.}~\bibnamefont {Howington}}, \bibinfo {author}
  {\bibfnamefont {S.}~\bibnamefont {Hu}}, \bibinfo {author} {\bibfnamefont
  {W.}~\bibnamefont {Hu}}, \bibinfo {author} {\bibfnamefont {C.-H.}\
  \bibnamefont {Huang}}, \bibinfo {author} {\bibfnamefont {J.}~\bibnamefont
  {Huang}}, \bibinfo {author} {\bibfnamefont {R.}~\bibnamefont {Huisman}},
  \bibinfo {author} {\bibfnamefont {H.}~\bibnamefont {Imai}}, \bibinfo {author}
  {\bibfnamefont {T.}~\bibnamefont {Imamichi}}, \bibinfo {author}
  {\bibfnamefont {K.}~\bibnamefont {Ishizaki}}, \bibinfo {author} {\bibnamefont
  {Ishwor}}, \bibinfo {author} {\bibfnamefont {R.}~\bibnamefont {Iten}},
  \bibinfo {author} {\bibfnamefont {T.}~\bibnamefont {Itoko}}, \bibinfo
  {author} {\bibfnamefont {A.}~\bibnamefont {Ivrii}}, \bibinfo {author}
  {\bibfnamefont {A.}~\bibnamefont {Javadi}}, \bibinfo {author} {\bibfnamefont
  {A.}~\bibnamefont {Javadi-Abhari}}, \bibinfo {author} {\bibfnamefont
  {W.}~\bibnamefont {Javed}}, \bibinfo {author} {\bibfnamefont
  {Q.}~\bibnamefont {Jianhua}}, \bibinfo {author} {\bibfnamefont
  {M.}~\bibnamefont {Jivrajani}}, \bibinfo {author} {\bibfnamefont
  {K.}~\bibnamefont {Johns}}, \bibinfo {author} {\bibfnamefont
  {S.}~\bibnamefont {Johnstun}}, \bibinfo {author} {\bibnamefont
  {Jonathan-Shoemaker}}, \bibinfo {author} {\bibnamefont {JosDenmark}},
  \bibinfo {author} {\bibnamefont {JoshDumo}}, \bibinfo {author} {\bibfnamefont
  {J.}~\bibnamefont {Judge}}, \bibinfo {author} {\bibfnamefont
  {T.}~\bibnamefont {Kachmann}}, \bibinfo {author} {\bibfnamefont
  {A.}~\bibnamefont {Kale}}, \bibinfo {author} {\bibfnamefont {N.}~\bibnamefont
  {Kanazawa}}, \bibinfo {author} {\bibfnamefont {J.}~\bibnamefont {Kane}},
  \bibinfo {author} {\bibnamefont {Kang-Bae}}, \bibinfo {author} {\bibfnamefont
  {A.}~\bibnamefont {Kapila}}, \bibinfo {author} {\bibfnamefont
  {A.}~\bibnamefont {Karazeev}}, \bibinfo {author} {\bibfnamefont
  {P.}~\bibnamefont {Kassebaum}}, \bibinfo {author} {\bibfnamefont
  {T.}~\bibnamefont {Kehrer}}, \bibinfo {author} {\bibfnamefont
  {J.}~\bibnamefont {Kelso}}, \bibinfo {author} {\bibfnamefont
  {S.}~\bibnamefont {Kelso}}, \bibinfo {author} {\bibfnamefont
  {H.}~\bibnamefont {van Kemenade}}, \bibinfo {author} {\bibfnamefont
  {V.}~\bibnamefont {Khanderao}}, \bibinfo {author} {\bibfnamefont
  {S.}~\bibnamefont {King}}, \bibinfo {author} {\bibfnamefont {Y.}~\bibnamefont
  {Kobayashi}}, \bibinfo {author} {\bibnamefont {Kovi11Day}}, \bibinfo {author}
  {\bibfnamefont {A.}~\bibnamefont {Kovyrshin}}, \bibinfo {author}
  {\bibfnamefont {R.}~\bibnamefont {Krishnakumar}}, \bibinfo {author}
  {\bibfnamefont {P.}~\bibnamefont {Krishnamurthy}}, \bibinfo {author}
  {\bibfnamefont {V.}~\bibnamefont {Krishnan}}, \bibinfo {author}
  {\bibfnamefont {K.}~\bibnamefont {Krsulich}}, \bibinfo {author}
  {\bibfnamefont {P.}~\bibnamefont {Kumkar}}, \bibinfo {author} {\bibfnamefont
  {G.}~\bibnamefont {Kus}}, \bibinfo {author} {\bibfnamefont {R.}~\bibnamefont
  {LaRose}}, \bibinfo {author} {\bibfnamefont {E.}~\bibnamefont {Lacal}},
  \bibinfo {author} {\bibfnamefont {R.}~\bibnamefont {Lambert}}, \bibinfo
  {author} {\bibfnamefont {H.}~\bibnamefont {Landa}}, \bibinfo {author}
  {\bibfnamefont {J.}~\bibnamefont {Lapeyre}}, \bibinfo {author} {\bibfnamefont
  {J.}~\bibnamefont {Latone}}, \bibinfo {author} {\bibfnamefont
  {S.}~\bibnamefont {Lawrence}}, \bibinfo {author} {\bibfnamefont
  {C.}~\bibnamefont {Lee}}, \bibinfo {author} {\bibfnamefont {G.}~\bibnamefont
  {Li}}, \bibinfo {author} {\bibfnamefont {T.~J.}\ \bibnamefont {Liang}},
  \bibinfo {author} {\bibfnamefont {J.}~\bibnamefont {Lishman}}, \bibinfo
  {author} {\bibfnamefont {D.}~\bibnamefont {Liu}}, \bibinfo {author}
  {\bibfnamefont {P.}~\bibnamefont {Liu}}, \bibinfo {author} {\bibnamefont
  {Lolcroc}}, \bibinfo {author} {\bibfnamefont {A.~K.}\ \bibnamefont {M}},
  \bibinfo {author} {\bibfnamefont {L.}~\bibnamefont {Madden}}, \bibinfo
  {author} {\bibfnamefont {Y.}~\bibnamefont {Maeng}}, \bibinfo {author}
  {\bibfnamefont {S.}~\bibnamefont {Maheshkar}}, \bibinfo {author}
  {\bibfnamefont {K.}~\bibnamefont {Majmudar}}, \bibinfo {author}
  {\bibfnamefont {A.}~\bibnamefont {Malyshev}}, \bibinfo {author}
  {\bibfnamefont {M.~E.}\ \bibnamefont {Mandouh}}, \bibinfo {author}
  {\bibfnamefont {J.}~\bibnamefont {Manela}}, \bibinfo {author} {\bibnamefont
  {Manjula}}, \bibinfo {author} {\bibfnamefont {J.}~\bibnamefont {Marecek}},
  \bibinfo {author} {\bibfnamefont {M.}~\bibnamefont {Marques}}, \bibinfo
  {author} {\bibfnamefont {K.}~\bibnamefont {Marwaha}}, \bibinfo {author}
  {\bibfnamefont {D.}~\bibnamefont {Maslov}}, \bibinfo {author} {\bibfnamefont
  {P.}~\bibnamefont {Maszota}}, \bibinfo {author} {\bibfnamefont
  {D.}~\bibnamefont {Mathews}}, \bibinfo {author} {\bibfnamefont
  {A.}~\bibnamefont {Matsuo}}, \bibinfo {author} {\bibfnamefont
  {F.}~\bibnamefont {Mazhandu}}, \bibinfo {author} {\bibfnamefont
  {D.}~\bibnamefont {McClure}}, \bibinfo {author} {\bibfnamefont
  {M.}~\bibnamefont {McElaney}}, \bibinfo {author} {\bibfnamefont
  {J.}~\bibnamefont {McElroy}}, \bibinfo {author} {\bibfnamefont
  {C.}~\bibnamefont {McGarry}}, \bibinfo {author} {\bibfnamefont
  {D.}~\bibnamefont {McKay}}, \bibinfo {author} {\bibfnamefont
  {D.}~\bibnamefont {McPherson}}, \bibinfo {author} {\bibfnamefont
  {S.}~\bibnamefont {Meesala}}, \bibinfo {author} {\bibfnamefont
  {D.}~\bibnamefont {Meirom}}, \bibinfo {author} {\bibfnamefont
  {C.}~\bibnamefont {Mendell}}, \bibinfo {author} {\bibfnamefont
  {T.}~\bibnamefont {Metcalfe}}, \bibinfo {author} {\bibfnamefont
  {M.}~\bibnamefont {Mevissen}}, \bibinfo {author} {\bibfnamefont
  {A.}~\bibnamefont {Meyer}}, \bibinfo {author} {\bibfnamefont
  {A.}~\bibnamefont {Mezzacapo}}, \bibinfo {author} {\bibfnamefont
  {R.}~\bibnamefont {Midha}}, \bibinfo {author} {\bibfnamefont
  {D.}~\bibnamefont {Millar}}, \bibinfo {author} {\bibfnamefont
  {D.}~\bibnamefont {Miller}}, \bibinfo {author} {\bibfnamefont
  {H.}~\bibnamefont {Miller}}, \bibinfo {author} {\bibfnamefont
  {Z.}~\bibnamefont {Minev}}, \bibinfo {author} {\bibfnamefont
  {A.}~\bibnamefont {Mitchell}}, \bibinfo {author} {\bibfnamefont
  {N.}~\bibnamefont {Moll}}, \bibinfo {author} {\bibfnamefont {A.}~\bibnamefont
  {Montanez}}, \bibinfo {author} {\bibfnamefont {G.}~\bibnamefont {Monteiro}},
  \bibinfo {author} {\bibfnamefont {M.~D.}\ \bibnamefont {Mooring}}, \bibinfo
  {author} {\bibfnamefont {R.}~\bibnamefont {Morales}}, \bibinfo {author}
  {\bibfnamefont {N.}~\bibnamefont {Moran}}, \bibinfo {author} {\bibfnamefont
  {D.}~\bibnamefont {Morcuende}}, \bibinfo {author} {\bibfnamefont
  {S.}~\bibnamefont {Mostafa}}, \bibinfo {author} {\bibfnamefont
  {M.}~\bibnamefont {Motta}}, \bibinfo {author} {\bibfnamefont
  {R.}~\bibnamefont {Moyard}}, \bibinfo {author} {\bibfnamefont
  {P.}~\bibnamefont {Murali}}, \bibinfo {author} {\bibfnamefont
  {D.}~\bibnamefont {Murata}}, \bibinfo {author} {\bibfnamefont
  {J.}~\bibnamefont {M{\"u}ggenburg}}, \bibinfo {author} {\bibfnamefont
  {T.}~\bibnamefont {NEMOZ}}, \bibinfo {author} {\bibfnamefont
  {D.}~\bibnamefont {Nadlinger}}, \bibinfo {author} {\bibfnamefont
  {K.}~\bibnamefont {Nakanishi}}, \bibinfo {author} {\bibfnamefont
  {G.}~\bibnamefont {Nannicini}}, \bibinfo {author} {\bibfnamefont
  {P.}~\bibnamefont {Nation}}, \bibinfo {author} {\bibfnamefont
  {E.}~\bibnamefont {Navarro}}, \bibinfo {author} {\bibfnamefont
  {Y.}~\bibnamefont {Naveh}}, \bibinfo {author} {\bibfnamefont {S.~W.}\
  \bibnamefont {Neagle}}, \bibinfo {author} {\bibfnamefont {P.}~\bibnamefont
  {Neuweiler}}, \bibinfo {author} {\bibfnamefont {A.}~\bibnamefont {Ngoueya}},
  \bibinfo {author} {\bibfnamefont {T.}~\bibnamefont {Nguyen}}, \bibinfo
  {author} {\bibfnamefont {J.}~\bibnamefont {Nicander}}, \bibinfo {author}
  {\bibnamefont {Nick-Singstock}}, \bibinfo {author} {\bibfnamefont
  {P.}~\bibnamefont {Niroula}}, \bibinfo {author} {\bibfnamefont
  {H.}~\bibnamefont {Norlen}}, \bibinfo {author} {\bibnamefont {NuoWenLei}},
  \bibinfo {author} {\bibfnamefont {L.~J.}\ \bibnamefont {O'Riordan}}, \bibinfo
  {author} {\bibfnamefont {O.}~\bibnamefont {Ogunbayo}}, \bibinfo {author}
  {\bibfnamefont {P.}~\bibnamefont {Ollitrault}}, \bibinfo {author}
  {\bibfnamefont {T.}~\bibnamefont {Onodera}}, \bibinfo {author} {\bibfnamefont
  {R.}~\bibnamefont {Otaolea}}, \bibinfo {author} {\bibfnamefont
  {S.}~\bibnamefont {Oud}}, \bibinfo {author} {\bibfnamefont {D.}~\bibnamefont
  {Padilha}}, \bibinfo {author} {\bibfnamefont {H.}~\bibnamefont {Paik}},
  \bibinfo {author} {\bibfnamefont {S.}~\bibnamefont {Pal}}, \bibinfo {author}
  {\bibfnamefont {Y.}~\bibnamefont {Pang}}, \bibinfo {author} {\bibfnamefont
  {A.}~\bibnamefont {Panigrahi}}, \bibinfo {author} {\bibfnamefont {V.~R.}\
  \bibnamefont {Pascuzzi}}, \bibinfo {author} {\bibfnamefont {S.}~\bibnamefont
  {Perriello}}, \bibinfo {author} {\bibfnamefont {E.}~\bibnamefont {Peterson}},
  \bibinfo {author} {\bibfnamefont {A.}~\bibnamefont {Phan}}, \bibinfo {author}
  {\bibfnamefont {K.}~\bibnamefont {Pilch}}, \bibinfo {author} {\bibfnamefont
  {F.}~\bibnamefont {Piro}}, \bibinfo {author} {\bibfnamefont {M.}~\bibnamefont
  {Pistoia}}, \bibinfo {author} {\bibfnamefont {C.}~\bibnamefont {Piveteau}},
  \bibinfo {author} {\bibfnamefont {J.}~\bibnamefont {Plewa}}, \bibinfo
  {author} {\bibfnamefont {P.}~\bibnamefont {Pocreau}}, \bibinfo {author}
  {\bibfnamefont {C.}~\bibnamefont {Possel}}, \bibinfo {author} {\bibfnamefont
  {A.}~\bibnamefont {Pozas-Kerstjens}}, \bibinfo {author} {\bibfnamefont
  {R.}~\bibnamefont {Pracht}}, \bibinfo {author} {\bibfnamefont
  {M.}~\bibnamefont {Prokop}}, \bibinfo {author} {\bibfnamefont
  {V.}~\bibnamefont {Prutyanov}}, \bibinfo {author} {\bibfnamefont
  {S.}~\bibnamefont {Puri}}, \bibinfo {author} {\bibfnamefont {D.}~\bibnamefont
  {Puzzuoli}}, \bibinfo {author} {\bibnamefont {Pythonix}}, \bibinfo {author}
  {\bibfnamefont {J.}~\bibnamefont {P{\'e}rez}}, \bibinfo {author}
  {\bibnamefont {Quant02}}, \bibinfo {author} {\bibnamefont {Quintiii}},
  \bibinfo {author} {\bibfnamefont {R.~I.}\ \bibnamefont {Rahman}}, \bibinfo
  {author} {\bibfnamefont {A.}~\bibnamefont {Raja}}, \bibinfo {author}
  {\bibfnamefont {R.}~\bibnamefont {Rajeev}}, \bibinfo {author} {\bibfnamefont
  {I.}~\bibnamefont {Rajput}}, \bibinfo {author} {\bibfnamefont
  {N.}~\bibnamefont {Ramagiri}}, \bibinfo {author} {\bibfnamefont
  {A.}~\bibnamefont {Rao}}, \bibinfo {author} {\bibfnamefont {R.}~\bibnamefont
  {Raymond}}, \bibinfo {author} {\bibfnamefont {O.}~\bibnamefont
  {Reardon-Smith}}, \bibinfo {author} {\bibfnamefont {R.~M.-C.}\ \bibnamefont
  {Redondo}}, \bibinfo {author} {\bibfnamefont {M.}~\bibnamefont {Reuter}},
  \bibinfo {author} {\bibfnamefont {J.}~\bibnamefont {Rice}}, \bibinfo {author}
  {\bibfnamefont {M.}~\bibnamefont {Riedemann}}, \bibinfo {author}
  {\bibnamefont {Rietesh}}, \bibinfo {author} {\bibfnamefont {D.}~\bibnamefont
  {Risinger}}, \bibinfo {author} {\bibfnamefont {P.}~\bibnamefont {Rivero}},
  \bibinfo {author} {\bibfnamefont {M.~L.}\ \bibnamefont {Rocca}}, \bibinfo
  {author} {\bibfnamefont {D.~M.}\ \bibnamefont {Rodr{\'\i}guez}}, \bibinfo
  {author} {\bibnamefont {RohithKarur}}, \bibinfo {author} {\bibfnamefont
  {B.}~\bibnamefont {Rosand}}, \bibinfo {author} {\bibfnamefont
  {M.}~\bibnamefont {Rossmannek}}, \bibinfo {author} {\bibfnamefont
  {M.}~\bibnamefont {Ryu}}, \bibinfo {author} {\bibfnamefont {T.}~\bibnamefont
  {SAPV}}, \bibinfo {author} {\bibfnamefont {N.~R.~C.}\ \bibnamefont {Sa}},
  \bibinfo {author} {\bibfnamefont {A.}~\bibnamefont {Saha}}, \bibinfo {author}
  {\bibfnamefont {A.}~\bibnamefont {Ash-Saki}}, \bibinfo {author}
  {\bibfnamefont {A.}~\bibnamefont {Salman}}, \bibinfo {author} {\bibfnamefont
  {S.}~\bibnamefont {Sanand}}, \bibinfo {author} {\bibfnamefont
  {M.}~\bibnamefont {Sandberg}}, \bibinfo {author} {\bibfnamefont
  {H.}~\bibnamefont {Sandesara}}, \bibinfo {author} {\bibfnamefont
  {R.}~\bibnamefont {Sapra}}, \bibinfo {author} {\bibfnamefont
  {H.}~\bibnamefont {Sargsyan}}, \bibinfo {author} {\bibfnamefont
  {A.}~\bibnamefont {Sarkar}}, \bibinfo {author} {\bibfnamefont
  {N.}~\bibnamefont {Sathaye}}, \bibinfo {author} {\bibfnamefont
  {N.}~\bibnamefont {Savola}}, \bibinfo {author} {\bibfnamefont
  {B.}~\bibnamefont {Schmitt}}, \bibinfo {author} {\bibfnamefont
  {C.}~\bibnamefont {Schnabel}}, \bibinfo {author} {\bibfnamefont
  {Z.}~\bibnamefont {Schoenfeld}}, \bibinfo {author} {\bibfnamefont {T.~L.}\
  \bibnamefont {Scholten}}, \bibinfo {author} {\bibfnamefont {E.}~\bibnamefont
  {Schoute}}, \bibinfo {author} {\bibfnamefont {M.}~\bibnamefont
  {Schulterbrandt}}, \bibinfo {author} {\bibfnamefont {J.}~\bibnamefont
  {Schwarm}}, \bibinfo {author} {\bibfnamefont {P.}~\bibnamefont {Schweigert}},
  \bibinfo {author} {\bibfnamefont {J.}~\bibnamefont {Seaward}}, \bibinfo
  {author} {\bibnamefont {Sergi}}, \bibinfo {author} {\bibfnamefont {I.~F.}\
  \bibnamefont {Sertage}}, \bibinfo {author} {\bibfnamefont {K.}~\bibnamefont
  {Setia}}, \bibinfo {author} {\bibfnamefont {F.}~\bibnamefont {Shah}},
  \bibinfo {author} {\bibfnamefont {N.}~\bibnamefont {Shammah}}, \bibinfo
  {author} {\bibfnamefont {W.}~\bibnamefont {Shanks}}, \bibinfo {author}
  {\bibfnamefont {R.}~\bibnamefont {Sharma}}, \bibinfo {author} {\bibfnamefont
  {P.}~\bibnamefont {Shaw}}, \bibinfo {author} {\bibfnamefont {Y.}~\bibnamefont
  {Shi}}, \bibinfo {author} {\bibfnamefont {J.}~\bibnamefont {Shoemaker}},
  \bibinfo {author} {\bibfnamefont {A.}~\bibnamefont {Silva}}, \bibinfo
  {author} {\bibfnamefont {A.}~\bibnamefont {Simonetto}}, \bibinfo {author}
  {\bibfnamefont {D.}~\bibnamefont {Singh}}, \bibinfo {author} {\bibfnamefont
  {D.}~\bibnamefont {Singh}}, \bibinfo {author} {\bibfnamefont
  {P.}~\bibnamefont {Singh}}, \bibinfo {author} {\bibfnamefont
  {P.}~\bibnamefont {Singkanipa}}, \bibinfo {author} {\bibfnamefont
  {Y.}~\bibnamefont {Siraichi}}, \bibinfo {author} {\bibnamefont {Siri}},
  \bibinfo {author} {\bibfnamefont {J.}~\bibnamefont {Sistos}}, \bibinfo
  {author} {\bibfnamefont {J.}~\bibnamefont {Sistos}}, \bibinfo {author}
  {\bibfnamefont {I.}~\bibnamefont {Sitdikov}}, \bibinfo {author}
  {\bibfnamefont {S.}~\bibnamefont {Sivarajah}}, \bibinfo {author}
  {\bibnamefont {Slavikmew}}, \bibinfo {author} {\bibfnamefont {M.~B.}\
  \bibnamefont {Sletfjerding}}, \bibinfo {author} {\bibfnamefont {J.~A.}\
  \bibnamefont {Smolin}}, \bibinfo {author} {\bibfnamefont {M.}~\bibnamefont
  {Soeken}}, \bibinfo {author} {\bibfnamefont {I.~O.}\ \bibnamefont {Sokolov}},
  \bibinfo {author} {\bibfnamefont {I.}~\bibnamefont {Sokolov}}, \bibinfo
  {author} {\bibfnamefont {V.~P.}\ \bibnamefont {Soloviev}}, \bibinfo {author}
  {\bibnamefont {SooluThomas}}, \bibinfo {author} {\bibnamefont {Starfish}},
  \bibinfo {author} {\bibfnamefont {D.}~\bibnamefont {Steenken}}, \bibinfo
  {author} {\bibfnamefont {M.}~\bibnamefont {Stypulkoski}}, \bibinfo {author}
  {\bibfnamefont {A.}~\bibnamefont {Suau}}, \bibinfo {author} {\bibfnamefont
  {S.}~\bibnamefont {Sun}}, \bibinfo {author} {\bibfnamefont {K.~J.}\
  \bibnamefont {Sung}}, \bibinfo {author} {\bibfnamefont {M.}~\bibnamefont
  {Suwama}}, \bibinfo {author} {\bibfnamefont {O.}~\bibnamefont {S{\l}owik}},
  \bibinfo {author} {\bibfnamefont {R.}~\bibnamefont {Taeja}}, \bibinfo
  {author} {\bibfnamefont {H.}~\bibnamefont {Takahashi}}, \bibinfo {author}
  {\bibfnamefont {T.}~\bibnamefont {Takawale}}, \bibinfo {author}
  {\bibfnamefont {I.}~\bibnamefont {Tavernelli}}, \bibinfo {author}
  {\bibfnamefont {C.}~\bibnamefont {Taylor}}, \bibinfo {author} {\bibfnamefont
  {P.}~\bibnamefont {Taylour}}, \bibinfo {author} {\bibfnamefont
  {S.}~\bibnamefont {Thomas}}, \bibinfo {author} {\bibfnamefont
  {K.}~\bibnamefont {Tian}}, \bibinfo {author} {\bibfnamefont {M.}~\bibnamefont
  {Tillet}}, \bibinfo {author} {\bibfnamefont {M.}~\bibnamefont {Tod}},
  \bibinfo {author} {\bibfnamefont {M.}~\bibnamefont {Tomasik}}, \bibinfo
  {author} {\bibfnamefont {C.}~\bibnamefont {Tornow}}, \bibinfo {author}
  {\bibfnamefont {E.}~\bibnamefont {de~la Torre}}, \bibinfo {author}
  {\bibfnamefont {J.~L.~S.}\ \bibnamefont {Toural}}, \bibinfo {author}
  {\bibfnamefont {K.}~\bibnamefont {Trabing}}, \bibinfo {author} {\bibfnamefont
  {M.}~\bibnamefont {Treinish}}, \bibinfo {author} {\bibfnamefont
  {D.}~\bibnamefont {Trenev}}, \bibinfo {author} {\bibnamefont {TrishaPe}},
  \bibinfo {author} {\bibfnamefont {F.}~\bibnamefont {Truger}}, \bibinfo
  {author} {\bibfnamefont {G.}~\bibnamefont {Tsilimigkounakis}}, \bibinfo
  {author} {\bibfnamefont {D.}~\bibnamefont {Tulsi}}, \bibinfo {author}
  {\bibfnamefont {D.}~\bibnamefont {Tuna}}, \bibinfo {author} {\bibfnamefont
  {W.}~\bibnamefont {Turner}}, \bibinfo {author} {\bibfnamefont
  {Y.}~\bibnamefont {Vaknin}}, \bibinfo {author} {\bibfnamefont {C.~R.}\
  \bibnamefont {Valcarce}}, \bibinfo {author} {\bibfnamefont {F.}~\bibnamefont
  {Varchon}}, \bibinfo {author} {\bibfnamefont {A.}~\bibnamefont {Vartak}},
  \bibinfo {author} {\bibfnamefont {A.~C.}\ \bibnamefont {Vazquez}}, \bibinfo
  {author} {\bibfnamefont {P.}~\bibnamefont {Vijaywargiya}}, \bibinfo {author}
  {\bibfnamefont {V.}~\bibnamefont {Villar}}, \bibinfo {author} {\bibfnamefont
  {B.}~\bibnamefont {Vishnu}}, \bibinfo {author} {\bibfnamefont
  {D.}~\bibnamefont {Vogt-Lee}}, \bibinfo {author} {\bibfnamefont
  {C.}~\bibnamefont {Vuillot}}, \bibinfo {author} {\bibnamefont {WQ}}, \bibinfo
  {author} {\bibfnamefont {J.}~\bibnamefont {Weaver}}, \bibinfo {author}
  {\bibfnamefont {J.}~\bibnamefont {Weidenfeller}}, \bibinfo {author}
  {\bibfnamefont {R.}~\bibnamefont {Wieczorek}}, \bibinfo {author}
  {\bibfnamefont {J.~A.}\ \bibnamefont {Wildstrom}}, \bibinfo {author}
  {\bibfnamefont {J.}~\bibnamefont {Wilson}}, \bibinfo {author} {\bibfnamefont
  {E.}~\bibnamefont {Winston}}, \bibinfo {author} {\bibnamefont
  {WinterSoldier}}, \bibinfo {author} {\bibfnamefont {J.~J.}\ \bibnamefont
  {Woehr}}, \bibinfo {author} {\bibfnamefont {S.}~\bibnamefont {Woerner}},
  \bibinfo {author} {\bibfnamefont {R.}~\bibnamefont {Woo}}, \bibinfo {author}
  {\bibfnamefont {C.~J.}\ \bibnamefont {Wood}}, \bibinfo {author}
  {\bibfnamefont {R.}~\bibnamefont {Wood}}, \bibinfo {author} {\bibfnamefont
  {S.}~\bibnamefont {Wood}}, \bibinfo {author} {\bibfnamefont {J.}~\bibnamefont
  {Wootton}}, \bibinfo {author} {\bibfnamefont {M.}~\bibnamefont {Wright}},
  \bibinfo {author} {\bibfnamefont {L.}~\bibnamefont {Xing}}, \bibinfo {author}
  {\bibfnamefont {J.}~\bibnamefont {YU}}, \bibinfo {author} {\bibnamefont
  {Yaiza}}, \bibinfo {author} {\bibfnamefont {B.}~\bibnamefont {Yang}},
  \bibinfo {author} {\bibfnamefont {U.}~\bibnamefont {Yang}}, \bibinfo {author}
  {\bibfnamefont {J.}~\bibnamefont {Yao}}, \bibinfo {author} {\bibfnamefont
  {D.}~\bibnamefont {Yeralin}}, \bibinfo {author} {\bibfnamefont
  {R.}~\bibnamefont {Yonekura}}, \bibinfo {author} {\bibfnamefont
  {D.}~\bibnamefont {Yonge-Mallo}}, \bibinfo {author} {\bibfnamefont
  {R.}~\bibnamefont {Yoshida}}, \bibinfo {author} {\bibfnamefont
  {R.}~\bibnamefont {Young}}, \bibinfo {author} {\bibfnamefont
  {J.}~\bibnamefont {Yu}}, \bibinfo {author} {\bibfnamefont {L.}~\bibnamefont
  {Yu}}, \bibinfo {author} {\bibnamefont {Yuma-Nakamura}}, \bibinfo {author}
  {\bibfnamefont {C.}~\bibnamefont {Zachow}}, \bibinfo {author} {\bibfnamefont
  {L.}~\bibnamefont {Zdanski}}, \bibinfo {author} {\bibfnamefont
  {H.}~\bibnamefont {Zhang}}, \bibinfo {author} {\bibfnamefont
  {I.}~\bibnamefont {Zidaru}}, \bibinfo {author} {\bibfnamefont
  {B.}~\bibnamefont {Zimmermann}}, \bibinfo {author} {\bibfnamefont
  {C.}~\bibnamefont {Zoufal}}, \bibinfo {author} {\bibnamefont {aeddins ibm}},
  \bibinfo {author} {\bibnamefont {alexzhang13}}, \bibinfo {author}
  {\bibnamefont {b63}}, \bibinfo {author} {\bibnamefont {bartek bartlomiej}},
  \bibinfo {author} {\bibnamefont {bcamorrison}}, \bibinfo {author}
  {\bibnamefont {brandhsn}}, \bibinfo {author} {\bibnamefont {nick bronn}},
  \bibinfo {author} {\bibnamefont {chetmurthy}}, \bibinfo {author}
  {\bibnamefont {choerst ibm}}, \bibinfo {author} {\bibnamefont {comet}},
  \bibinfo {author} {\bibnamefont {dalin27}}, \bibinfo {author} {\bibnamefont
  {deeplokhande}}, \bibinfo {author} {\bibnamefont {dekel.meirom}}, \bibinfo
  {author} {\bibnamefont {derwind}}, \bibinfo {author} {\bibnamefont {dime10}},
  \bibinfo {author} {\bibnamefont {dlasecki}}, \bibinfo {author} {\bibnamefont
  {ehchen}}, \bibinfo {author} {\bibnamefont {ewinston}}, \bibinfo {author}
  {\bibnamefont {fanizzamarco}}, \bibinfo {author} {\bibnamefont {fs1132429}},
  \bibinfo {author} {\bibnamefont {gadial}}, \bibinfo {author} {\bibnamefont
  {galeinston}}, \bibinfo {author} {\bibnamefont {georgezhou20}}, \bibinfo
  {author} {\bibnamefont {georgios ts}}, \bibinfo {author} {\bibnamefont
  {gruu}}, \bibinfo {author} {\bibnamefont {hhorii}}, \bibinfo {author}
  {\bibnamefont {hhyap}}, \bibinfo {author} {\bibnamefont {hykavitha}},
  \bibinfo {author} {\bibnamefont {itoko}}, \bibinfo {author} {\bibnamefont
  {jeppevinkel}}, \bibinfo {author} {\bibnamefont {jessica angel7}}, \bibinfo
  {author} {\bibnamefont {jezerjojo14}}, \bibinfo {author} {\bibnamefont
  {jliu45}}, \bibinfo {author} {\bibnamefont {johannesgreiner}}, \bibinfo
  {author} {\bibnamefont {jscott2}}, \bibinfo {author} {\bibnamefont
  {kUmezawa}}, \bibinfo {author} {\bibnamefont {klinvill}}, \bibinfo {author}
  {\bibnamefont {krutik2966}}, \bibinfo {author} {\bibnamefont {ma5x}},
  \bibinfo {author} {\bibnamefont {michelle4654}}, \bibinfo {author}
  {\bibnamefont {msuwama}}, \bibinfo {author} {\bibnamefont {nico lgrs}},
  \bibinfo {author} {\bibnamefont {nrhawkins}}, \bibinfo {author} {\bibnamefont
  {ntgiwsvp}}, \bibinfo {author} {\bibnamefont {ordmoj}}, \bibinfo {author}
  {\bibnamefont {sagar pahwa}}, \bibinfo {author} {\bibnamefont
  {pritamsinha2304}}, \bibinfo {author} {\bibnamefont {rithikaadiga}}, \bibinfo
  {author} {\bibnamefont {ryancocuzzo}}, \bibinfo {author} {\bibnamefont
  {saktar unr}}, \bibinfo {author} {\bibnamefont {saswati qiskit}}, \bibinfo
  {author} {\bibnamefont {sebastian mair}}, \bibinfo {author} {\bibnamefont
  {septembrr}}, \bibinfo {author} {\bibnamefont {sethmerkel}}, \bibinfo
  {author} {\bibnamefont {sg495}}, \bibinfo {author} {\bibnamefont
  {shaashwat}}, \bibinfo {author} {\bibnamefont {smturro2}}, \bibinfo {author}
  {\bibnamefont {sternparky}}, \bibinfo {author} {\bibnamefont {strickroman}},
  \bibinfo {author} {\bibnamefont {tigerjack}}, \bibinfo {author} {\bibnamefont
  {tsura crisaldo}}, \bibinfo {author} {\bibnamefont {upsideon}}, \bibinfo
  {author} {\bibnamefont {vadebayo49}}, \bibinfo {author} {\bibnamefont
  {welien}}, \bibinfo {author} {\bibnamefont {willhbang}}, \bibinfo {author}
  {\bibnamefont {wmurphy collabstar}}, \bibinfo {author} {\bibnamefont
  {yang.luh}}, \bibinfo {author} {\bibnamefont {yuri@FreeBSD}},\ and\ \bibinfo
  {author} {\bibfnamefont {M.}~\bibnamefont {{\v{C}}epulkovskis}},\ }\href
  {https://doi.org/10.5281/zenodo.2573505} {\enquote {\bibinfo {title}
  {{Qiskit}: An open-source framework for quantum computing},}\ } (\bibinfo
  {year} {2021})\BibitemShut {NoStop}%
\bibitem [{\citenamefont {Seeley}, \citenamefont {Richard},\ and\ \citenamefont
  {Love}(2012)}]{seel12a}%
  \BibitemOpen
  \bibfield  {author} {\bibinfo {author} {\bibfnamefont {J.~T.}\ \bibnamefont
  {Seeley}}, \bibinfo {author} {\bibfnamefont {M.~J.}\ \bibnamefont
  {Richard}},\ and\ \bibinfo {author} {\bibfnamefont {P.~J.}\ \bibnamefont
  {Love}},\ }\bibfield  {title} {\enquote {\bibinfo {title} {The
  {Bravyi-Kitaev} transformation for quantum computation of electronic
  structure},}\ }\href {https://doi.org/10.1063/1.4768229} {\bibfield
  {journal} {\bibinfo  {journal} {J.~Chem.~Phys.}\ }\textbf {\bibinfo {volume}
  {137}},\ \bibinfo {pages} {224109} (\bibinfo {year} {2012})}\BibitemShut
  {NoStop}%
\bibitem [{\citenamefont {Nyk\"{a}nen}\ \emph {et~al.}(2022)\citenamefont
  {Nyk\"{a}nen}, \citenamefont {Rossi}, \citenamefont {Borrelli}, \citenamefont
  {Maniscalco},\ and\ \citenamefont {Garc{\'i}a-P{\'e}rez}}]{nyka22a}%
  \BibitemOpen
  \bibfield  {author} {\bibinfo {author} {\bibfnamefont {A.}~\bibnamefont
  {Nyk\"{a}nen}}, \bibinfo {author} {\bibfnamefont {M.~A.~C.}\ \bibnamefont
  {Rossi}}, \bibinfo {author} {\bibfnamefont {E.-M.}\ \bibnamefont {Borrelli}},
  \bibinfo {author} {\bibfnamefont {S.}~\bibnamefont {Maniscalco}},\ and\
  \bibinfo {author} {\bibfnamefont {G.}~\bibnamefont {Garc{\'i}a-P{\'e}rez}},\
  }\href {https://doi.org/10.48550/ARXIV.2212.09719} {\enquote {\bibinfo
  {title} {Mitigating the measurement overhead of {ADAPT-VQE} with optimised
  informationally complete generalised measurements},}\ } (\bibinfo {year}
  {2022}),\ \bibinfo {note} {arXiv: 2212.09719}\BibitemShut {NoStop}%
\bibitem [{aur(2022)}]{aurora}%
  \BibitemOpen
  \href@noop {} {} (\bibinfo {year} {2022}),\ \bibinfo {note} {\textsc{Aurora}
  v.0.1, Algorithmiq Ltd.}\BibitemShut {Stop}%
\bibitem [{\citenamefont {Gonthier}\ \emph {et~al.}(2022)\citenamefont
  {Gonthier}, \citenamefont {Radin}, \citenamefont {Buda}, \citenamefont
  {Doskocil}, \citenamefont {Abuan},\ and\ \citenamefont {Romero}}]{gont22a}%
  \BibitemOpen
  \bibfield  {author} {\bibinfo {author} {\bibfnamefont {J.~F.}\ \bibnamefont
  {Gonthier}}, \bibinfo {author} {\bibfnamefont {M.~D.}\ \bibnamefont {Radin}},
  \bibinfo {author} {\bibfnamefont {C.}~\bibnamefont {Buda}}, \bibinfo {author}
  {\bibfnamefont {E.~J.}\ \bibnamefont {Doskocil}}, \bibinfo {author}
  {\bibfnamefont {C.~M.}\ \bibnamefont {Abuan}},\ and\ \bibinfo {author}
  {\bibfnamefont {J.}~\bibnamefont {Romero}},\ }\bibfield  {title} {\enquote
  {\bibinfo {title} {Measurements as a roadblock to near-term practical quantum
  advantage in chemistry: Resource analysis},}\ }\href
  {https://doi.org/10.1103/physrevresearch.4.033154} {\bibfield  {journal}
  {\bibinfo  {journal} {Phys.~Rev.~Res.}\ }\textbf {\bibinfo {volume} {4}}
  (\bibinfo {year} {2022}),\ 10.1103/physrevresearch.4.033154}\BibitemShut
  {NoStop}%
\bibitem [{\citenamefont {{Dunning Jr.}}(1989)}]{dunn89}%
  \BibitemOpen
  \bibfield  {author} {\bibinfo {author} {\bibfnamefont {T.~H.}\ \bibnamefont
  {{Dunning Jr.}}},\ }\bibfield  {title} {\enquote {\bibinfo {title} {Gaussian
  basis sets for use in correlated molecular calculations. {I}. the atoms boron
  through neon and hydrogen},}\ }\href {https://doi.org/10.1063/1.456153}
  {\bibfield  {journal} {\bibinfo  {journal} {J. Chem. Phys.}\ }\textbf
  {\bibinfo {volume} {90}},\ \bibinfo {pages} {1007} (\bibinfo {year}
  {1989})}\BibitemShut {NoStop}%
\bibitem [{\citenamefont {Levine}\ \emph {et~al.}(2021)\citenamefont {Levine},
  \citenamefont {Durden}, \citenamefont {Esch}, \citenamefont {Liang},\ and\
  \citenamefont {Shu}}]{levi21a}%
  \BibitemOpen
  \bibfield  {author} {\bibinfo {author} {\bibfnamefont {B.~G.}\ \bibnamefont
  {Levine}}, \bibinfo {author} {\bibfnamefont {A.~S.}\ \bibnamefont {Durden}},
  \bibinfo {author} {\bibfnamefont {M.~P.}\ \bibnamefont {Esch}}, \bibinfo
  {author} {\bibfnamefont {F.}~\bibnamefont {Liang}},\ and\ \bibinfo {author}
  {\bibfnamefont {Y.}~\bibnamefont {Shu}},\ }\bibfield  {title} {\enquote
  {\bibinfo {title} {{CAS} without {SCF}{\textemdash}why to use {CASCI} and
  where to get the orbitals},}\ }\href {https://doi.org/10.1063/5.0042147}
  {\bibfield  {journal} {\bibinfo  {journal} {J.~Chem.~Phys.}\ }\textbf
  {\bibinfo {volume} {154}},\ \bibinfo {pages} {090902} (\bibinfo {year}
  {2021})}\BibitemShut {NoStop}%
\bibitem [{\citenamefont {Curutchet}\ and\ \citenamefont
  {Mennucci}(2017)}]{curu17a}%
  \BibitemOpen
  \bibfield  {author} {\bibinfo {author} {\bibfnamefont {C.}~\bibnamefont
  {Curutchet}}\ and\ \bibinfo {author} {\bibfnamefont {B.}~\bibnamefont
  {Mennucci}},\ }\bibfield  {title} {\enquote {\bibinfo {title} {{Quantum
  Chemical Studies of Light Harvesting}},}\ }\href
  {https://doi.org/10.1021/acs.chemrev.5b00700} {\bibfield  {journal} {\bibinfo
   {journal} {Chem. Rev.}\ }\textbf {\bibinfo {volume} {117}},\ \bibinfo
  {pages} {294--343} (\bibinfo {year} {2017})}\BibitemShut {NoStop}%
\bibitem [{nis()}]{nistO2}%
  \BibitemOpen
  \href@noop {} {}\bibinfo {note} {NIST Chemistry WebBook,
  https://webbook.nist.gov, accessed 13.12.2022}\BibitemShut {NoStop}%
\bibitem [{\citenamefont {Talarico}\ \emph {et~al.}()\citenamefont {Talarico},
  \citenamefont {Fitzpatrick}, \citenamefont {Maniscalco}, \citenamefont
  {Garc{\'\i}a-P{\'e}rez},\ and\ \citenamefont {Knecht}}]{vqe_pt2}%
  \BibitemOpen
  \bibfield  {author} {\bibinfo {author} {\bibfnamefont {W.}~\bibnamefont
  {Talarico}}, \bibinfo {author} {\bibfnamefont {A.}~\bibnamefont
  {Fitzpatrick}}, \bibinfo {author} {\bibfnamefont {S.}~\bibnamefont
  {Maniscalco}}, \bibinfo {author} {\bibfnamefont {G.}~\bibnamefont
  {Garc{\'\i}a-P{\'e}rez}},\ and\ \bibinfo {author} {\bibfnamefont
  {S.}~\bibnamefont {Knecht}},\ }\href@noop {} {\enquote {\bibinfo {title} {An
  efficient implementation of a multireference perturbation method for
  near-term quantum devices},}\ }\bibinfo {note} {Manuscript in preparation,
  2022.}\BibitemShut {Stop}%
\bibitem [{\citenamefont {Harding}\ \emph {et~al.}(2007)\citenamefont
  {Harding}, \citenamefont {Metzroth}, \citenamefont {Gauss},\ and\
  \citenamefont {Auer}}]{hard07a}%
  \BibitemOpen
  \bibfield  {author} {\bibinfo {author} {\bibfnamefont {M.~E.}\ \bibnamefont
  {Harding}}, \bibinfo {author} {\bibfnamefont {T.}~\bibnamefont {Metzroth}},
  \bibinfo {author} {\bibfnamefont {J.}~\bibnamefont {Gauss}},\ and\ \bibinfo
  {author} {\bibfnamefont {A.~A.}\ \bibnamefont {Auer}},\ }\bibfield  {title}
  {\enquote {\bibinfo {title} {Parallel calculation of {{CCSD}} and {{CCSD}(T)}
  analytic first and second derivatives},}\ }\href
  {https://doi.org/10.1021/ct700152c} {\bibfield  {journal} {\bibinfo
  {journal} {J.~Chem.~Theory~Comput.}\ }\textbf {\bibinfo {volume} {4}},\
  \bibinfo {pages} {64--74} (\bibinfo {year} {2007})}\BibitemShut {NoStop}%
\bibitem [{\citenamefont {Sayfutyarova}\ \emph {et~al.}(2017)\citenamefont
  {Sayfutyarova}, \citenamefont {Sun}, \citenamefont {Chan},\ and\
  \citenamefont {Knizia}}]{sayf17a}%
  \BibitemOpen
  \bibfield  {author} {\bibinfo {author} {\bibfnamefont {E.~R.}\ \bibnamefont
  {Sayfutyarova}}, \bibinfo {author} {\bibfnamefont {Q.}~\bibnamefont {Sun}},
  \bibinfo {author} {\bibfnamefont {G.~K.-L.}\ \bibnamefont {Chan}},\ and\
  \bibinfo {author} {\bibfnamefont {G.}~\bibnamefont {Knizia}},\ }\bibfield
  {title} {\enquote {\bibinfo {title} {Automated construction of molecular
  active spaces from atomic valence orbitals},}\ }\href
  {https://doi.org/10.1021/acs.jctc.7b00128} {\bibfield  {journal} {\bibinfo
  {journal} {J.~Chem.~Theory~Comput.}\ }\textbf {\bibinfo {volume} {13}},\
  \bibinfo {pages} {4063--4078} (\bibinfo {year} {2017})}\BibitemShut {NoStop}%
\bibitem [{\citenamefont {Liu}\ and\ \citenamefont {Peng}(2006)}]{liu2006}%
  \BibitemOpen
  \bibfield  {author} {\bibinfo {author} {\bibfnamefont {W.}~\bibnamefont
  {Liu}}\ and\ \bibinfo {author} {\bibfnamefont {D.}~\bibnamefont {Peng}},\
  }\bibfield  {title} {\enquote {\bibinfo {title} {Infinite-order
  quasirelativistic density functional method based on the exact matrix
  quasirelativistic theory},}\ }\href
  {https://doi.org/http://dx.doi.org/10.1063/1.2222365} {\bibfield  {journal}
  {\bibinfo  {journal} {J. Chem. Phys.}\ }\textbf {\bibinfo {volume} {125}},\
  \bibinfo {pages} {044102} (\bibinfo {year} {2006})}\BibitemShut {NoStop}%
\bibitem [{\citenamefont {Ilia{\v{s}}}\ and\ \citenamefont
  {Saue}(2007)}]{ilia07}%
  \BibitemOpen
  \bibfield  {author} {\bibinfo {author} {\bibfnamefont {M.}~\bibnamefont
  {Ilia{\v{s}}}}\ and\ \bibinfo {author} {\bibfnamefont {T.}~\bibnamefont
  {Saue}},\ }\bibfield  {title} {\enquote {\bibinfo {title} {An infinite-order
  two-component relativistic {Hamiltonian} by a simple one-step
  transformation},}\ }\href {https://doi.org/10.1063/1.2436882} {\bibfield
  {journal} {\bibinfo  {journal} {J.~Chem.Phys.}\ }\textbf {\bibinfo {volume}
  {126}},\ \bibinfo {pages} {064102} (\bibinfo {year} {2007})}\BibitemShut
  {NoStop}%
\bibitem [{\citenamefont {Balabanov}\ and\ \citenamefont
  {Peterson}(2005)}]{bala05a}%
  \BibitemOpen
  \bibfield  {author} {\bibinfo {author} {\bibfnamefont {N.~B.}\ \bibnamefont
  {Balabanov}}\ and\ \bibinfo {author} {\bibfnamefont {K.}~\bibnamefont
  {Peterson}},\ }\bibfield  {title} {\enquote {\bibinfo {title} {Systematically
  convergent basis sets for transition metals. {I}. all-electron correlation
  consistent basis sets for the 3d elements sc-zn},}\ }\href
  {https://doi.org/10.1063/1.1998907} {\bibfield  {journal} {\bibinfo
  {journal} {J.~Chem.~Phys.}\ }\textbf {\bibinfo {volume} {123}},\ \bibinfo
  {pages} {064107} (\bibinfo {year} {2005})}\BibitemShut {NoStop}%
\bibitem [{\citenamefont {Schleich}, \citenamefont {Kottmann},\ and\
  \citenamefont {Aspuru-Guzik}(2022)}]{schl22a}%
  \BibitemOpen
  \bibfield  {author} {\bibinfo {author} {\bibfnamefont {P.}~\bibnamefont
  {Schleich}}, \bibinfo {author} {\bibfnamefont {J.~S.}\ \bibnamefont
  {Kottmann}},\ and\ \bibinfo {author} {\bibfnamefont {A.}~\bibnamefont
  {Aspuru-Guzik}},\ }\bibfield  {title} {\enquote {\bibinfo {title} {Improving
  the accuracy of the variational quantum eigensolver for molecular systems by
  the explicitly-correlated perturbative [2]$_{\rm r12}$ correction},}\ }\href
  {https://doi.org/10.1039/d2cp00247g} {\bibfield  {journal} {\bibinfo
  {journal} {Phys. Chem. Chem. Phys.}\ }\textbf {\bibinfo {volume} {24}},\
  \bibinfo {pages} {13550--13564} (\bibinfo {year} {2022})}\BibitemShut
  {NoStop}%
\bibitem [{\citenamefont {Fromager}, \citenamefont {Toulouse},\ and\
  \citenamefont {Jensen}(2007)}]{fromager07}%
  \BibitemOpen
  \bibfield  {author} {\bibinfo {author} {\bibfnamefont {E.}~\bibnamefont
  {Fromager}}, \bibinfo {author} {\bibfnamefont {J.}~\bibnamefont {Toulouse}},\
  and\ \bibinfo {author} {\bibfnamefont {H.~J.~A.}\ \bibnamefont {Jensen}},\
  }\bibfield  {title} {\enquote {\bibinfo {title} {On the universality of the
  long-/short-range separation in multiconfigurational density functional
  theory},}\ }\href {https://doi.org/10.1063/1.2566459} {\bibfield  {journal}
  {\bibinfo  {journal} {J.~Chem.~Phys.}\ }\textbf {\bibinfo {volume} {126}},\
  \bibinfo {pages} {074111} (\bibinfo {year} {2007})}\BibitemShut {NoStop}%
\bibitem [{\citenamefont {Hedeg{\aa}rd}\ \emph {et~al.}(2015)\citenamefont
  {Hedeg{\aa}rd}, \citenamefont {Knecht}, \citenamefont {Kielberg},
  \citenamefont {Jensen},\ and\ \citenamefont {Reiher}}]{hede15b}%
  \BibitemOpen
  \bibfield  {author} {\bibinfo {author} {\bibfnamefont {E.~D.}\ \bibnamefont
  {Hedeg{\aa}rd}}, \bibinfo {author} {\bibfnamefont {S.}~\bibnamefont
  {Knecht}}, \bibinfo {author} {\bibfnamefont {J.~S.}\ \bibnamefont
  {Kielberg}}, \bibinfo {author} {\bibfnamefont {H.~J.~A.}\ \bibnamefont
  {Jensen}},\ and\ \bibinfo {author} {\bibfnamefont {M.}~\bibnamefont
  {Reiher}},\ }\bibfield  {title} {\enquote {\bibinfo {title} {Density matrix
  renormalization group with efficient dynamical electron correlation through
  range separation},}\ }\href {https://doi.org/10.1063/1.4922295} {\bibfield
  {journal} {\bibinfo  {journal} {J.~Chem.~Phys.}\ }\textbf {\bibinfo {volume}
  {142}},\ \bibinfo {pages} {224108} (\bibinfo {year} {2015})}\BibitemShut
  {NoStop}%
\bibitem [{\citenamefont {Rossmannek}\ \emph {et~al.}(2021)\citenamefont
  {Rossmannek}, \citenamefont {Barkoutsos}, \citenamefont {Ollitrault},\ and\
  \citenamefont {Tavernelli}}]{ross21a}%
  \BibitemOpen
  \bibfield  {author} {\bibinfo {author} {\bibfnamefont {M.}~\bibnamefont
  {Rossmannek}}, \bibinfo {author} {\bibfnamefont {P.~K.}\ \bibnamefont
  {Barkoutsos}}, \bibinfo {author} {\bibfnamefont {P.~J.}\ \bibnamefont
  {Ollitrault}},\ and\ \bibinfo {author} {\bibfnamefont {I.}~\bibnamefont
  {Tavernelli}},\ }\bibfield  {title} {\enquote {\bibinfo {title} {Quantum
  {HF}/{DFT}-embedding algorithms for electronic structure calculations:
  Scaling up to complex molecular systems},}\ }\href
  {https://doi.org/10.1063/5.0029536} {\bibfield  {journal} {\bibinfo
  {journal} {J. Chem. Phys.}\ }\textbf {\bibinfo {volume} {154}},\ \bibinfo
  {pages} {114105} (\bibinfo {year} {2021})}\BibitemShut {NoStop}%
\bibitem [{\citenamefont {Garc{\'i}a-P{\'e}rez}\ \emph
  {et~al.}(2022)\citenamefont {Garc{\'i}a-P{\'e}rez}, \citenamefont {Borrelli},
  \citenamefont {Leahy}, \citenamefont {Malmi}, \citenamefont {Maniscalco},
  \citenamefont {Rossi}, \citenamefont {Sokolov},\ and\ \citenamefont
  {Cavalcanti}}]{garc22b}%
  \BibitemOpen
  \bibfield  {author} {\bibinfo {author} {\bibfnamefont {G.}~\bibnamefont
  {Garc{\'i}a-P{\'e}rez}}, \bibinfo {author} {\bibfnamefont {E.-M.}\
  \bibnamefont {Borrelli}}, \bibinfo {author} {\bibfnamefont {M.}~\bibnamefont
  {Leahy}}, \bibinfo {author} {\bibfnamefont {J.}~\bibnamefont {Malmi}},
  \bibinfo {author} {\bibfnamefont {S.}~\bibnamefont {Maniscalco}}, \bibinfo
  {author} {\bibfnamefont {M.~A.~C.}\ \bibnamefont {Rossi}}, \bibinfo {author}
  {\bibfnamefont {B.}~\bibnamefont {Sokolov}},\ and\ \bibinfo {author}
  {\bibfnamefont {D.}~\bibnamefont {Cavalcanti}},\ }\href
  {https://doi.org/10.48550/ARXIV.2207.01360} {\enquote {\bibinfo {title}
  {Virtual linear map algorithm for classical boost in near-term quantum
  computing},}\ } (\bibinfo {year} {2022}),\ \bibinfo {note} {arXiv:
  2207.01360}\BibitemShut {NoStop}%
\bibitem [{\citenamefont {Filippov}\ \emph {et~al.}(2022)\citenamefont
  {Filippov}, \citenamefont {Sokolov}, \citenamefont {Rossi}, \citenamefont
  {Malmi}, \citenamefont {Borrelli}, \citenamefont {Cavalcanti}, \citenamefont
  {Maniscalco},\ and\ \citenamefont {García-Pérez}}]{vilma22a}%
  \BibitemOpen
  \bibfield  {author} {\bibinfo {author} {\bibfnamefont {S.}~\bibnamefont
  {Filippov}}, \bibinfo {author} {\bibfnamefont {B.}~\bibnamefont {Sokolov}},
  \bibinfo {author} {\bibfnamefont {M.~A.~C.}\ \bibnamefont {Rossi}}, \bibinfo
  {author} {\bibfnamefont {J.}~\bibnamefont {Malmi}}, \bibinfo {author}
  {\bibfnamefont {E.-M.}\ \bibnamefont {Borrelli}}, \bibinfo {author}
  {\bibfnamefont {D.}~\bibnamefont {Cavalcanti}}, \bibinfo {author}
  {\bibfnamefont {S.}~\bibnamefont {Maniscalco}},\ and\ \bibinfo {author}
  {\bibfnamefont {G.}~\bibnamefont {García-Pérez}},\ }\href
  {https://doi.org/10.48550/ARXIV.2212.10225} {\enquote {\bibinfo {title}
  {Matrix product channel: Variationally optimized quantum tensor network to
  mitigate noise and reduce errors for the variational quantum eigensolver},}\
  } (\bibinfo {year} {2022}),\ \bibinfo {note} {arXiv: 2212.10225}\BibitemShut
  {NoStop}%
\end{thebibliography}%

\end{document}